\documentclass[aps,pra,pdf,superscriptaddress,twocolumn,showpacs,nofootinbib]{revtex4-2}

\usepackage{amsmath,amssymb,amsfonts}
\usepackage{eqnarray}
\usepackage[newcommands]{ragged2e}
\usepackage{graphicx,caption}
\usepackage{float}
\usepackage{subfig}
\usepackage{bm}
\usepackage{braket}
\usepackage{amsmath}
\usepackage{physics}
\usepackage{mathtools}
\usepackage{graphicx,color,xcolor,colortbl}
\usepackage{tikz}
\usepackage{microtype}
\usepackage[utf8]{inputenc}
\usepackage[title]{appendix}
\usetikzlibrary{shapes}
\usepackage[colorlinks=true,
linkcolor=blue,
filecolor=magenta,      
urlcolor=blue,
citecolor=blue]{hyperref}

\DeclareUnicodeCharacter{2009}{\,}
\definecolor{Gray}{gray}{0.85}
\definecolor{LightCyan}{rgb}{0.25,0.9,0.95}
\definecolor{LightMagenta}{rgb}{0.91,0.21,0.96}

\graphicspath{{images/}}
\makeatletter
\newcommand*{\rom}[1]{\expandafter\@slowromancap\romannumeral #1@}
\makeatother

\begin{document}
\title{{\bf Sinusoidal magnetic field induced topological excitations \\ in a spin-orbit coupled spinor condensate} }
\author{Arpana Saboo}%
\email{arpana.saboo@gmail.com}
\affiliation{%
    Department of Physics, Indian Institute of Technology Kharagpur, Kharagpur, West Bengal 721302, India
}%
\author{Soumyadeep Halder}
\affiliation{%
    Department of Physics, Indian Institute of Technology Kharagpur, Kharagpur, West Bengal 721302, India
}%
\author{Subrata Das}
\affiliation{%
    Department of Physics, Indian Institute of Technology Kharagpur, Kharagpur, West Bengal 721302, India
}%
\affiliation{%
    Department of Physics, Virginia Tech, Blacksburg, Virginia 24061, USA
}%
\author {Sonjoy Majumder}
\email{sonjoym@phy.iitkgp.ac.in}
\affiliation{%
    Department of Physics, Indian Institute of Technology Kharagpur, Kharagpur, West Bengal 721302, India
}%
 
\begin{abstract}
    We explore topological excitations in a spin-1 Bose-Einstein condensate subjected to an in-plane sinusoidally varying magnetic field and Rashba spin-orbit coupling (SOC). In the absence of SOC, the periodic magnetic field induces vortex-anti-vortex structures in the $\ket{F=1, m_F=\pm1}$ condensates at saddle-points, such that the net topological charge remains zero. The introduction of Rashba SOC breaks the system's symmetry, leading to non-conservation of overall angular momentum in the spin-1 condensate. This anisotropy results in the emergence of certain skyrmion spin textures. We provide a comparative study for various in-plane magnetic field configurations while keeping the SOC strength constant. Our numerical simulations within the mean-field framework reveal the potential to engineer diverse topological excitations controlled by the interplay between spin-orbit coupling and in-plane magnetic field in a spinor condensate.
\end{abstract}
\maketitle   
\section{Introduction}    
\label{sec:intro}      
The experimental achievement of introducing spin-orbit coupling (SOC) in ultracold atomic physics  \cite{lin_spinorbit-coupled_2011,wang_spinorbit_2010,Hui_soc_2012, wang_spin-orbit_2012,anderson_synthetic_2012,wu_realization_2016, huang_experimental_2016,campbell_realistic_2011, wu2011unconventional} marks a transformative milestone, opening up new frontiers for exploration in the quantum realm. This groundbreaking progress propels the field into unexplored territories, allowing for precise control and manipulation of the intricate interplay between particle spins and orbital motion \cite{peng_exotic_2020, zhu_spin_2020, wen_ground_2012,  mardonov_dynamics_2015, sun_interacting_2016, su_rashba-type_2016}. Spin-orbit (SO) coupled trapped Bose-Einstein condensates (BECs) typically exhibit a diverse range of exotic quantum states in their order parameter, including vortex lattice \cite{wang_vortex_2017, wang_vortex_2020, liu_vortex_2013}, half-quantum vortex \cite{Hui_soc_2012, ramachandhran_half-quantum_2012, gautam_fractional_2016}, vortex neclace \cite{white_odd-petal-number_2017}, skyrmion \cite{al2001skyrmions, kawakami_stable_2012, luo_three-dimensional_2019, su_crystallized_2012}, domain wall \cite{wang_spinorbit_2010}, solitons \cite{gautam_vortex_2017, gautam_three_2018, adhikari_symbiotic_2021, meng_spin_2022, chai_magnetic_2020}, knots \cite{hall_tying_2016, kawaguchi_knots_2008}, supersolid stripe phase \cite{Ho_bose_2011,li_quantum_2012,li_stripe_2017, zhao_magnetic_2020, geier_dynamics_2023}, SO-induced supersolidity \cite{italo_supersolid_2021, geier_dynamics_2023}, spin and field  squeezing \cite{huang_spin_2015, chen_spin_2020}, etc. Furthermore, recent research has extended into novel directions such as spin-tensor momentum coupling \cite{luo_spin_2017, Hu_topological_2018, lei_symmetry_2020, sun_bright_2020, chen_spin_nematic_2020, lei_unpaired_2022, Qiu_dynamics_2023}, spin-orbital angular momentum coupling \cite{DeMarco_angular_2015, sun_spin_2015, chen_spin_2016,chen_spinorbital_2018, zhang_ground_2019,li_phase_2022, bidasyuk_fine_2022, han_molecular_2022, peng2022spin, NG_spin_2023}, and more. Notably, there have been recent experimental observations of the splitting of vortex cores \cite{xiao_controlled_2021} and the first-order phase separations \cite{li_quantum_2012,martone_Tricriticalities_2016, gui_spin_2023,wang_ground-state_2017}. These topological defects in ultracold condensates play a crucial role in drawing analogies with those found in condensed matter systems.         
      
In this context, spinor condensates \cite{ho_spinor_1998, ohmi_bose_1998, stamper-kurn_optical_1998, Lan_raman_2014, yu_phase_2016, symes_dynamics_2018} emerge as exceptional platforms within ultracold atomic physics, serving as a copious area of research for the investigation of solid-state-like properties \cite{cheng_Localization_2014, oztas_spin_2019}. These remarkable quantum systems, exhibiting a multitude of spin degrees of freedom, provide an intriguing bridge between the microscopic world of ultracold atoms and the condensed matter physics of solids. Spinor condensates offer a controlled environment to explore phenomena such as magnetic ordering \cite{symes_nematic_2017}, topological excitations and defects \cite{Ueda_topological_2014,meng_spin_2022} turbulence \cite{hong_spin_2023,jung_random_2023, saboo_rayleigh_2023}, shedding light on the intricate connection between quantum gases and condensed matter systems \cite{Ueda_topological_2014}. This synergy of ultracold atomic physics and solid-state physics not only enriches our understanding of fundamental physics but also holds the promise of unveiling new quantum phenomena with potential applications in quantum information processing and beyond \cite{BYRNES2015102}.   
   
On the other hand, recent literature shows that in-plane magnetic fields \cite{Yang_2022,yang_dynamics_2022, chen_spin_2021, LIU2024107263, ray_monopoles_2015, luo_tunable_2016, anderson_magnetically_2013, zhao_magnon_2022, simkin_magnetic_1999, chen_anisotropic_2020} can be significantly used in investigating the equisite many-body physics of ultracold atoms. Relevant studies demonstrate that in-plane gradient magnetic fields can be used to achieve SOC and spin Hall states in BECs \cite{oshima_spin_2016}. These magnetic fields play a vital role in creating synthetic non-Abelian gauge fields \cite{gorini_non_abelian_2010, Farajollahpour_synthetic_2020}. Varying magnetic fields can be accredited for the observation of a wide assortment of topological defects, such as magnetic monopoles \cite{ ray_monopoles_2015, chang_method_2002}, quantum knots \cite{hall_tying_2016}, skyrmions \cite{luo_three-dimensional_2019}. The ground state of spinor BECs in the gradient field has been investigated to exhibit the central Mermin-Ho vortex with symmetrical vortex lattices \cite{mermin_circulation_1976, jin_gauge-potential-induced_2019}. The combined effect of SOC and gradient magnetic field on such systems result in skyrmion chains \cite{luo_three-dimensional_2019, zhang_manipulation_2018} in the ground state. 

Previous investigations on SO-coupled spinor BECs have primarily focused on the effects of either a uniform magnetic field or a polarized-gradient magnetic field. However, the influence of other spatially varying magnetic fields have not been addressed adequately. The inclusion of sinusoidal magnetic fields, although less common than uniform or gradient fields, the experiment \cite{campbell_magnetic_2016} provides a unique protocol to induce spatially varying magnetic interactions. These fields can create periodic modulations in the spin-dependent interactions, leading to the emergence of novel quantum states and enhanced control over spin textures.

In this work, we investigate the ground state of a spin-1 BEC subjected to diverse combinations of sinusoidally varying magnetic fields within the $x$-$y$ plane so as to explore the synergistic impact of a periodic magnetic field and the SOC. The intricate interplay between the external magnetic field and SOC yields topological exciations in the spin-1 BEC in the form of exotic vortices and spin textures. We offer a comparative study between the various choices of sinusoidally varying magnetic fields, thereby highlighting the different topological structures so obtained for each case. Our study aims to leverage the sinusoidal magnetic fields to explore and stabilize complex spin textures such as skyrmions, which are topologically protected spin configurations characterized by a whirling pattern of spins. Skyrmions are particularly compelling due to their robustness against perturbations and their potential as information carriers for next-generation spintronic memory and logic computing devices with high-density, offering non-volatile storage, low power consumption \cite{zhou_magnetic_2019,Zhang_magnetic_2023}.
 
This paper is organised as follows: In section \ref{sec:model}, we introduce our model system and discuss its mean-field ground state under different situations for the different combinations of sinusoidally varying in-plane magnetic field, and thereby, casting these into the Gross-Pitaevskii (GP) equations. In section \ref{sec:results}, we discuss the results obtained by numerically solving the GP equations and investigate the topological structures (vortices and spin textures) generated in the system as a consequence of the interplay between the SOC and the oscillating magnetic field with the spin-1 order parameter. In section \ref{sec:conclusion}, we summarize our findings and provide insights for future scope. We discuss the impact of varying the SOC strength in appendix \ref{apx:seca}.

\section{Model}
\label{sec:model}
In the mean-field, we consider a Rashba SO-coupled ferromagnetic spin-1 BEC of $N$ atoms in a quasi-2D harmonic trap in a spatially varying magnetic field. The effective Hamiltonian \cite{kawaguchi_spinor_2012} in the Gross-Pitaevskii form can be given as $H= H_0 + H_{\rm int}$,
  
\begin{equation}
  \label{eqn:1}
 H_0 = \int d\mathbf{r}  \Psi^{\dagger} \left[-\frac{\hbar^2 \nabla^2}{2M} + V(\mathbf{r}) + \nu_{\rm{soc}} + g_F\mu_B \mathbf{B(\mathbf{r}).f} \right]\Psi
\end{equation}
\begin{equation}
  \label{eqn:2}
H_{\rm int} = \int d\mathbf{r}\left( \frac{1}{2}c_0 n^2 + \frac{1}{2}c_2|\mathbf{F}|^2\right)
\end{equation}
where $\Psi =[\Psi_1(\mathbf{r}), \Psi_0(\mathbf{r}), \Psi_{-1}(\mathbf{r})]^T$ with $\mathbf{r}=x\hat{x} + y\hat{y}$ is the spinor wave function which satisfies the normalization criteria, $\int d\mathbf{r}\Psi^{\dagger}\Psi = N$, and $M$ is the atomic mass. $V(\mathbf{r})=\frac{1}{2}M\omega_{\perp}^2(x^2 + y^2)$ is the 2D harmonic trap with $\omega_{\perp}$ being the radial trapping frequency and $l=\sqrt{\hbar/M\omega_{\perp}}$ is the harmonic oscillator length. The quasi-2D geometry is achieved by imposing a tight binding along the axial direction with trap aspect ratio, $ \lambda =\omega_z/ \omega_{\perp} \gg 1$. Here, $n=n_1 + n_0 + n_{-1} = \sum_m \Psi^{\dagger}_m \Psi_m$, $(m=\pm1,0)$ is the atomic density. $\mathbf{F} = (F_x, F_y, F_z)$ represents the spin density vector defined as $F_{\alpha} = \Psi^{\dagger}f_{\alpha}\Psi \quad(\alpha = x,y,z)$ where $\mathbf{f}=(f_x, f_y, f_z)$ is the irreducible representation of the $3 \times 3$ Pauli spin matrices.  $H_{\rm int}$ denotes the standard density-density and spin-exchange interactions characterised by $c_0$ and $c_2$ parameters, respectively. For the ferromagnetic ground state, $c_2 < 0$, while $c_2 > 0$ stands for the polar ground state. The coupling parameters are given as $c_0 = \frac{4\pi \hbar^2(a_0 + 2a_2)}{3M}$ and $c_2 = \frac{4\pi \hbar^2(a_2 - a_0)}{3M}$, where $a_{0}$, $a_2$ are the $s$-wave scattering lengths for the spin channels with total spins $0$ and $2$, respectively.

The Rashba SOC interaction \cite{lin_spinorbit-coupled_2011,Hui_soc_2012} is given as $\nu_{\rm SOC} = \gamma (f_xp_y - f_yp_x)$, where $\gamma$ is the SOC strength and $(p_x,p_y)$ represents the momentum in the quasi-2D space. Note that this SOC is SU(2) type, therefore it cannot couple thr states $\Psi_1$ and $\Psi_{-1}$, directly. In real experiments, the SU(2) SOC Hamiltonian can be achieved by Raman coupling three laser beams of different polarizations and frequencies \cite{mu_realization_2016}. In addition, the external magnetic field is given as $B(\mathbf{r})={(B_x \hat{x} - B_y \hat{y})}$ in the 2D plane. Here, $B_x, B_y$ are spatially varying magnetic fields along the $x$ and $y$ directions. We absorb the negative sign in the direction and denote $B(\mathbf{r}) = {(B_x, B_y)}$. The Lande's g-factor $g_F=-\frac{1}{2}$ and $\mu_B$ is the Bohr magneton.

The ground state and dynamics of the system can be described by the following non-dimensionalised coupled GP equations,
\begin{align}
	\label{eqn:3}
  {i}\frac{\partial{\psi_1}}{\partial{t}} & =\bigg[ -\frac{1}{2}\nabla^2 + V + c_0|\psi|^2 + c_2\bigg(|\psi_1|^2 + |\psi_0|^2  \nonumber \\ &- |\psi_{{-}1}|^2\bigg)\bigg]\psi_1 + c_2\psi_{{-}1}^*\psi_0^2 + ({B_x} + {i}B_y)\psi_0 \nonumber \\ &-{i}{\gamma}\left(\frac{\partial}{\partial y} + {i}\frac{\partial}{\partial x}\right)\psi_0
\end{align}
\begin{align} 
	\label{eqn:4}
  i\frac{\partial{\psi_0}}{\partial{t}} & =\left[ -\frac{1}{2}\nabla^2 + V + c_0|\psi|^2 + c_2\left(|\psi_1|^2 + |\psi_{{-}1}|^2\right)\right]\psi_{0} \nonumber \\ &+ 2c_2\psi_{1}\psi_0^*\psi_{{-}1}   + (B_x - iB_y)\psi_1 +(B_x + iB_y)\psi_{{-}1}  \nonumber \\ &-i{\gamma}\left(\frac{\partial}{\partial y} - i\frac{\partial}{\partial x}\right)\psi_1 -i{\gamma}\left(\frac{\partial}{\partial y} + i\frac{\partial}{\partial x}\right)\psi_{{-}1}
\end{align}
\begin{align}
	\label{eqn:5}
  i\frac{\partial{\psi_{\rm{-}1}}}{\partial{t}} & = \bigg[ -\frac{1}{2}\nabla^2 + V + c_0|\psi|^2 + c_2\bigg(-|\psi_1|^2 + |\psi_0|^2   \nonumber \\ &+ |\psi_{{-}1}|^2\bigg)\bigg]\psi_{{-}1} + c_2\psi_{1}^*\psi_0^2  + ({B_x} - iB_y)\psi_0   \nonumber \\ &-i{\gamma}\left(\frac{\partial}{\partial y} - i\frac{\partial}{\partial x}\right)\psi_0
\end{align}
where $\psi_m= N^{-1/2}l\Psi_m \quad (m=\pm 1,0)$ denotes the dimensionless $m$-th component wave function and $|\psi|^2 = \sum_{m} |\psi_m|^2$ is the total particle density confined in dimensionless external potential $V= (x^2 + y^2)/2$. The contact interaction parameters in the dimensionless form are given as $c_0= 2N\sqrt{2\pi \lambda}(a_0 + 2a_2)/3l$ and $c_2= 2N \sqrt{2\pi \lambda}(a_2 - a_0)/3l$, respectively. In our numerical calculations, the lengths, time, energy (interaction and SOC) and magnetic field are measured in the units of $l, 1/\omega_{\perp}, \hbar\omega_{\perp}$ and $(g_F \mu_B l)/\hbar \omega_{\perp}$, respectively.

The vorticity of the $m$-th component can be measured as \cite{andersen_quantized_2006, ruben_vortex_2008} 
\begin{equation}
  \label{eqn:6}
  \Omega_m = \nabla \times J_m, \quad m=\pm1, 0
\end{equation}
with the probability current density,
\begin{equation}
  \label{eqn:7}
  J_m= \frac{{i}\hbar}{2 M} (\psi_m \nabla \psi^*_{m} - \psi^*_m \nabla \psi_{m}).
\end{equation}

The topological excitation of the system arising resulting in the formation of vortices, is characterised by the spin texture \cite{liu_circular-hyperbolic_2012}. In a spinor BEC, the spin texture describes the spatial variation of the spin degrees of freedom across the condensate, ranging from complex configurations like skyrmions and spin vortices. Grasping these spin textures is key to delving into the intricate physics of spinor BECs. For a spin-1 BEC, the spin texture can be written as \cite{Takeshi_coreless_2004, Kasamatsu_spin_2005},
\begin{equation}
	\label{eqn:8}
  \mathbf{S_{\alpha}} = \sum_{m,n=0,\pm 1} \psi^*_m (f_{\alpha})_{m,n}\psi_n/|\psi|^2 \quad (\alpha=x,y,z).
\end{equation} 

The topological charge density which characterises the topological spin structure of the system is given as,
\begin{equation}
	\label{eqn:9}
	q(\mathbf{r}) = \frac{1}{4\pi} \mathbf{s} \cdot \left(\frac{\partial \mathbf{s}}{\partial x} \times \frac{\partial \mathbf{s}}{\partial y} \right).
\end{equation}
Here $\mathbf{s}= \mathbf{S/|S|}$  and the total topological charge $Q$ is defined as 
\begin{equation}
  \label{eqn:10}
	Q =\int q(\mathbf{r}) dx dy.
\end{equation}

The topological charge $|Q|$ is unchanged, however one exchanges the components ${\mathbf{S}_x, \mathbf{S}_y, \mathbf{S}_z}$. 

\section{Results and Discussions}
\label{sec:results}
We investigate the ground state structures of the spin-orbit coupled ferromagnetic spin-1 BEC under the influence of external in-plane magnetic field.  In this section, we first discuss the effect of a polarized gradient magnetic field in the $x$ or $y$ direction. We show the ground state density profiles and the spin textures for cases both with and without SOC. Thereafter, we discuss cases for sinusoidally varying in-plane magnetic fields.  Further, we present the ground state structures with and without SOC for each combination of external magnetic field $B(\mathbf{r})={(B_x, B_y)} = {B_0 (f(x), g(y))}$, where ${B_0}$ is the strength of external magnetic field in dimensionless units. 

We offer a detailed numerical study to analyze how the system parameters affect the ground state structures. The richness of the present system lies in the large number of tunable parameters, which includes the spin-independent and spin-exchange contact interactions $c_0$ and $c_2$ respectively, SOC strength, $\gamma$ and magnetic field strength, ${B_0}$. Here, we have considered $10^4$ number of $\rm{^{87}{Rb}}$ atoms confined in a quasi-2D harmonic trap $V(\mathbf{r})$ with transverse trapping frequency, $\omega_{\perp}=2\pi \times 20$ Hz with trap aspect ratio, $\lambda = \omega_z/\omega_{\perp} = 20$ \cite{lim_sumerical_2008} and the characteristic oscillator length, $l=2.41\mu m$. The $s$-wave scattering lengths for channels of total spin 0 and 2 are  $a_0 = 101.8  a_B$ and $a_2 = 100.4 a_B$, respectively, where $a_B$ is the Bohr radius. We keep the magnetic field strength ${B_0} =0.25$ for all choices of magnetic fields. And for all the cases with SOC, we keep the SOC strength fixed at $\gamma=0.5$. With these parameters, using the split-step Crank-Nicolson scheme \cite{crank_nicolson_1947,ANTOINE20132621,muruganandam_2009_fortranprogramstimedependent}, we numerically solve the governing GP equations (\ref{eqn:3}-\ref{eqn:5}) for imaginary time propagation to generate the ground state solutions in the presence of various in-plane polarized and cross-polarized magnetic fields for a fixed SO coupling strength. All of our simulations run from a spatial extent of $-20 l$ to $20 l$ in both $x$ and $y$ directions with $2001 \times 2001$ grid points. The employed spatial discretization refers to $\Delta x = \Delta y = 0.02 l$, with a time step $\delta t = (2 \times 10^{-4})/\omega_{\perp}$. 

\subsection{In-plane polarized gradient magnetic field } 
\label{subsection:gradient} 
\begin{figure}[t]
  \centering
  \includegraphics[width=0.48\textwidth]{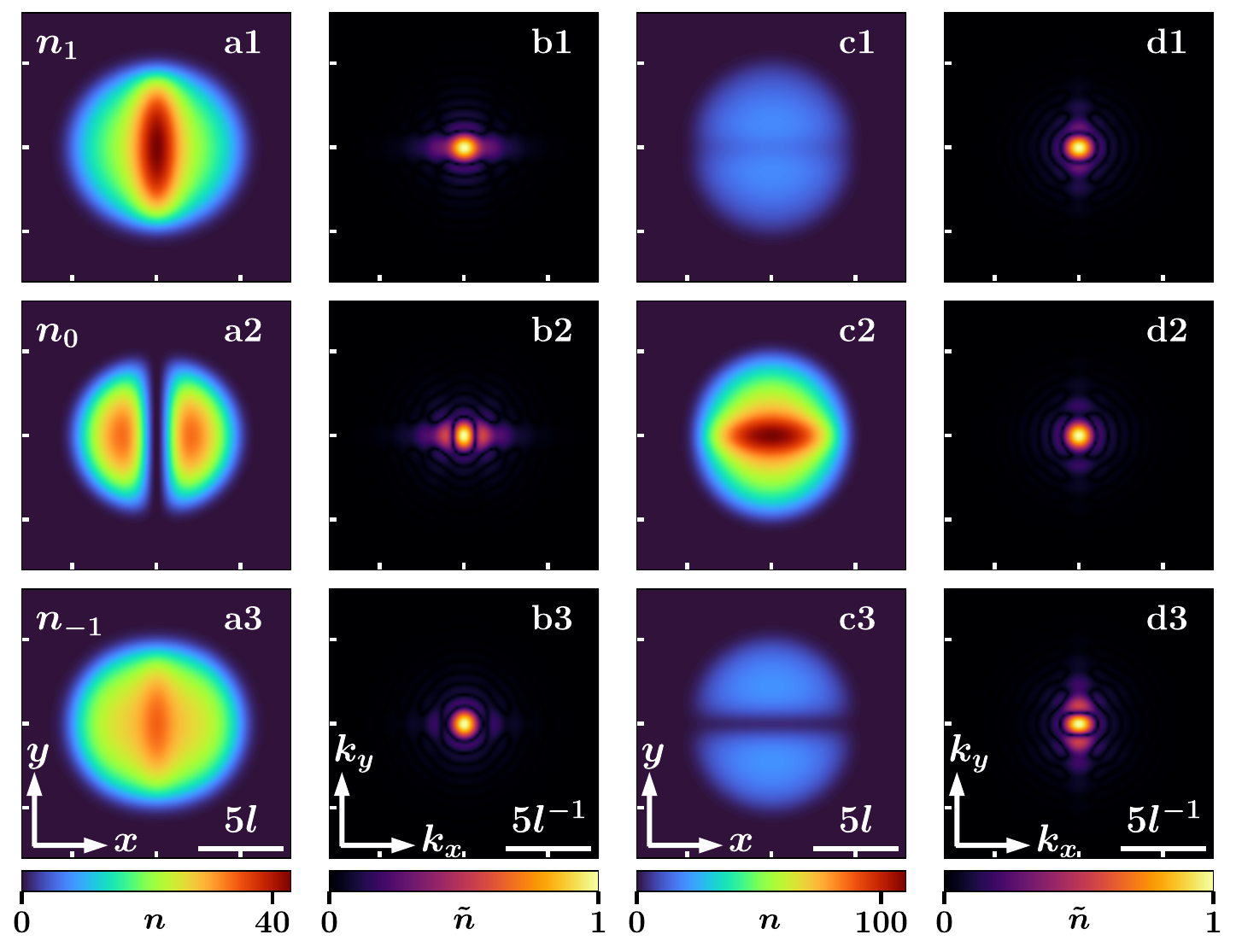}
  \caption{The ground state densities $n$ and momentum distributions $\tilde{n}$ for the spin-1 BEC of $10^4$ $\rm{^{87}{Rb}}$ atoms without SOC. The first and second columns show the density (a1-a3) and momentum (b1-b3) distributions of the $m_F=1$, $m_F=0$ and $m_F=-1$ components for the $x$-polarised magnetic field $B=B_0 x$. The third and fourth columns represent the density (c1-c3) and momentum (d1-d3) distributions for the $y$-polarised magnetic field $B=B_0 y$. The magnetic field strength $B_0 = 0.25$ for both cases. The momentum space densities have been normalised by dividing the maximum value. The densities $n$ and $\tilde{n}$ are expressed in units of $l^{-2}$ and $l^2$, respectively, where $l=2.41 \mu m$ is the characteristic length.}
  \label{fig:1}
\end{figure} 
In this section, we consider a polarized gradient magnetic field either in $x$ or $y$-direction,  \textit{i.e.,} $x$-polarised ${B_x=B_0 x}$ or $y$-polarised ${B_y= B_0 y}$. To ensure that the magnetic field is divergenceless,
\begin{figure}[t] 
  \centering
  \includegraphics[width=0.48\textwidth]{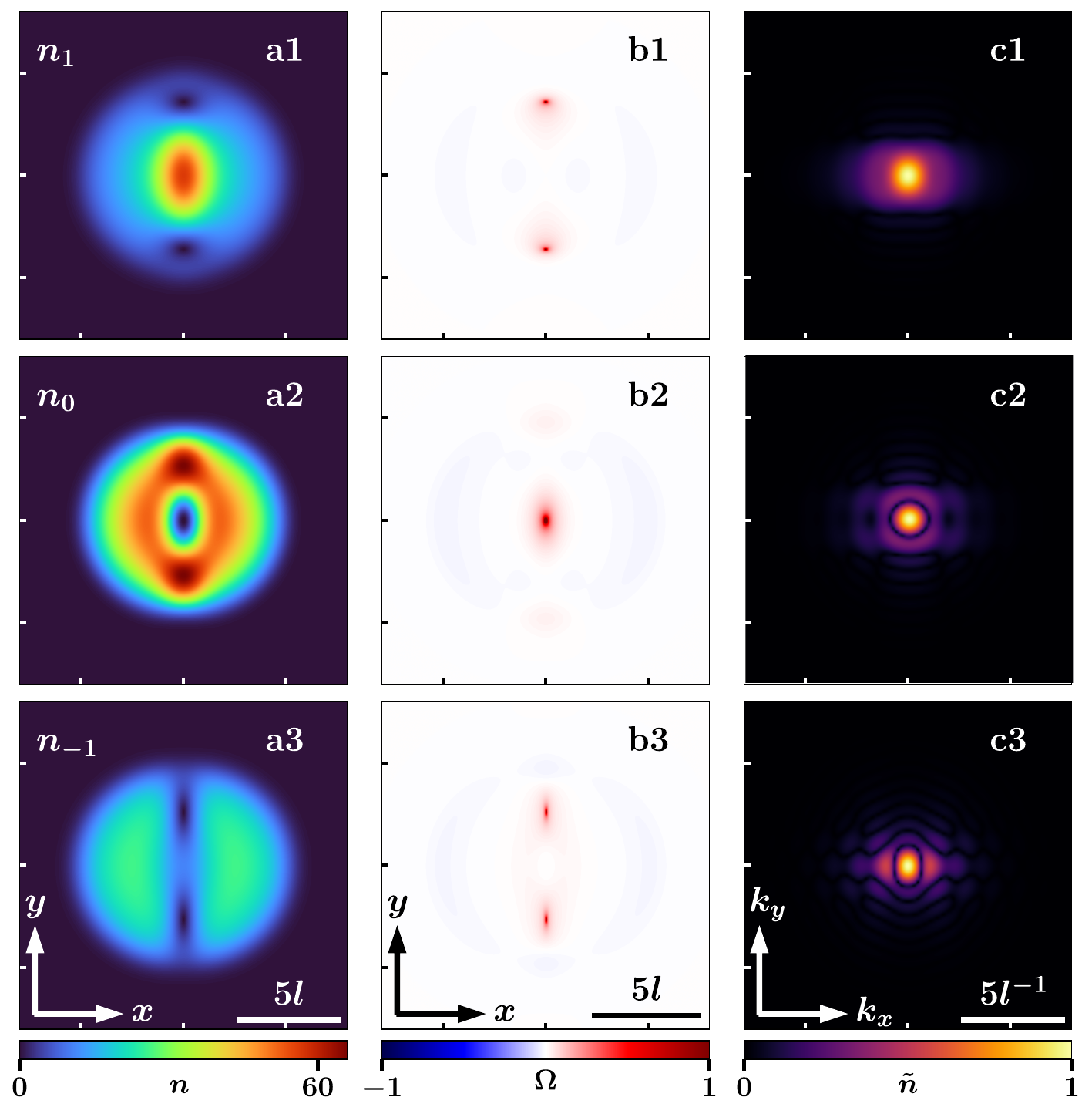}
  \caption{The ground state structures for the spin-orbit coupled spin-1 BEC of $10^4$ $\rm{^{87}{Rb}}$ atoms with SOC strength $\gamma=0.5$ in the the $x$-polarised magnetic field $B=B_0 x$ with strength $B_0 = 0.25$. The first column (a1-a3) shows the density distributions $n$, second column (b1-b3) shows the vorticity profiles and the third column (c1-c3) shows the momentum distributions $\tilde{n}$ of the $m_F=1$, $m_F=0$ and $m_F=-1$ components. The momentum space densities have been normalised by dividing the maximum value. The densities $n$ and $\tilde{n}$ are expressed in units of $l^{-2}$ and $l^2$, respectively, where $l=2.41 \mu m$ is the characteristic length. The vorticity $\Omega$ is expressed in dimensionless units.}
  \label{fig:2}
\end{figure}  
\noindent an additional magnetic field along the $z$-direction, \textit{i.e.,} $B_z = -B_0 z $ is required. However, as we ensure a very tight binding along the $z$-direction, the magnetic field along this direction will not affect the quasi-2D condensates. The motivation to explore this case is to underline the impact of such direction-dependent in-plane field on the SO-coupled BEC. To begin with, we first study the case in the absence of SOC. The ground state density distributions $n_1=|\psi_1|^2$, $n_0=|\psi_0|^2$ and $n_{-1}=|\psi_{-1}|^2$ for the $x-$polarised magnetic field are shown in Fig. \ref{fig:1} (a1-a3) and that for the $y-$polarised case are shown in Fig. \ref{fig:1} (c1-c3). In the absence of SOC, the density $n_0$ is  fragmented into two halves about the $x$-axis with both fragments being phase coherent individually, while the densities $n_1$ and $n_{-1}$ have a overall disk-shaped condensate with high densities at the $x=0$ line. However, for the $y$-polarised case, the density modulations are observed along the $y=0$ line. Due to the intricate interplay between the spin-$1$ order parameter and the magnetic field bias, the density profiles $n_{\pm 1}$ and $n_0$ show reverse features, \textit{i.e.,} the central high density peak is observed in $n_0$ while the fragments are observed in $n_1, n_{-1}$. In the absence of SOC, there are no vortices formed in the system as the polarised gradient magnetic field doesn't have any saddle-point. However, the momentum space density profiles (as shown in Fig. \ref{fig:1} (b1-b3) and Fig. \ref{fig:1} (d1-d3) for the $x$-polarised and $y$-polarised case, respectively), emphasizes the unique structure of the condensates owing to the direction-dependent density modulations. These density modulations can be associated with the momentum distribution. Along with the central peak at $k_x, k_y = 0$, there exists density peaks at some higher $k_x$ value for the $x$-polarised magnetic field and similarly, some higher $k_y$ value for the $y$-polarised case. These $k_i$ values could be associated to a density wave with wavelength $\lambda_i = 2\pi/ k_i, \quad i=x, y$. In other words, extra peaks in momemtum distribution at $k_x, k_y \neq 0$ corresponds to density modulations in the real space, such that the wavelength $\lambda_i$ of this modulation is $\propto 1/k_i$.

Now, with SOC in play, our investigation centers to unravel the competition between the SOC and external magnetic field. Depending on the choices of the external magnetic field applied, the ground state structures for the SO-coupled spin-1 BEC are different. We show the SO-coupled ground state structures for the $x$-polarised magnetic field in Fig. \ref{fig:2}. The ground state densities $n_{\pm 1}, n_0$ exhibit modulations along the $x=0$ line as shown in Fig. \ref{fig:2} (a1-a3).
\begin{figure}[t]
  \centering
  \includegraphics[width=0.48\textwidth]{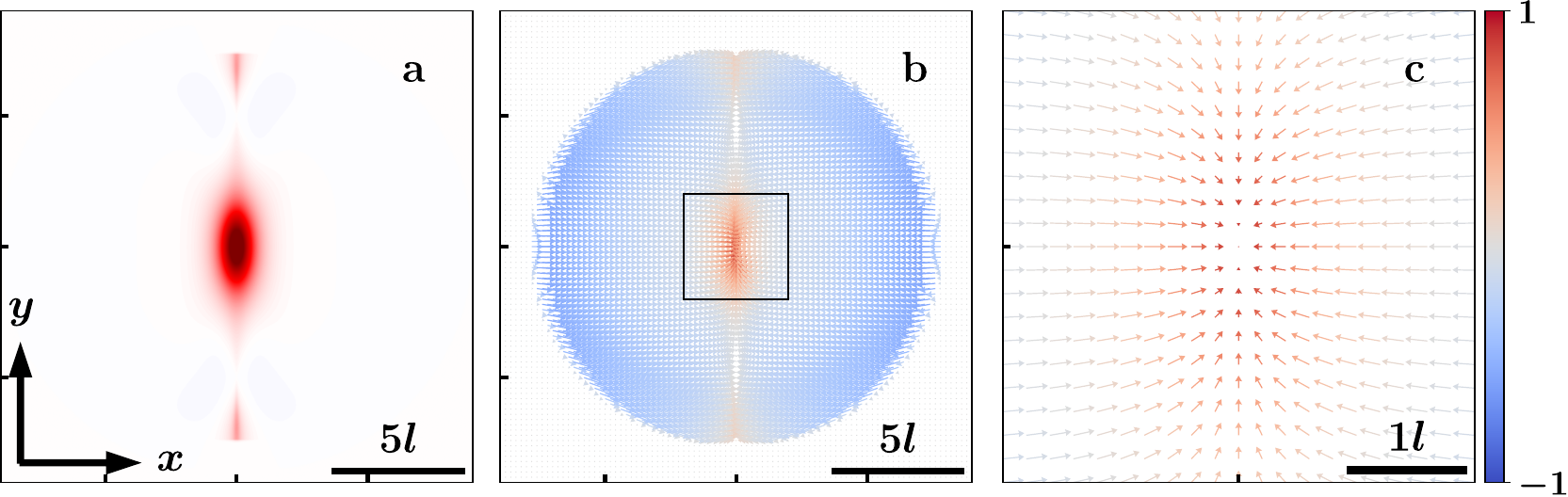}
  \caption{(a) shows the topological charge distribution, (b) shows the spin texture of the system and (c) shows the local amplification of the spin texture, marked by the black square in (b) for the SO-coupled spin-1 BEC shown in Fig. \ref{fig:2}. The color of the arrows represent the $S_z$ values.  Units are dimensionless.} 
  \label{fig:3}
\end{figure}
\noindent  Our numerical simulations show that the formation of vortices in the three components is pertinent to the interplay of SOC and the external magnetic field. As the SOC couples the intrinsic spin to the particles' motion, it imparts some extra angular momentum to the system which results in the formation of vortices in the system. The SOC influences with the magnetic field in such a way that the effective magnetic field is modified. In addition to this, as there are no saddle-points in the magnetic field, these vortices appear at the $x = 0$ line under the influence of SOC. The vorticity profile associated to each component is shown in Fig. \ref{fig:2} (b1-b3). Due to the anisotropic nature of the SOC, the vortices are slightly deformed along the $y$-direction as the component densities are modulated along the $x=0$ line. The k-space density profiles (as shown in Fig. \ref{fig:2} (c1-c3)) depict that apart from the central peaks, some extra density peaks predominantly along the $k_x$-direction are observed. These extra peaks at $k_x \neq 0$ corresponds to the density modulations along the $x$-direction. This is due to the fact that SOC influences the distribution of particles in the k-space, similar to the impact of SOC as seen in the distribution of electronic states as a function of momentum in solid-state physics \cite{winkler_book_2003}.

The combined effect of the in-plane direction dependent gradient field and the SOC yield intricate spin textures in the spin-1 BEC. The topological charge distribution $q(\mathbf{r})$ and the corresponding spin texture is shown in Fig. \ref{fig:3} (a) and Fig. \ref{fig:3} (b), respectively. We can identify the presence of a stable spin configuration characterized by a converging pattern centred at a point carrying +ve charge, called skyrmion, from the spin texture. The local amplification of the skyrmion spin texture is shown in Fig. \ref{fig:3} (c). The spin texture is elongated along the $x=0$ line due to the SOC-induced anisotropy in the system.
\begin{figure}[t]
  \centering
  \includegraphics[width=0.48\textwidth]{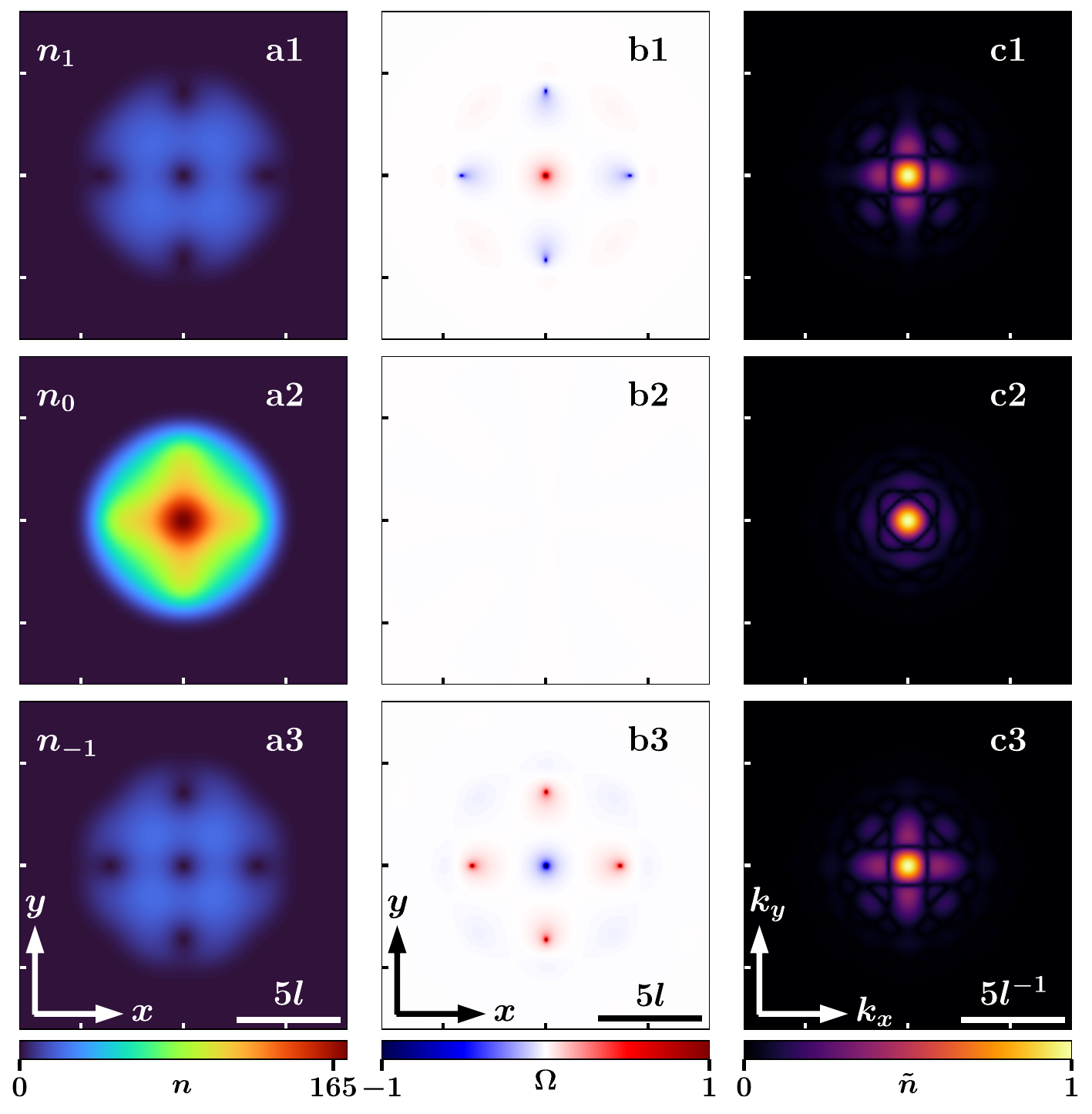}
  \caption{The ground state structures for spin-1 BEC of $10^4$ $\rm{^{87}{Rb}}$ atoms without SOC in the \textit{sine-sine} magnetic field $B_x =B_0 \sin x$ and $B_y = B_0 \sin y$ with strength $B_0 = 0.25$. The first column (a1-a3) shows the density distributions, second column (b1-b3) shows the vorticity profiles and the third column (c1-c3) shows the momentum distributions of the $m_F=1$, $m_F=0$ and $m_F=-1$ components. The momentum space densities have been normalised by dividing the maximum value.  The densities $n$ and $\tilde{n}$ are expressed in units of $l^{-2}$ and $l^2$, respectively, where $l=2.41 \mu m$ is the characteristic length. The vorticity $\Omega$ is expressed in dimensionless units.}
  \label{fig:4}
\end{figure}  
\noindent

For the SO-coupled spin-1 BEC in $y$-polarised gradient magnetic field, we can analogously understand the density modulations along the $y=0$ line for all the three components. However, these density modulations converge into the formation of anti-vortices in the system, and the corresponding spin textures have anti-skyrmions along the $x$-axis. It is interesting to note that the state which had fragments in the absence of SOC, now has more vortices or anti-vortices as we observe for the case with SOC. The number of vortices increases as we increase the SOC strength $\gamma$ (see appendix \ref{apx:seca}). However, for the polarised magnetic field, they appear in the same alignment, with the vortex spacing being reduced and more deformed. 

\begin{figure}[t]
  \centering
  \includegraphics[width=0.48\textwidth]{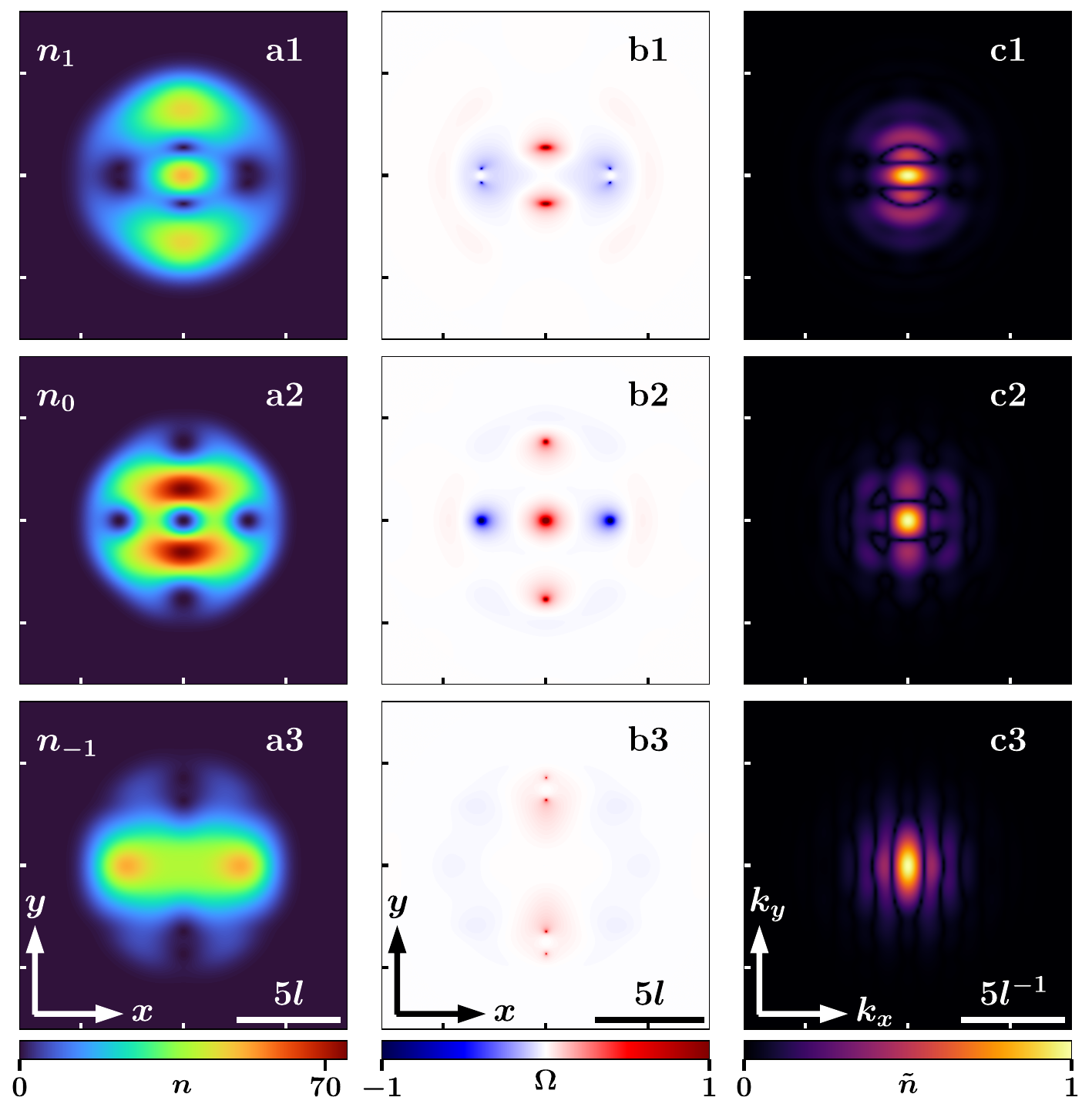}
  \caption{The ground state structures for spin-orbit coupled spin-1 BEC of $10^4$ $\rm{^{87}{Rb}}$ atoms with SOC strength $\gamma=0.5$ in the \textit{sine-sine} magnetic field $B_x =B_0 \sin x$ and $B_y = B_0 \sin y$ with strength $B_0 = 0.25$. The first column (a1-a3) shows the density distributions, second column (b1-b3) shows the vorticity profiles and the third column (c1-c3) shows the momentum distributions of the $m_F=1$, $m_F=0$ and $m_F=-1$ components. The momentum space densities have been normalised by dividing the maximum value. The densities $n$ and $\tilde{n}$ are expressed in units of $l^{-2}$ and $l^2$, respectively, where $l=2.41 \mu m$ is the characteristic length. The vorticity $\Omega$ is expressed in dimensionless units.}
  \label{fig:5} 
\end{figure}

\subsection{Cross-polarised sinusoidally varying in-plane magnetic fields}  
\label{subsection:cross-field}
Thus far, we have discussed how impactful the polarised gradient field can be on the ground state structure of a SO-coupled spin-1 BEC. To further exploit this direction-dependent interplay of external magnetic field and SOC, we now choose sinusoidally varying in-plane magnetic fields in both $x$- and $y$-directions. In general, to make the magnetic field divergenceless, one has to apply a magnetic field $B_z = -B_0 z f(x,y)$ in the $z$-direction, where $f(x,y)$ are sinusoidal functions in the $x$-$y$ plane. However, as we consider a quasi-2D geometry, for tight binding along the $z-$direction, $B_z$ will not affect the spinor condensate. In this section, we center our focus only on the in-plane sinusoidal magnetic field. Experimentally, a polarised sinusoidal field can be realised using a similar technique discussed in \cite{luo_three-dimensional_2019}. Campbell et al., $2016$ has also demonstrated a similar in-plane sinusoidal field in their work \cite{campbell_magnetic_2016}. We discuss below various combinations of $B(\mathbf{r}) = (B_x, B_y)$ in different cases.  

\begin{figure}[t]
	\centering
	\includegraphics[width=0.48\textwidth]{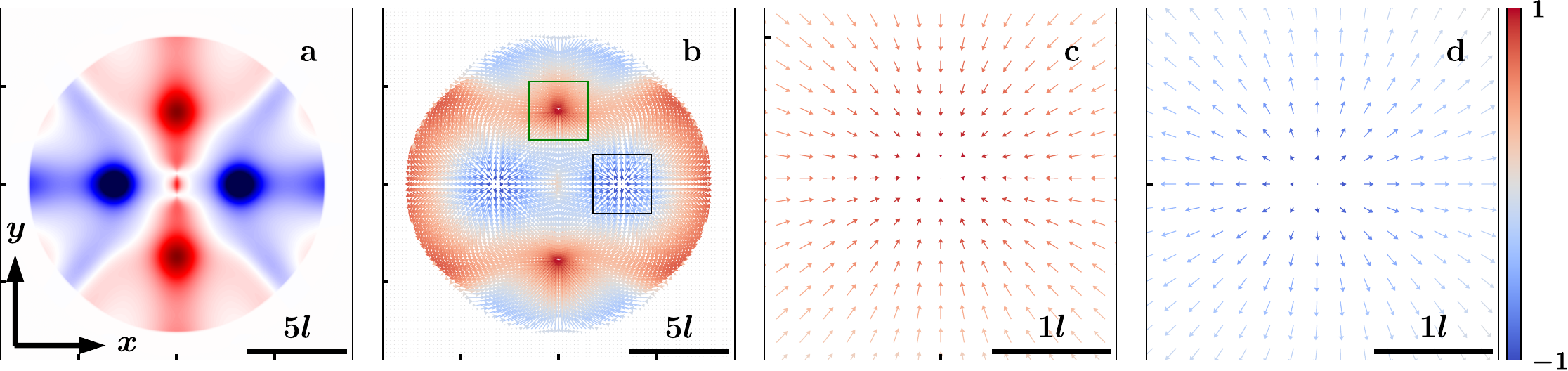}
	\caption{(a) shows the topological charge distribution, (b) shows the spin texture of the system and (c),(d) shows the local amplification of the spin texture, marked by green and black squares, respectively, for the SO-coupled spin-1 BEC shown in Fig. \ref{fig:5}. The color of the arrows represent the $S_z$ values.  Units are dimensionless.}
	\label{fig:6}
\end{figure}

\subsubsection{Case-\rom{1}: Sine-Sine magnetic field ${B}(\mathbf{r}) = B_0( \sin x, \sin y)$ } 
\label{subsubsection:case1} 
For the present case, we choose the magnetic fields as sine functions in both directions, \textit{i.e.,} ${B_x}= B_0 \sin x$ and ${B_y}= B_0 \sin y$. We first present the ground state structures for the no SOC case in Fig. \ref{fig:4}. The density profiles $n_1, n_0$ and $n_{-1}$ of the condensates $m_F=1, m_F=0$ and $m_F=-1$, respectively are shown in Fig. \ref{fig:4} (a1-a3). The \textit{sine-sine} type of magnetic field results in the density modulations in $m_F = \pm 1$ components with the appearance of topological exciations in the form of a triad of alternating vortex-anti-vortex along both $x$- and $y$-axes. This is due to the fact that the applied magnetic fields from both directions generate certain saddle-point structures such that the nature of the triads in $m_F=1$ and $m_F=-1$ components are complementary to each other. The contrasting nature of the triads is due to the fact that the magnetic field appears as conjugate for the $m_F=1$ and $m_F=-1$ components in the GP equations (\ref{eqn:3}-\ref{eqn:5}). Furthermore, combining with the spin-mixing term, the magnetic field yields vortex brights in the $m_F = 0$ component at the loci (saddle-points of $B(\rm{r})$) of vortices in the $m_F = \pm 1$ components so as to conserve the overall angular momentum of the spin-1 BEC in the absence of SOC. Thus, the $m_F=0$ component has no vortices, instead a high density peak is observed at the center of the condensate.  The vorticity profiles associated to the condensate densities are depicted in Fig. \ref{fig:4} (b1-b3). The k-space density profiles (see Fig. \ref{fig:4} (c1-c3)) support observed density modulations in the system. For the $m_F=1$ and $m_F=-1$ components, apart from the central peak, we observe some additional density peaks both in the $k_x$ and $k_y$ directions which correspond to the density modulations with wavelength inversely $\propto k_x, k_y$. In the absence of SOC, the total angular momentum of the spin-1 system remains conserved, resulting in a topological charge of $|Q| = 0$. Consequently, vortices or anti-vortices within the system do not play an active role in shaping the spin texture.

\begin{figure}[t]
	\centering
	\includegraphics[width=0.48\textwidth]{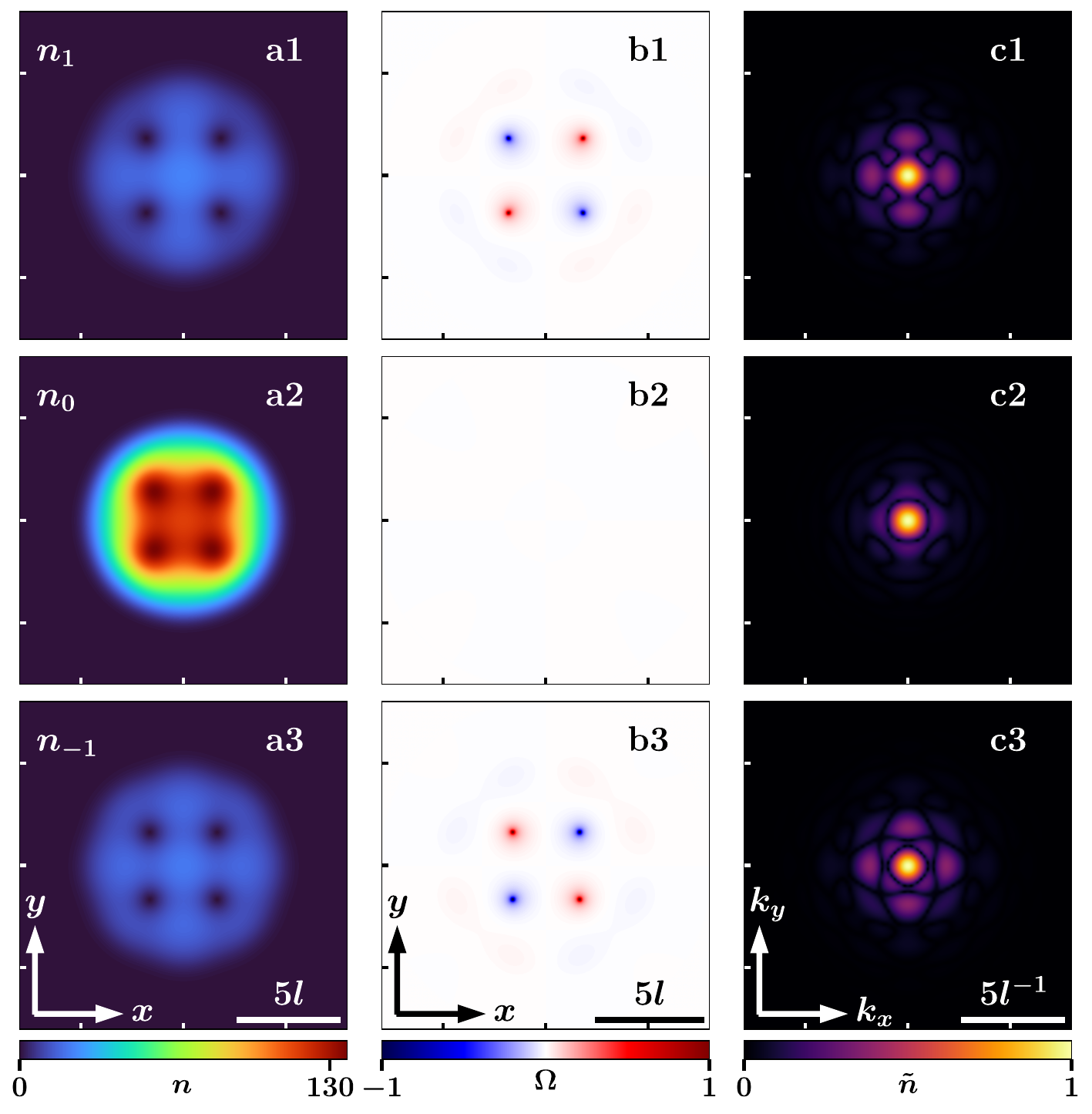}
	\caption{The ground state structures for spin-1 BEC of $10^4$ $\rm{^{87}{Rb}}$ atoms without SOC in the \textit{cosine-cosine} magnetic field $B_x =B_0 \cos x$ and $B_y = B_0 \cos y$ with strength $B_0 = 0.25$. The first column (a1-a3) shows the density distributions, second column (b1-b3) shows the vorticity profiles and the third column (c1-c3) shows the momentum distributions of the $m_F=1$, $m_F=0$ and $m_F=-1$ components. The momentum space densities have been normalised by dividing the maximum value.  The densities $n$ and $\tilde{n}$ are expressed in units of $l^{-2}$ and $l^2$, respectively, where $l=2.41 \mu m$ is the characteristic length. The vorticity $\Omega$ is expressed in dimensionless units.}
	\label{fig:7}
\end{figure}

In the presence of SOC, the ground state structures emerge as a result of the interplay among the contributions from the magnetic field, spin-orbit coupling, and the spin-mixing dynamics. We show the density profiles of the $m_F=1, m_F=0$ and $m_F=-1$ components in Fig. \ref{fig:5} (a1-a3). The system experiences an induced anisotropy as a result of rotational symmetry breaking due to SOC. The vorticity profiles (see Fig. \ref{fig:5} (b1-b3)) depict the existence of vortices, anti-vortices in the spin-1 condensate. Due to the induced anisotropy, the vortices or anti-vortices are not only deformed in shape, but also are biased in opposite axes $x$ and $y$, for the $m_F=1$ and $m_F=-1$ components, respectively. In other words, for this case the vortices and anti-vortices are respectively nucleated at the two orthogonal axes. The $m_F=0$ component experiences density modulations in both directions as a consequence of the spin-exchange interaction, SOC and external magnetic field. The k-space density structures (as shown in Fig.\ref{fig:5} (c1-c3)) have extra peaks in the preferred direction of high spatial densities, in addition to the anisotropically elongated central peak in the $m_F = \pm 1$ components. Analogous to the previously discussed situations, here also we can associate a density modulation, arising due to the combined outcome of SOC and the \textit{sine-sine} magnetic field, with wavelength $\lambda_i$ for additional momentum peak at $k_i \neq 0, \quad i=(x,y)$.

We show for the above discussed case, the topological charge distribution in Fig. \ref{fig:6} (a) and associated spin texture in Fig. \ref{fig:6} (b). The overall spin texture has both anti-skyrmions \footnote{Here, we characterise anti-skyrmions on the basis of their $S_z$ values and the -ve topological charge they carry.} and skyrmions along the two orthogonal axes $x$ and $y$, respectively. Fig. \ref{fig:6} (c-d) are the local amplification of the spin textures showing a skyrmion and anti-skyrmion, respectively. The spin textures arise in the system due to the SOC, however the arrangement of various spin structures depend on the choice of external in-plane magnetic field.

\begin{figure}[t]  
  \centering
  \includegraphics[width=0.48\textwidth]{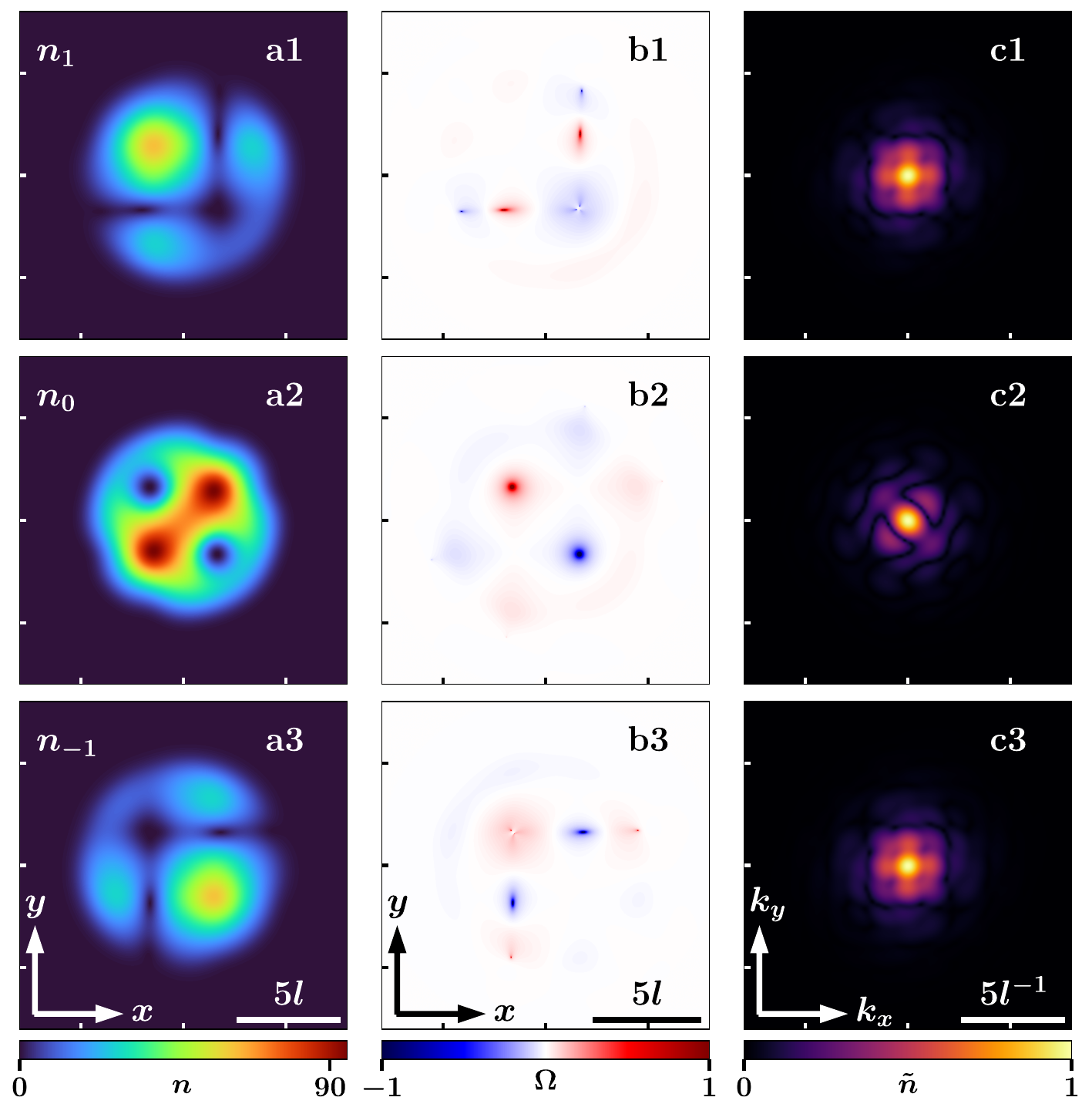}
  \caption{The ground state structures for spin-orbit coupled spin-1 BEC of $10^4$ $\rm{^{87}{Rb}}$ atoms with SOC strength $\gamma=0.5$ in the \textit{cosine-cosine} magnetic field $B_x =B_0 \cos x$ and $B_y = B_0 \cos y$ with strength $B_0 = 0.25$. The first column (a1-a3) shows the density distributions, second column (b1-b3) shows the vorticity profiles and the third column (c1-c3) shows the momentum distributions of the $m_F=1$, $m_F=0$ and $m_F=-1$ components. The momentum space densities have been normalised by dividing the maximum value.  The densities $n$ and $\tilde{n}$ are expressed in units of $l^{-2}$ and $l^2$, respectively, where $l=2.41 \mu m$ is the characteristic length. The vorticity $\Omega$ is expressed in dimensionless units.}
  \label{fig:8}  
\end{figure}

\subsubsection{Case-\rom{2}: Cosine-Cosine magnetic field ${B}(\mathbf{r}) = B_0( \cos x, \cos y)$ }
\label{subsubsection:case2} 

Here, we choose the magnetic fields as cosine functions in both directions, \textit{i.e.,} ${B_x}= B_0 \cos x$ and ${B_y}= B_0 \cos y$. The component-wise density profiles of the ground state due to the magnetic field independent of the SO-coupling are shown in Fig. \ref{fig:7} (a1-a3). Comparable to the \textit{sine-sine} magnetic field case, here also  a certain stable saddle-point structure is formed due to the interaction of periodic magnetic fields in both directions. 
Due to the presence of a saddle-point structure, there emerges a configuration wherein an equal number of vortices and anti-vortices are formed within the $m_F=1$ and $m_F=-1$ components, respectively. This arrangement ensures the preservation of angular momentum conservation within the spin-1 system. The spin-exchange interaction along with the effect of magnetic field, yields  vortex brights in the $m_F=0$ component. These local high densities in the $m_F=0$ component compensate for the vortices or anti-vortices in the adjacent $m_F=\pm1$ components. The vorticity profiles associated with the component densities are presented in Fig. \ref{fig:7} (b1-b3), which clearly explains the angular momentum conservation in the system. The topological excitations arise in the system in the form of exotic vortex-anti-vortex configuration at the saddle-points of the external magnetic field. The density modulations due to such exotic structures can be explained by the k-space density profiles (see Fig. \ref{fig:7} (c1-c3)). For this case also, apart from the central peak, there exists extra density peaks in the momentum space both in $k_x, k_y$ conforming the density modulations with wavlength inversely proportional to non-zero momentum peaks in both directions. The topological charge is zero for the non SOC case to ensure the angular momentum conservation.

However, the coopetition between the \textit{cosine-cosine} magnetic field and the SOC results in the re-distribution of particles in the system. The density profiles of the SO-coupled spin-1 BEC are shown in Fig. \ref{fig:8} (a1-a3). Trivially, the spin-mixing interaction ensures that the density profiles of $m_F=1$ and $m_F=-1$ components complement each other, with the presence of vortices in one at the loci of high density for the other component. Moreover, the vortices in one are compensated by the anti-vortices in the other. The $m_F=0$ component experiences topological excitations in the form of vortices arranged at the diagonal saddle-points along with local high densities at the opposite diagonal. Also, the criss-cross vortex and vortex-brights in the $m_F=0$ component complements the local high densities and vortices in the $m_F = \pm 1$ components. The vorticity profiles as shown in Fig. \ref{fig:8} (b1-b3) give a clear picture about the topological excitations in the system. Remarkably, the magnetic field influences the vortices in the $m_F=0$ component to have opposite vorticity in the ground state. In Fig. \ref{fig:8} (c1-c3), we present the k-space densities of the three components of the spin-1 BEC. Similar to the previously discussed cases, here also, apart from the central peak, extra density peaks along the $k_x, k_y$ directions are observed, which corresponds to the density modulations. Interestingly, for the $m_F=0$ component, the k-space density is modulated obliquely as the local high densities are in that direction.
\begin{figure}[t]
  \centering
   \includegraphics[width=0.48\textwidth]{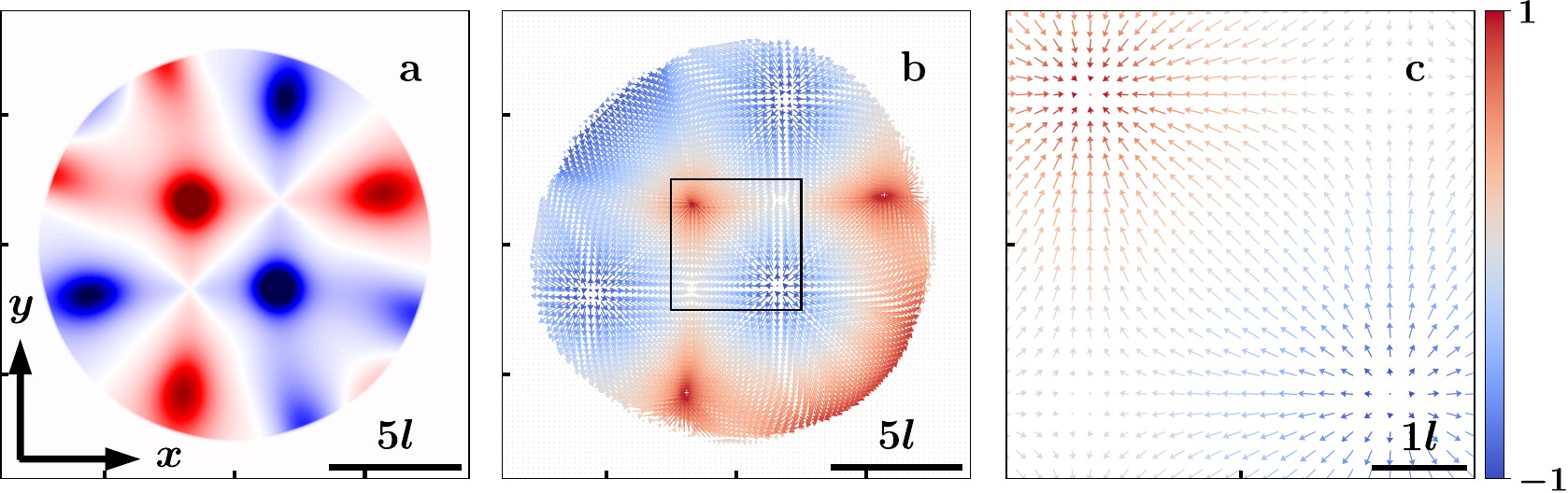}
  \caption{(a) shows the topological charge distribution, (b) shows the spin texture of the system and (c) shows the local amplification of the spin texture, marked by black square for the SO-coupled spin-1 BEC shown in Fig. \ref{fig:8}. The color of the arrows represent the $S_z$ values.  Units are dimensionless.}
  \label{fig:9}
\end{figure}

We present the associated topological charge distribution in Fig. \ref{fig:9} (a) and the spin texture is shown in Fig. \ref{fig:9} (b). For the diagonal vortices in the $m_F =0$ component, as the cores are symmetric, their respective skyrmion and anti-skyrmion structures are circular. However, for elongated vortices or anti-vortices, the corresponding topological charge distributions are also elliptical. We show in Fig. \ref{fig:9} (c), the local amplification of the spin texture for the region of the diagonal skyrmion and anti-skyrmion. The spin texture due to the \textit{cosine-cosine} magnetic field has skyrmion structures along the oblique diagonal. Here, the anisotropy is induced in the direction corresponding to the phase difference in the applied magnetic field, as compared to the \textit{sine-sine} case.

It is paramount to note that in the absence of SOC, the ground state structures for the \textit{sine-sine} case and the \textit{cosine-cosine} case can be analogously explained by taking into account the fact that the topological excitations are observed for certain saddle-points in the magnetic fields. In this light, it becomes imperative to study the ground state structures for the \textit{sine-cosine} and/or \textit{cosine-sine} fields.

\begin{figure}[t]
  \centering
  \includegraphics[width=0.48\textwidth]{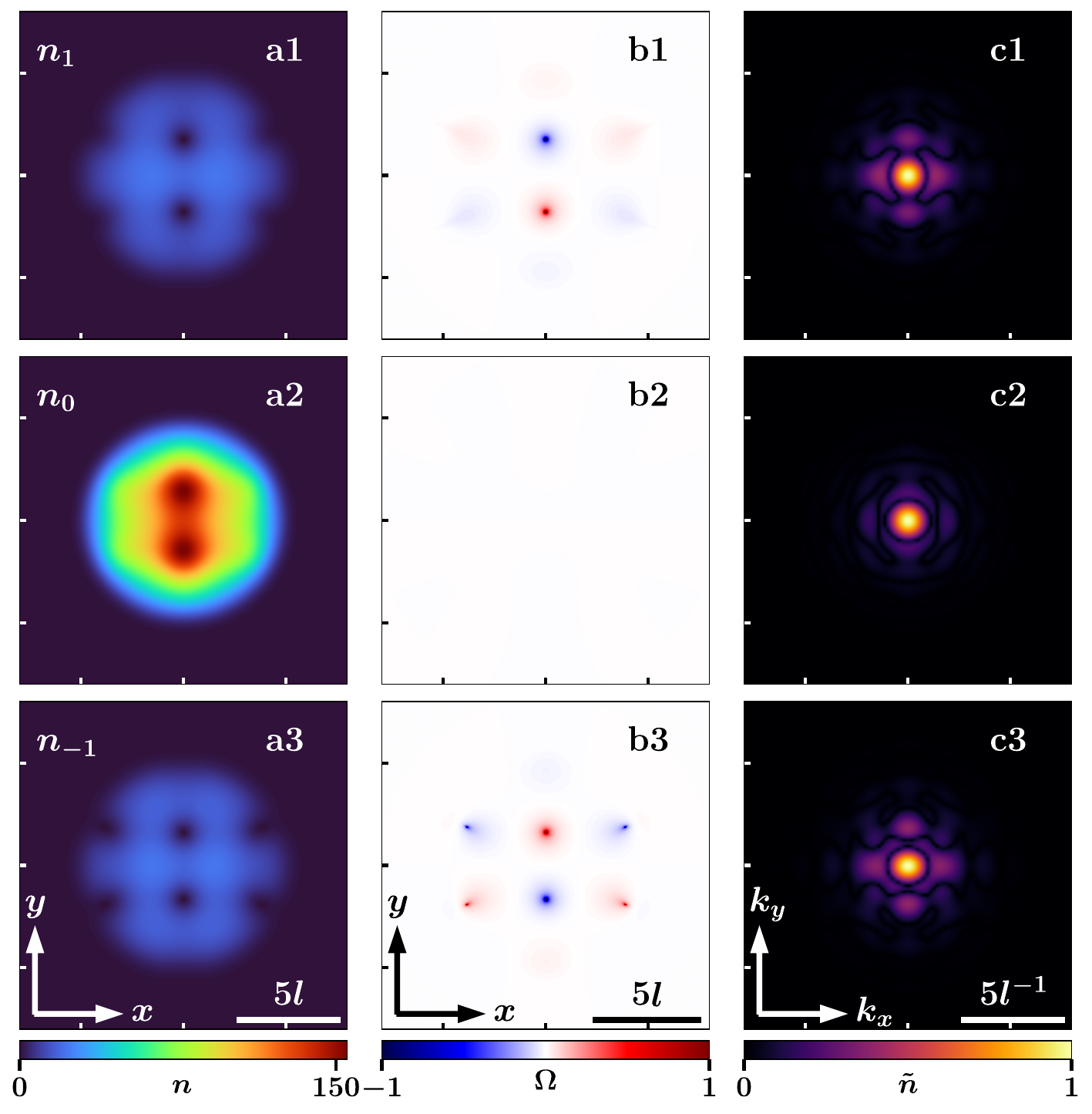}
  \caption{The ground state structures for spin-1 BEC of $10^4$ $\rm{^{87}{Rb}}$ atoms without SOC in the \textit{sine-cosine} magnetic field $B_x =B_0 \sin x$ and $B_y = B_0 \cos y$ with strength $B_0 = 0.25$. The first column (a1-a3) shows the density distributions, second column (b1-b3) shows the vorticity profiles and the third column (c1-c3) shows the momentum distributions of the $m_F=1$, $m_F=0$ and $m_F=-1$ components. The momentum space densities have been normalised by dividing the maximum value. The densities $n$ and $\tilde{n}$ are expressed in units of $l^{-2}$ and $l^2$, respectively, where $l=2.41 \mu m$ is the characteristic length. The vorticity $\Omega$ is expressed in dimensionless units.}
  \label{fig:10} 
\end{figure}
 
\subsubsection{Case-\rom{3}: Sine-Cosine magnetic field $B(\mathbf{r}) = B_0( \sin x, \cos y)$ }
\label{subsubsection:case3}
In this case, we study the ground state structures for the spin-1 BEC with the external magnetic field chosen as $B_x= B_0 \sin x$ and $B_y= B_0 \cos y$. For this combination of magnetic fields, in the absence of SOC, the ground state structures exhibit density modulations due to the resultant interaction of the \textit{sine-cosine} magnetic fields in the $x,y$ directions, respectively. We present the ground state density profiles of the $m_F=1$, $m_F=0$ and $m_F=-1$ components in Fig. \ref{fig:10}

\begin{figure}[t]
  \centering
  \includegraphics[width=0.48\textwidth]{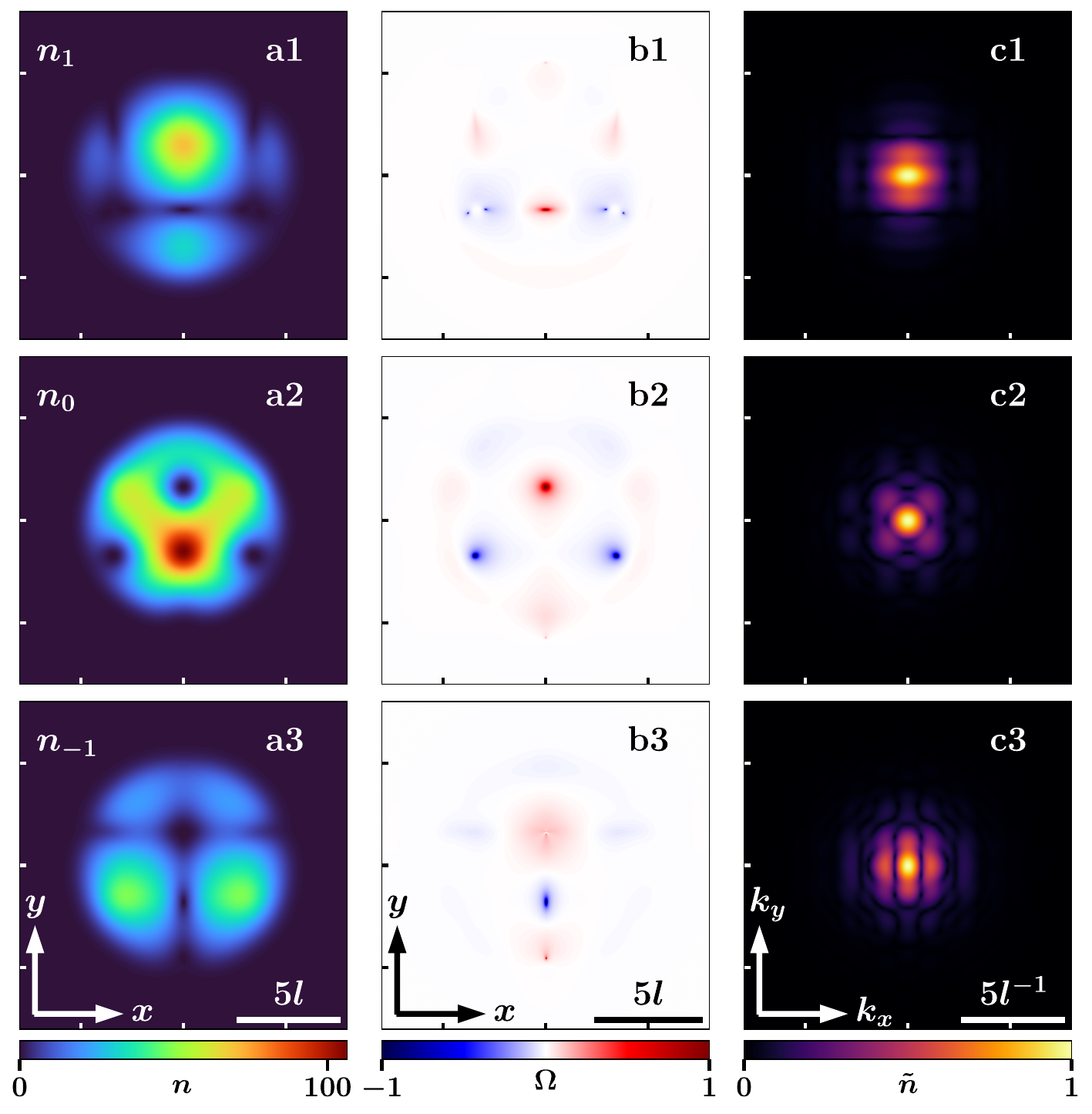}
  \caption{The ground state structures for spin-orbit coupled spin-1 BEC of $10^4$ $\rm{^{87}{Rb}}$ atoms with SOC strength $\gamma=0.5$ in the \textit{sine-cosine} magnetic field $B_x =B_0 \sin x$ and $B_y = B_0 \cos y$ with strength $B_0 = 0.25$. The first column (a1-a3) shows the density distributions, second column (b1-b3) shows the vorticity profiles and the third column (c1-c3) shows the momentum distributions of the $m_F=1$, $m_F=0$ and $m_F=-1$ components. The momentum space densities have been normalised by dividing the maximum value.  The densities $n$ and $\tilde{n}$ are expressed in units of $l^{-2}$ and $l^2$, respectively, where $l=2.41 \mu m$ is the characteristic length. The vorticity $\Omega$ is expressed in dimensionless units.}
  \label{fig:11}
\end{figure}

\noindent (a1-a3). In the absence of SOC, topological excitations in the form of vortices and anti-vortices are observed only in the $m_F=1$ and $m_F=-1$ components. These are arranged in pairs of alternating vorticity along the $y$-direction and as alternating vortex-anti-vortex triads in the $x$-direction. Due to different magnetic fields in different directions, the local high densities in the $m_F=0$ component (which arrise as an outcome of the spin-exchange interaction of $m_F =0$ component with the $m_F= \pm 1$ components), are anisotropic in nature. The vorticity profile (see Fig. \ref{fig:10} (b1-b3)) confirms the complementary response of the $m_F=1$ and $m_F=-1$ components to the external magnetic field with alternating vortices and anti-vortices. The momentum distribution is shown in Fig. \ref{fig:10} (c1-c3). For this case also, additional density peaks are observed in both the $k_x$ and $k_y$ directions in momentum space, alongside the central peak. However, as their magnitudes are not same in orthogonal directions, the density modulations have different wavelengths along different axes. 
 
In the presence of spin-orbit coupling, the system undergoes symmetry breaking, manifesting in density modulations in opposing directions. Notably, these modulations exhibit distinct characteristics across different components of the system. The density as well as vortex distributions are highly anisotropic in nature due to the combined effect of SOC-induced anisotropy and different sinusoidal fileds in different directions. We present the component-wise density profiles in Fig. \ref{fig:11} (a1-a3). In the $m_F=1$ component, there is a local high density surrounded by vortices, which is compensated by vortices and vortex brights in the $m_F=-1$ component. The $m_F=0$ component has a unique configuration for which a local high density is observed, which is surrounded by anti-vortices along the $x$-direction and an adjacent vortex along the $y$-direction. The corresponding vorticity profiles are given in Fig. \ref{fig:11} (b1-b3). For this case also, the coreless vortices have deformed elliptical shape as a consequence of induced anisotropy in the system. The momentum distributions as shown in Fig. \ref{fig:11} (c1-c3) conform the effect of anisotropy in the density modulations in the spin-1 system. The k-space densities are elongated in opposite directions for the $m_F=1$ and $m_F=-1$ components, respectively, \textit{i.e.,} for the $m_F=1$ case, the central peak is slightly elongated along the $x-$axis while for the $m_F=-1$ component, the central density peak is slightly elongated in the $y-$direction. This is to respect the density modulations with different wavelengths in the anisotropic system. For the $m_F=0$ component, the momentum distribution is symmetric however not the same about the $k_x, k_y$ axis. 
  
\begin{figure}[t]
  \centering
  \includegraphics[width=0.48\textwidth]{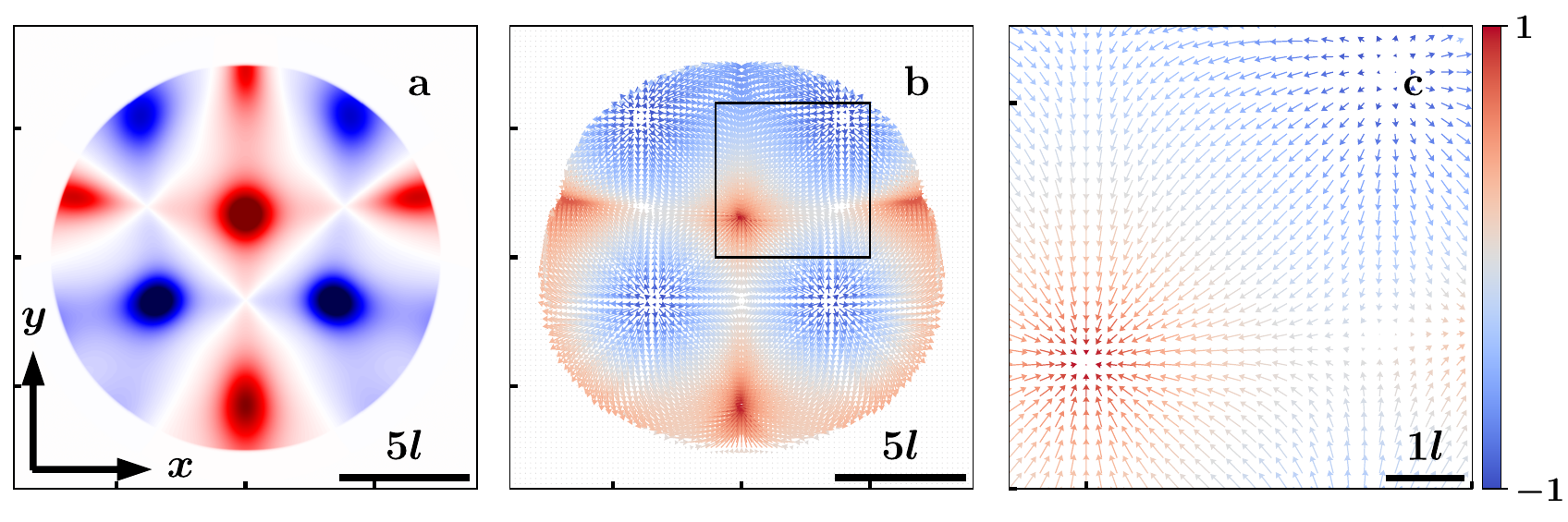}
  \caption{(a) shows the topological charge distribution, (b) shows the spin texture of the system and (c) shows the local amplification of the spin texture, marked by black square, for the SO-coupled spin-1 BEC shown in Fig. \ref{fig:11}. The color of the arrows represent the $S_z$ values.  Units are dimensionless.}
  \label{fig:12} 
\end{figure}

The topological charge distribution demonstrated in Fig. \ref{fig:12} (a) reveals the presence of skyrmions and anti-skyrmions, which is confirmed by the spin texture of the system shown in Fig. \ref{fig:12} (b). Because of the combined effect of anisotropy and different magnetic fields along different directions, the spin texture has both skyrmions and anti-skyrmions. Owing to the vortex-anti-vortex triad, the skyrmion-anti-skyrmions are arranged in a triangular fashion. Fig. \ref{fig:12} (c) shows the local amplification of the spin texture showcasing the central vortex-anti-vortex triad.
 
\subsubsection{Case-\rom{4}: Cosine-Sine magnetic field $B(\rm{r}) = B_0( \cos x, \sin y)$ }
\label{subsubsection:case4}
For the present case, we propose an alternate choice of cross-magnetic field in the $x$-$y$ plane, \textit{i.e.,} $B_x = B_0 \cos x$ and $B_y = B_0 \sin x$. Inspite of the striking similarities with the \textit{sine-cosine} case (discussed in section \ref{subsubsection:case3}), there are subtle differences in the present case. It is trivial to understand that for this case also, the ground state structures are direction-dependent as a result of choosing different magnetic fields in different directions. Importantly, as the magnetic fields along the $x,y$-directions are swapped, their effect on the spin-1 ground state will therefore, be also swapped. In the absence of SOC, the density profiles, associated vorticities and momentum 
\begin{figure}[t]   
  \centering 
  \includegraphics[width=0.48\textwidth]{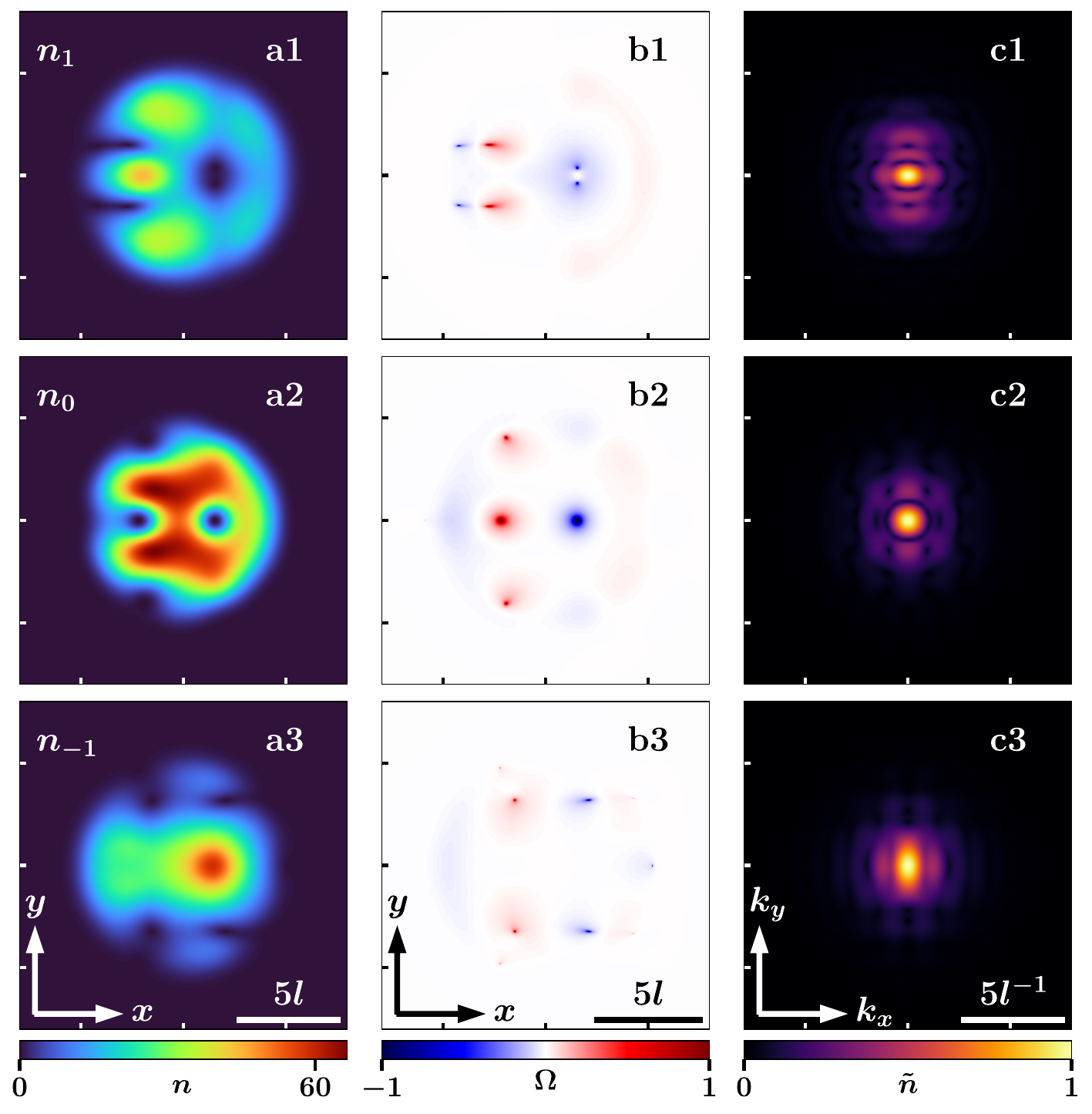}
  \caption{The ground state structures for spin-orbit coupled spin-1 BEC of $10^4$ $\rm{^{87}{Rb}}$ atoms with SOC strength $\gamma=0.5$ in the \textit{sine-cosine} magnetic field $B_x =B_0 \cos x$ and $B_y = B_0 \sin y$ with strength $B_0 = 0.25$. The first column (a1-a3) shows the density distributions, second column (b1-b3) shows the vorticity profiles and the third column (c1-c3) shows the momentum distributions of the $m_F=1$, $m_F=0$ and $m_F=-1$ components. The momentum space densities have been normalised by dividing the maximum value.  The densities $n$ and $\tilde{n}$ are expressed in units of $l^{-2}$ and $l^2$, respectively, where $l=2.41 \mu m$ is the characteristic length. The vorticity $\Omega$ is expressed in dimensionless units.}
  \label{fig:13}
\end{figure} 
\begin{figure}[t] 
  \centering
  \includegraphics[width=0.48\textwidth]{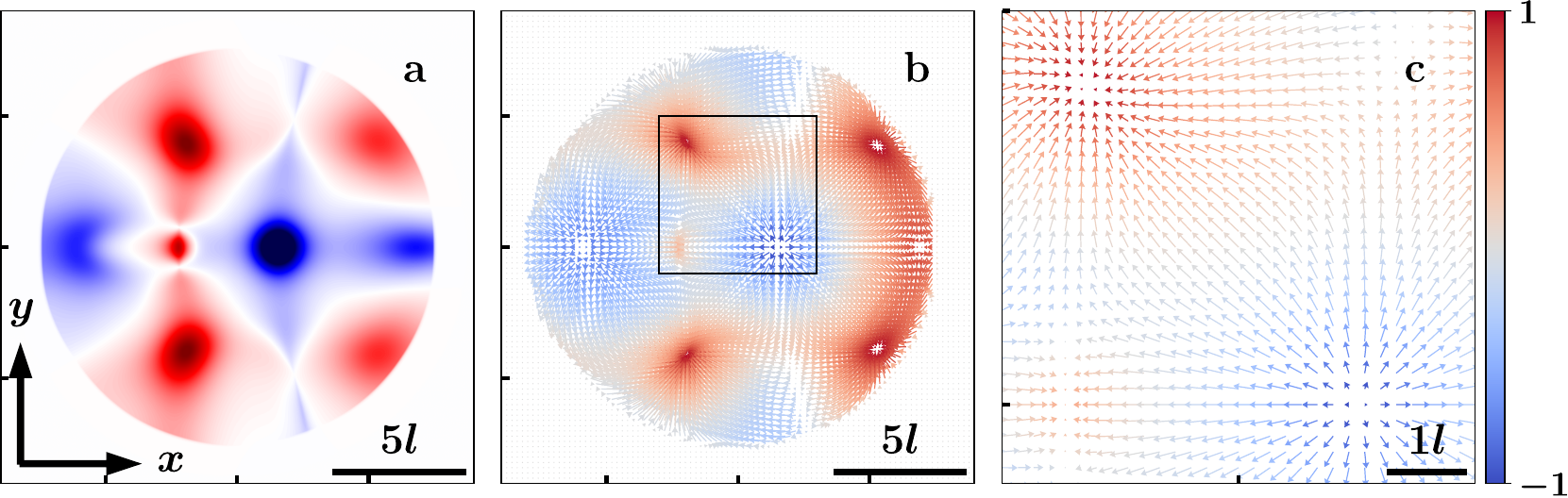}
  \caption{(a) shows the topological charge distribution, (b) shows the spin texture of the system and (c) shows the local amplification of the spin texture, marked by black square, for the SO-coupled spin-1 BEC shown in Fig. \ref{fig:13}. The color of the arrows represent the $S_z$ values.  Units are dimensionless.}
  \label{fig:14}
\end{figure}
\noindent space densities of the three components are interchanged along the $x$-$y$ axes. 

However, for the SO-coupled spin-1 BEC in the \textit{cosine-sine} magnetic field, the ground state structures are not simply the transpose of that of the \textit{sine-cosine} case. The ground state structures which are an outcome of the intricate interplay between the external magnetic field, SOC and the spinor order parameter is being presented in Fig. \ref{fig:13}.  The density profiles (see Fig. \ref{fig:13} (a1-a3)) vouch for the experienced anisotropy in the system due to SOC. The $m_F=1$ component has a unique configuration of vortices along with adjacent local high densities in the system. The $m_F=0$ component has a linear chain of vortices in the $y$-direction with an additional anti-vortex in the $x$-direction. The $m_F=-1$ component has a local high density surrounded by vortex structures on adjacent sides. The vorticity profile as shown in Fig. \ref{fig:13} (b1-b3) depicts the distribution of vortices or anti-vortices in the system. The coreless vortices have a slightly distorted shape due to the symmetry breaking. Fig. \ref{fig:13} (c1-c3) displays the momentum distributions. The presence of induced anisotropy in the system leads to observable elongations in the k-space densities along specific directions, owing to the varying periodicities of in-plane magnetic fields in the $x$ and $y$-directions. Moreover, notable density peaks are evident at certain non-zero $k_x, k_y$.

The topological charge distribution shown in Fig. \ref{fig:14} (a). The skyrmions and anti-skyrmions are observed in different axes. The spin texture is shown in Fig. \ref{fig:14} (b). Aside from the central triad formed with alternating vortex-anti-vortex configuration, a few other spin textures can be observed owing to the coreless vortices in the system. Fig. \ref{fig:14} (c) shows the local amplification of the spin texture showing the central triad in the system. As the \textit{sine-cosine} and \textit{cosine-sine} cases are not just mere transpose of each other in the presence of SOC, the skyrmion arrangement although in the form of triads are distinct from one another.

In this work, we have focussed our results for a weak SOC with strength $\gamma=0.5$, only. However, as we increase the $\gamma$ values, a very intricate arrangement of skyrmions and anti-skyrmions is formed due to the highly induced anisotropy in the system. As a consequence, the density modulations with short wavelengths are observed in the ground state configurations giving rise to complicated vortex-anti-vortex structures depending upon the choice of external in-plane magnetic field. We discuss the impact of variation of $\gamma$ on the spin texture in appendix \ref{apx:seca}.
 
\section{Conclusion}
\label{sec:conclusion}

In summary, the interaction between spin-orbit coupling and an external in-plane sinusoidal magnetic field in a spin-1 BEC introduces additional complexity to the system. The periodic modulation of the magnetic field in the $x$-$y$ plane creates a distinct spatial variation that influences the condensate's ground state. Vortex formations occur at saddle-points of the in-plane magnetic field, adding to the system's topological excitations. Coupled with spin-exchange interactions and spin-mixing dynamics, the magnetic field induces conjugate vortices and anti-vortices in the $m_F=\pm 1$ components, and vortex brights in the $m_F=0$ component. These density modulations correspond to a wavelength $\lambda$ inversely proportional to the non-zero momentum values at which additional density peaks emerge in the k-space.

In the presence of SOC, the rotational symmetry of the system breaks giving rise to certain topological excitations in the system. Due to the symmetry-breaking, an anisotropy is induced in the system which renders a very unique density distribution to three components of the spin-1 BEC. As a consequence, the overall angular momentum is not conserved in the system. Furthermore, the combination of SOC and the spatially varying magnetic field gives rise to spatially modulated spin textures, including the formation of unique configuration of skyrmions and anti-skyrmions. These topological defects represent fascinating quantum states within the condensate, showcasing the intricate interplay between the periodicity induced by the sinusoidal magnetic field and the coupling of spin and motion.

The sinusoidal variation introduces periodicity into the system, creating an opportunity to engineer and manipulate the condensate's structure and properties. Beyond this, the dynamics of a SO-coupled spin-1 BEC in a polarised sinusoidal magnetic field can be explored. Moreover, it will be interesting to look into the skyrmion dynamics of the SO-coupled system. The ability to tune the sinusoidal variation of the magnetic field in a controlled environment not only provides a deeper understanding of the quantum phases emerging from the interplay but also opens up possibilities for designing novel quantum devices (\textit{e.g.,} skyrmion qubits \cite{Psaroudaki_skyrmion_2021}) and applications. 

Understanding the interplay between SOC and sinusoidal magnetic fields in spin-1 BECs not only advances fundamental physics but also paves the way for new quantum technologies. By investigating these systems, researchers can develop more sophisticated methods for manipulating quantum states, explore the stability and dynamics of topologically non-trivial excitations, and potentially discover new quantum phases with practical applications. Our findings suggests ways to engineer diverse topological excitations in a spin-1 BEC employing a periodically varying magnetic field. This may serve as an important benchmark for future theoretical and experimental studies aimed at harnessing quantum mechanics for technological innovation. 

\section*{Acknowledgements}
We thank Hari Sadhan Ghosh for many insightful
discussions. We acknowledge the National Supercomputing Mission (NSM) for providing computing resources of
PARAM Shakti at IIT Kharagpur, which is implemented
by C-DAC and supported by the Ministry of Electronics
and Information Technology (MeitY) and Department of
Science and Technology (DST), Government of India. AS gratefully acknowledges the support from the Prime Minister's Research Fellowship (PMRF), India. SH acknowledges the MHRD, Govt. of India for the research fellowship. SD acknowledges support from AFOSR FA9550-23-1-0034.

\appendix
\section{Impact of varying spin-orbit coupling strength on the spin texture}
\label{apx:seca}
\begin{figure}[t]   
	\centering 
	\includegraphics[width=0.48\textwidth]{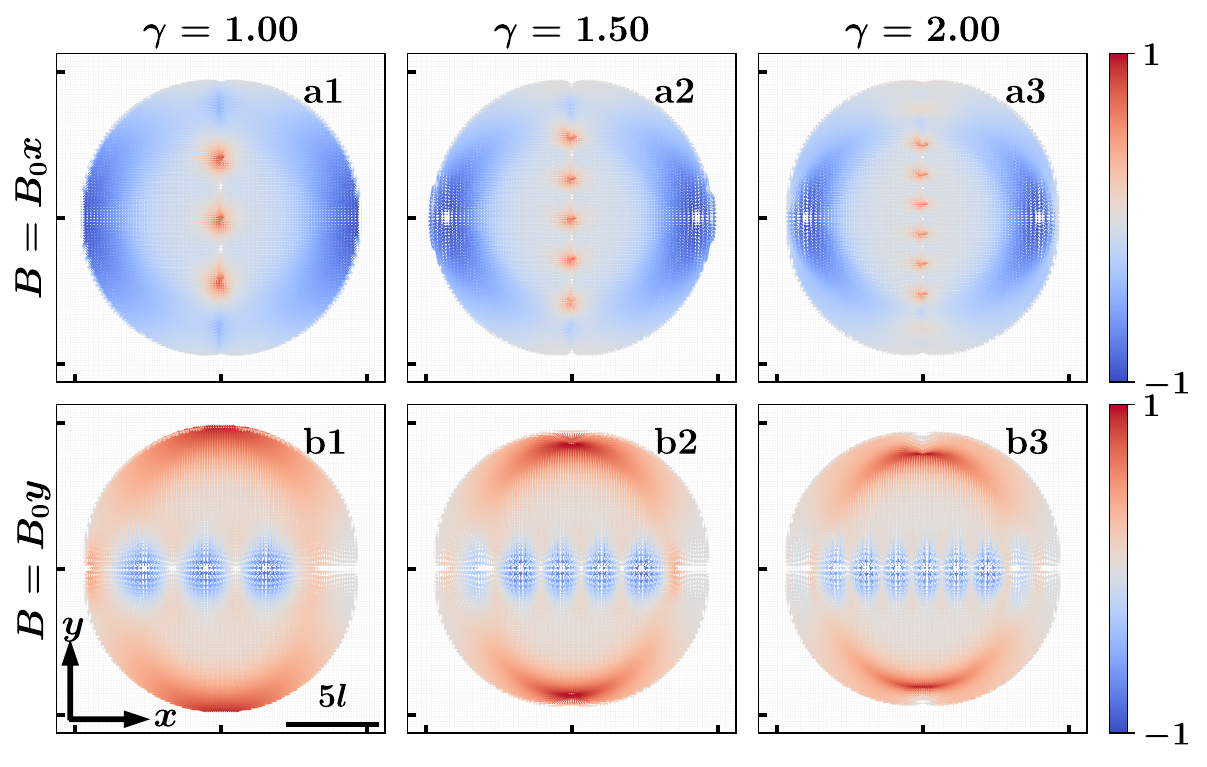}
	\caption{(a1-a3) shows the spin texture for the $x$-polarised gradient magnetic field for varying SOC strength $\gamma=1.00,1.50,2.00$ respectively. (b1-b3) shows the same for the $y$-polarised gradient magnetic field. Units are dimensionless.}
	\label{fig:15}
\end{figure}

In section \ref{sec:results} of the main text, we discussed the interplay of the SOC and magnetic field on the spin-1 BEC for a fixed SOC strength $\gamma = 0.5$ for various cases of gradient and sinusoidally varying in-plane magnetic field. Here, we show the impact of variation of the SOC strength on the spin texture. Fig. \ref{fig:15} displays the spatial distribution of the spin texture under the influence of the polarized gradient magnetic field (as discussed in \ref{subsection:gradient}) with varying SOC strengths. The polarised gradient magnetic field applied along $x$-axis ($y$-axis) renders skyrmions (anti-skyrmions) along the $x=0$ ($y=0$) line in presence of the SOC. The number of such structures increases while the wavelength of density modulation decreases with increasing SOC strength, $\gamma$.

\begin{figure}[b] 
  \centering
  \includegraphics[width=0.48\textwidth]{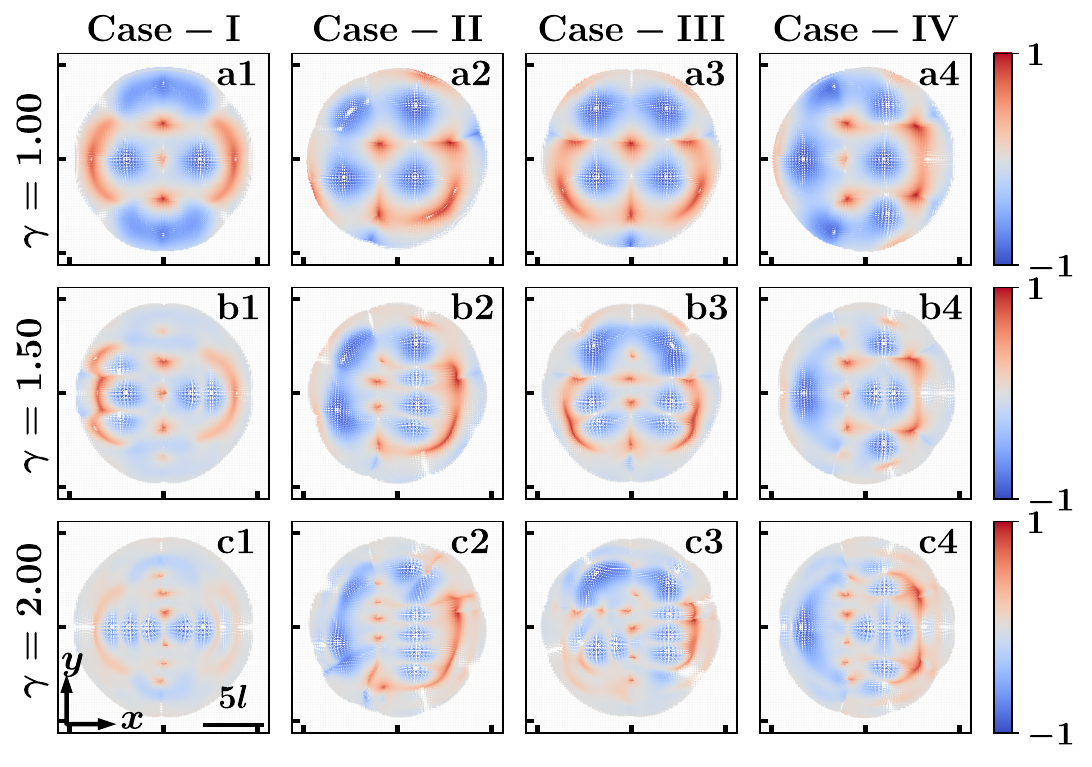}
  \caption{The spin texture with various SOC strengths $\gamma$ for the cases of sinusoidal magnetic fields discussed in section \ref{subsection:cross-field}. The columns present the case specific results while the rows represent different $\gamma$ values. Units are dimensionless.}
  \label{fig:16}
\end{figure}

For the cross-polarised sinusoidally varying in-plane magnetic fields discussed in section \ref{subsection:cross-field}, we present the spin textures for varying SOC strengths in Fig. \ref{fig:16}.  We observe that for different SOC strengths, the spin texture follows a similar trend in both $x$ and $y$-directions for each case. However, the density modulations increase with the $\gamma$ values. As a result of the grwoing anisotropy due to increased SOC strength, the system exhibits a complicated spin texture with more number of skyrmions and anti-skyrmions. Moreover, the interaction between these skyrmions and anti-skyrmions affect their arrangement giving rise to a complex eolotropic structure. As investigations into this intricate interplay continue, we anticipate the discovery of new spatially modulated quantum states, offering exciting prospects for both fundamental research and technological advancements.

\bibliographystyle{apsrev4-2}
\bibliography{spin.bib}

\begin{thebibliography}{108}%
\makeatletter
\providecommand \@ifxundefined [1]{%
 \@ifx{#1\undefined}
}%
\providecommand \@ifnum [1]{%
 \ifnum #1\expandafter \@firstoftwo
 \else \expandafter \@secondoftwo
 \fi
}%
\providecommand \@ifx [1]{%
 \ifx #1\expandafter \@firstoftwo
 \else \expandafter \@secondoftwo
 \fi
}%
\providecommand \natexlab [1]{#1}%
\providecommand \enquote  [1]{``#1''}%
\providecommand \bibnamefont  [1]{#1}%
\providecommand \bibfnamefont [1]{#1}%
\providecommand \citenamefont [1]{#1}%
\providecommand \href@noop [0]{\@secondoftwo}%
\providecommand \href [0]{\begingroup \@sanitize@url \@href}%
\providecommand \@href[1]{\@@startlink{#1}\@@href}%
\providecommand \@@href[1]{\endgroup#1\@@endlink}%
\providecommand \@sanitize@url [0]{\catcode `\\12\catcode `\$12\catcode
  `\&12\catcode `\#12\catcode `\^12\catcode `\_12\catcode `\%12\relax}%
\providecommand \@@startlink[1]{}%
\providecommand \@@endlink[0]{}%
\providecommand \url  [0]{\begingroup\@sanitize@url \@url }%
\providecommand \@url [1]{\endgroup\@href {#1}{\urlprefix }}%
\providecommand \urlprefix  [0]{URL }%
\providecommand \Eprint [0]{\href }%
\providecommand \doibase [0]{https://doi.org/}%
\providecommand \selectlanguage [0]{\@gobble}%
\providecommand \bibinfo  [0]{\@secondoftwo}%
\providecommand \bibfield  [0]{\@secondoftwo}%
\providecommand \translation [1]{[#1]}%
\providecommand \BibitemOpen [0]{}%
\providecommand \bibitemStop [0]{}%
\providecommand \bibitemNoStop [0]{.\EOS\space}%
\providecommand \EOS [0]{\spacefactor3000\relax}%
\providecommand \BibitemShut  [1]{\csname bibitem#1\endcsname}%
\let\auto@bib@innerbib\@empty
\bibitem [{\citenamefont {Lin}\ \emph {et~al.}(2011)\citenamefont {Lin},
  \citenamefont {Jiménez-García},\ and\ \citenamefont
  {Spielman}}]{lin_spinorbit-coupled_2011}%
  \BibitemOpen
  \bibfield  {author} {\bibinfo {author} {\bibfnamefont {Y.-J.}\ \bibnamefont
  {Lin}}, \bibinfo {author} {\bibfnamefont {K.}~\bibnamefont
  {Jiménez-García}},\ and\ \bibinfo {author} {\bibfnamefont {I.~B.}\
  \bibnamefont {Spielman}},\ }\href {https://doi.org/10.1038/nature09887}
  {\bibfield  {journal} {\bibinfo  {journal} {Nature}\ }\textbf {\bibinfo
  {volume} {471}},\ \bibinfo {pages} {83} (\bibinfo {year} {2011})}\BibitemShut
  {NoStop}%
\bibitem [{\citenamefont {Wang}\ \emph {et~al.}(2010)\citenamefont {Wang},
  \citenamefont {Gao}, \citenamefont {Jian},\ and\ \citenamefont
  {Zhai}}]{wang_spinorbit_2010}%
  \BibitemOpen
  \bibfield  {author} {\bibinfo {author} {\bibfnamefont {C.}~\bibnamefont
  {Wang}}, \bibinfo {author} {\bibfnamefont {C.}~\bibnamefont {Gao}}, \bibinfo
  {author} {\bibfnamefont {C.-M.}\ \bibnamefont {Jian}},\ and\ \bibinfo
  {author} {\bibfnamefont {H.}~\bibnamefont {Zhai}},\ }\href
  {https://doi.org/10.1103/PhysRevLett.105.160403} {\bibfield  {journal}
  {\bibinfo  {journal} {Phys. Rev. Lett.}\ }\textbf {\bibinfo {volume} {105}},\
  \bibinfo {pages} {160403} (\bibinfo {year} {2010})}\BibitemShut {NoStop}%
\bibitem [{\citenamefont {Hu}\ \emph {et~al.}(2012)\citenamefont {Hu},
  \citenamefont {Ramachandhran}, \citenamefont {Pu},\ and\ \citenamefont
  {Liu}}]{Hui_soc_2012}%
  \BibitemOpen
  \bibfield  {author} {\bibinfo {author} {\bibfnamefont {H.}~\bibnamefont
  {Hu}}, \bibinfo {author} {\bibfnamefont {B.}~\bibnamefont {Ramachandhran}},
  \bibinfo {author} {\bibfnamefont {H.}~\bibnamefont {Pu}},\ and\ \bibinfo
  {author} {\bibfnamefont {X.-J.}\ \bibnamefont {Liu}},\ }\href
  {https://doi.org/10.1103/PhysRevLett.108.010402} {\bibfield  {journal}
  {\bibinfo  {journal} {Phys. Rev. Lett.}\ }\textbf {\bibinfo {volume} {108}},\
  \bibinfo {pages} {010402} (\bibinfo {year} {2012})}\BibitemShut {NoStop}%
\bibitem [{\citenamefont {Wang}\ \emph {et~al.}(2012)\citenamefont {Wang},
  \citenamefont {Yu}, \citenamefont {Fu}, \citenamefont {Miao}, \citenamefont
  {Huang}, \citenamefont {Chai}, \citenamefont {Zhai},\ and\ \citenamefont
  {Zhang}}]{wang_spin-orbit_2012}%
  \BibitemOpen
  \bibfield  {author} {\bibinfo {author} {\bibfnamefont {P.}~\bibnamefont
  {Wang}}, \bibinfo {author} {\bibfnamefont {Z.-Q.}\ \bibnamefont {Yu}},
  \bibinfo {author} {\bibfnamefont {Z.}~\bibnamefont {Fu}}, \bibinfo {author}
  {\bibfnamefont {J.}~\bibnamefont {Miao}}, \bibinfo {author} {\bibfnamefont
  {L.}~\bibnamefont {Huang}}, \bibinfo {author} {\bibfnamefont
  {S.}~\bibnamefont {Chai}}, \bibinfo {author} {\bibfnamefont {H.}~\bibnamefont
  {Zhai}},\ and\ \bibinfo {author} {\bibfnamefont {J.}~\bibnamefont {Zhang}},\
  }\href {https://doi.org/10.1103/PhysRevLett.109.095301} {\bibfield  {journal}
  {\bibinfo  {journal} {Physical Review Letters}\ }\textbf {\bibinfo {volume}
  {109}},\ \bibinfo {pages} {095301} (\bibinfo {year} {2012})}\BibitemShut
  {NoStop}%
\bibitem [{\citenamefont {Anderson}\ \emph {et~al.}(2012)\citenamefont
  {Anderson}, \citenamefont {Juzeliūnas}, \citenamefont {Galitski},\ and\
  \citenamefont {Spielman}}]{anderson_synthetic_2012}%
  \BibitemOpen
  \bibfield  {author} {\bibinfo {author} {\bibfnamefont {B.~M.}\ \bibnamefont
  {Anderson}}, \bibinfo {author} {\bibfnamefont {G.}~\bibnamefont
  {Juzeliūnas}}, \bibinfo {author} {\bibfnamefont {V.~M.}\ \bibnamefont
  {Galitski}},\ and\ \bibinfo {author} {\bibfnamefont {I.~B.}\ \bibnamefont
  {Spielman}},\ }\href {https://doi.org/10.1103/PhysRevLett.108.235301}
  {\bibfield  {journal} {\bibinfo  {journal} {Physical Review Letters}\
  }\textbf {\bibinfo {volume} {108}},\ \bibinfo {pages} {235301} (\bibinfo
  {year} {2012})}\BibitemShut {NoStop}%
\bibitem [{\citenamefont {Wu}\ \emph {et~al.}(2016{\natexlab{a}})\citenamefont
  {Wu}, \citenamefont {Zhang}, \citenamefont {Sun}, \citenamefont {Xu},
  \citenamefont {Wang}, \citenamefont {Ji}, \citenamefont {Deng}, \citenamefont
  {Chen}, \citenamefont {Liu},\ and\ \citenamefont
  {Pan}}]{wu_realization_2016}%
  \BibitemOpen
  \bibfield  {author} {\bibinfo {author} {\bibfnamefont {Z.}~\bibnamefont
  {Wu}}, \bibinfo {author} {\bibfnamefont {L.}~\bibnamefont {Zhang}}, \bibinfo
  {author} {\bibfnamefont {W.}~\bibnamefont {Sun}}, \bibinfo {author}
  {\bibfnamefont {X.-T.}\ \bibnamefont {Xu}}, \bibinfo {author} {\bibfnamefont
  {B.-Z.}\ \bibnamefont {Wang}}, \bibinfo {author} {\bibfnamefont {S.-C.}\
  \bibnamefont {Ji}}, \bibinfo {author} {\bibfnamefont {Y.}~\bibnamefont
  {Deng}}, \bibinfo {author} {\bibfnamefont {S.}~\bibnamefont {Chen}}, \bibinfo
  {author} {\bibfnamefont {X.-J.}\ \bibnamefont {Liu}},\ and\ \bibinfo {author}
  {\bibfnamefont {J.-W.}\ \bibnamefont {Pan}},\ }\href
  {https://doi.org/10.1126/science.aaf6689} {\bibfield  {journal} {\bibinfo
  {journal} {Science}\ }\textbf {\bibinfo {volume} {354}},\ \bibinfo {pages}
  {83} (\bibinfo {year} {2016}{\natexlab{a}})}\BibitemShut {NoStop}%
\bibitem [{\citenamefont {Huang}\ \emph {et~al.}(2016)\citenamefont {Huang},
  \citenamefont {Meng}, \citenamefont {Wang}, \citenamefont {Peng},
  \citenamefont {Zhang}, \citenamefont {Chen}, \citenamefont {Li},
  \citenamefont {Zhou},\ and\ \citenamefont {Zhang}}]{huang_experimental_2016}%
  \BibitemOpen
  \bibfield  {author} {\bibinfo {author} {\bibfnamefont {L.}~\bibnamefont
  {Huang}}, \bibinfo {author} {\bibfnamefont {Z.}~\bibnamefont {Meng}},
  \bibinfo {author} {\bibfnamefont {P.}~\bibnamefont {Wang}}, \bibinfo {author}
  {\bibfnamefont {P.}~\bibnamefont {Peng}}, \bibinfo {author} {\bibfnamefont
  {S.-L.}\ \bibnamefont {Zhang}}, \bibinfo {author} {\bibfnamefont
  {L.}~\bibnamefont {Chen}}, \bibinfo {author} {\bibfnamefont {D.}~\bibnamefont
  {Li}}, \bibinfo {author} {\bibfnamefont {Q.}~\bibnamefont {Zhou}},\ and\
  \bibinfo {author} {\bibfnamefont {J.}~\bibnamefont {Zhang}},\ }\href
  {https://doi.org/10.1038/nphys3672} {\bibfield  {journal} {\bibinfo
  {journal} {Nature Physics}\ }\textbf {\bibinfo {volume} {12}},\ \bibinfo
  {pages} {540} (\bibinfo {year} {2016})}\BibitemShut {NoStop}%
\bibitem [{\citenamefont {Campbell}\ \emph {et~al.}(2011)\citenamefont
  {Campbell}, \citenamefont {Juzeliūnas},\ and\ \citenamefont
  {Spielman}}]{campbell_realistic_2011}%
  \BibitemOpen
  \bibfield  {author} {\bibinfo {author} {\bibfnamefont {D.~L.}\ \bibnamefont
  {Campbell}}, \bibinfo {author} {\bibfnamefont {G.}~\bibnamefont
  {Juzeliūnas}},\ and\ \bibinfo {author} {\bibfnamefont {I.~B.}\ \bibnamefont
  {Spielman}},\ }\href {https://doi.org/10.1103/PhysRevA.84.025602} {\bibfield
  {journal} {\bibinfo  {journal} {Physical Review A}\ }\textbf {\bibinfo
  {volume} {84}},\ \bibinfo {pages} {025602} (\bibinfo {year}
  {2011})}\BibitemShut {NoStop}%
\bibitem [{\citenamefont {Wu}\ \emph {et~al.}(2011)\citenamefont {Wu},
  \citenamefont {Mondragon-Shem},\ and\ \citenamefont
  {Zhou}}]{wu2011unconventional}%
  \BibitemOpen
  \bibfield  {author} {\bibinfo {author} {\bibfnamefont {C.-J.}\ \bibnamefont
  {Wu}}, \bibinfo {author} {\bibfnamefont {I.}~\bibnamefont {Mondragon-Shem}},\
  and\ \bibinfo {author} {\bibfnamefont {X.-F.}\ \bibnamefont {Zhou}},\ }\href
  {https://doi.org/10.1088/0256-307X/28/9/097102} {\bibfield  {journal}
  {\bibinfo  {journal} {Chinese Physics Letters}\ }\textbf {\bibinfo {volume}
  {28}},\ \bibinfo {pages} {097102} (\bibinfo {year} {2011})}\BibitemShut
  {NoStop}%
\bibitem [{\citenamefont {Peng}\ \emph {et~al.}(2020)\citenamefont {Peng},
  \citenamefont {Li}, \citenamefont {Wang},\ and\ \citenamefont
  {Cao}}]{peng_exotic_2020}%
  \BibitemOpen
  \bibfield  {author} {\bibinfo {author} {\bibfnamefont {P.}~\bibnamefont
  {Peng}}, \bibinfo {author} {\bibfnamefont {G.-Q.}\ \bibnamefont {Li}},
  \bibinfo {author} {\bibfnamefont {B.-H.}\ \bibnamefont {Wang}},\ and\
  \bibinfo {author} {\bibfnamefont {Z.-Z.}\ \bibnamefont {Cao}},\ }\href
  {https://doi.org/https://doi.org/10.1016/j.chaos.2020.110332} {\bibfield
  {journal} {\bibinfo  {journal} {Chaos, Solitons \& Fractals}\ }\textbf
  {\bibinfo {volume} {141}},\ \bibinfo {pages} {110332} (\bibinfo {year}
  {2020})}\BibitemShut {NoStop}%
\bibitem [{\citenamefont {Zhu}\ \emph {et~al.}(2020)\citenamefont {Zhu},
  \citenamefont {Pan},\ and\ \citenamefont {An}}]{zhu_spin_2020}%
  \BibitemOpen
  \bibfield  {author} {\bibinfo {author} {\bibfnamefont {Q.-L.}\ \bibnamefont
  {Zhu}}, \bibinfo {author} {\bibfnamefont {L.}~\bibnamefont {Pan}},\ and\
  \bibinfo {author} {\bibfnamefont {J.}~\bibnamefont {An}},\ }\href
  {https://doi.org/10.1103/PhysRevA.102.053320} {\bibfield  {journal} {\bibinfo
   {journal} {Phys. Rev. A}\ }\textbf {\bibinfo {volume} {102}},\ \bibinfo
  {pages} {053320} (\bibinfo {year} {2020})}\BibitemShut {NoStop}%
\bibitem [{\citenamefont {Wen}\ \emph {et~al.}(2012)\citenamefont {Wen},
  \citenamefont {Sun}, \citenamefont {Wang}, \citenamefont {Ji},\ and\
  \citenamefont {Liu}}]{wen_ground_2012}%
  \BibitemOpen
  \bibfield  {author} {\bibinfo {author} {\bibfnamefont {L.}~\bibnamefont
  {Wen}}, \bibinfo {author} {\bibfnamefont {Q.}~\bibnamefont {Sun}}, \bibinfo
  {author} {\bibfnamefont {H.~Q.}\ \bibnamefont {Wang}}, \bibinfo {author}
  {\bibfnamefont {A.~C.}\ \bibnamefont {Ji}},\ and\ \bibinfo {author}
  {\bibfnamefont {W.~M.}\ \bibnamefont {Liu}},\ }\href
  {https://doi.org/10.1103/PhysRevA.86.043602} {\bibfield  {journal} {\bibinfo
  {journal} {Phys. Rev. A}\ }\textbf {\bibinfo {volume} {86}},\ \bibinfo
  {pages} {043602} (\bibinfo {year} {2012})}\BibitemShut {NoStop}%
\bibitem [{\citenamefont {Mardonov}\ \emph {et~al.}(2015)\citenamefont
  {Mardonov}, \citenamefont {Modugno},\ and\ \citenamefont
  {Sherman}}]{mardonov_dynamics_2015}%
  \BibitemOpen
  \bibfield  {author} {\bibinfo {author} {\bibfnamefont {S.}~\bibnamefont
  {Mardonov}}, \bibinfo {author} {\bibfnamefont {M.}~\bibnamefont {Modugno}},\
  and\ \bibinfo {author} {\bibfnamefont {E.~Y.}\ \bibnamefont {Sherman}},\
  }\href {https://doi.org/10.1103/PhysRevLett.115.180402} {\bibfield  {journal}
  {\bibinfo  {journal} {Physical Review Letters}\ }\textbf {\bibinfo {volume}
  {115}},\ \bibinfo {pages} {180402} (\bibinfo {year} {2015})}\BibitemShut
  {NoStop}%
\bibitem [{\citenamefont {Sun}\ \emph {et~al.}(2016)\citenamefont {Sun},
  \citenamefont {Qu}, \citenamefont {Xu}, \citenamefont {Zhang},\ and\
  \citenamefont {Zhang}}]{sun_interacting_2016}%
  \BibitemOpen
  \bibfield  {author} {\bibinfo {author} {\bibfnamefont {K.}~\bibnamefont
  {Sun}}, \bibinfo {author} {\bibfnamefont {C.}~\bibnamefont {Qu}}, \bibinfo
  {author} {\bibfnamefont {Y.}~\bibnamefont {Xu}}, \bibinfo {author}
  {\bibfnamefont {Y.}~\bibnamefont {Zhang}},\ and\ \bibinfo {author}
  {\bibfnamefont {C.}~\bibnamefont {Zhang}},\ }\href
  {https://doi.org/10.1103/PhysRevA.93.023615} {\bibfield  {journal} {\bibinfo
  {journal} {Phys. Rev. A}\ }\textbf {\bibinfo {volume} {93}},\ \bibinfo
  {pages} {023615} (\bibinfo {year} {2016})}\BibitemShut {NoStop}%
\bibitem [{\citenamefont {Su}\ \emph {et~al.}(2016)\citenamefont {Su},
  \citenamefont {Gou}, \citenamefont {Sun}, \citenamefont {Wen}, \citenamefont
  {Liu}, \citenamefont {Ji}, \citenamefont {Ruseckas},\ and\ \citenamefont
  {Juzeliūnas}}]{su_rashba-type_2016}%
  \BibitemOpen
  \bibfield  {author} {\bibinfo {author} {\bibfnamefont {S.-W.}\ \bibnamefont
  {Su}}, \bibinfo {author} {\bibfnamefont {S.-C.}\ \bibnamefont {Gou}},
  \bibinfo {author} {\bibfnamefont {Q.}~\bibnamefont {Sun}}, \bibinfo {author}
  {\bibfnamefont {L.}~\bibnamefont {Wen}}, \bibinfo {author} {\bibfnamefont
  {W.-M.}\ \bibnamefont {Liu}}, \bibinfo {author} {\bibfnamefont {A.-C.}\
  \bibnamefont {Ji}}, \bibinfo {author} {\bibfnamefont {J.}~\bibnamefont
  {Ruseckas}},\ and\ \bibinfo {author} {\bibfnamefont {G.}~\bibnamefont
  {Juzeliūnas}},\ }\href {https://doi.org/10.1103/PhysRevA.93.053630}
  {\bibfield  {journal} {\bibinfo  {journal} {Physical Review A}\ }\textbf
  {\bibinfo {volume} {93}},\ \bibinfo {pages} {053630} (\bibinfo {year}
  {2016})}\BibitemShut {NoStop}%
\bibitem [{\citenamefont {Wang}\ \emph
  {et~al.}(2017{\natexlab{a}})\citenamefont {Wang}, \citenamefont {Wen},
  \citenamefont {Yang}, \citenamefont {Shi},\ and\ \citenamefont
  {Li}}]{wang_vortex_2017}%
  \BibitemOpen
  \bibfield  {author} {\bibinfo {author} {\bibfnamefont {H.}~\bibnamefont
  {Wang}}, \bibinfo {author} {\bibfnamefont {L.}~\bibnamefont {Wen}}, \bibinfo
  {author} {\bibfnamefont {H.}~\bibnamefont {Yang}}, \bibinfo {author}
  {\bibfnamefont {C.}~\bibnamefont {Shi}},\ and\ \bibinfo {author}
  {\bibfnamefont {J.}~\bibnamefont {Li}},\ }\href
  {https://doi.org/10.1088/1361-6455/aa7afd} {\bibfield  {journal} {\bibinfo
  {journal} {Journal of Physics B: Atomic, Molecular and Optical Physics}\
  }\textbf {\bibinfo {volume} {50}},\ \bibinfo {pages} {155301} (\bibinfo
  {year} {2017}{\natexlab{a}})}\BibitemShut {NoStop}%
\bibitem [{\citenamefont {Wang}\ and\ \citenamefont
  {Li}(2020)}]{wang_vortex_2020}%
  \BibitemOpen
  \bibfield  {author} {\bibinfo {author} {\bibfnamefont {J.-G.}\ \bibnamefont
  {Wang}}\ and\ \bibinfo {author} {\bibfnamefont {Y.-Q.}\ \bibnamefont {Li}},\
  }\href {https://doi.org/10.1016/j.rinp.2020.103099} {\bibfield  {journal}
  {\bibinfo  {journal} {Results in Physics}\ }\textbf {\bibinfo {volume}
  {17}},\ \bibinfo {pages} {103099} (\bibinfo {year} {2020})}\BibitemShut
  {NoStop}%
\bibitem [{\citenamefont {Liu}\ \emph {et~al.}(2013)\citenamefont {Liu},
  \citenamefont {Yu}, \citenamefont {Gou},\ and\ \citenamefont
  {Liu}}]{liu_vortex_2013}%
  \BibitemOpen
  \bibfield  {author} {\bibinfo {author} {\bibfnamefont {C.-F.}\ \bibnamefont
  {Liu}}, \bibinfo {author} {\bibfnamefont {Y.-M.}\ \bibnamefont {Yu}},
  \bibinfo {author} {\bibfnamefont {S.-C.}\ \bibnamefont {Gou}},\ and\ \bibinfo
  {author} {\bibfnamefont {W.-M.}\ \bibnamefont {Liu}},\ }\href
  {https://doi.org/10.1103/PhysRevA.87.063630} {\bibfield  {journal} {\bibinfo
  {journal} {Physical Review A}\ }\textbf {\bibinfo {volume} {87}},\ \bibinfo
  {pages} {063630} (\bibinfo {year} {2013})}\BibitemShut {NoStop}%
\bibitem [{\citenamefont {Ramachandhran}\ \emph {et~al.}(2012)\citenamefont
  {Ramachandhran}, \citenamefont {Opanchuk}, \citenamefont {Liu}, \citenamefont
  {Pu}, \citenamefont {Drummond},\ and\ \citenamefont
  {Hu}}]{ramachandhran_half-quantum_2012}%
  \BibitemOpen
  \bibfield  {author} {\bibinfo {author} {\bibfnamefont {B.}~\bibnamefont
  {Ramachandhran}}, \bibinfo {author} {\bibfnamefont {B.}~\bibnamefont
  {Opanchuk}}, \bibinfo {author} {\bibfnamefont {X.-J.}\ \bibnamefont {Liu}},
  \bibinfo {author} {\bibfnamefont {H.}~\bibnamefont {Pu}}, \bibinfo {author}
  {\bibfnamefont {P.~D.}\ \bibnamefont {Drummond}},\ and\ \bibinfo {author}
  {\bibfnamefont {H.}~\bibnamefont {Hu}},\ }\href
  {https://doi.org/10.1103/PhysRevA.85.023606} {\bibfield  {journal} {\bibinfo
  {journal} {Physical Review A}\ }\textbf {\bibinfo {volume} {85}},\ \bibinfo
  {pages} {023606} (\bibinfo {year} {2012})}\BibitemShut {NoStop}%
\bibitem [{\citenamefont {Gautam}\ and\ \citenamefont
  {Adhikari}(2016)}]{gautam_fractional_2016}%
  \BibitemOpen
  \bibfield  {author} {\bibinfo {author} {\bibfnamefont {S.}~\bibnamefont
  {Gautam}}\ and\ \bibinfo {author} {\bibfnamefont {S.~K.}\ \bibnamefont
  {Adhikari}},\ }\href {https://doi.org/10.1103/PhysRevA.93.013630} {\bibfield
  {journal} {\bibinfo  {journal} {Phys. Rev. A}\ }\textbf {\bibinfo {volume}
  {93}},\ \bibinfo {pages} {013630} (\bibinfo {year} {2016})}\BibitemShut
  {NoStop}%
\bibitem [{\citenamefont {White}\ \emph {et~al.}(2017)\citenamefont {White},
  \citenamefont {Zhang},\ and\ \citenamefont
  {Busch}}]{white_odd-petal-number_2017}%
  \BibitemOpen
  \bibfield  {author} {\bibinfo {author} {\bibfnamefont {A.~C.}\ \bibnamefont
  {White}}, \bibinfo {author} {\bibfnamefont {Y.}~\bibnamefont {Zhang}},\ and\
  \bibinfo {author} {\bibfnamefont {T.}~\bibnamefont {Busch}},\ }\href
  {https://doi.org/10.1103/PhysRevA.95.041604} {\bibfield  {journal} {\bibinfo
  {journal} {Physical Review A}\ }\textbf {\bibinfo {volume} {95}},\ \bibinfo
  {pages} {041604} (\bibinfo {year} {2017})}\BibitemShut {NoStop}%
\bibitem [{\citenamefont {Al~Khawaja}\ and\ \citenamefont
  {Stoof}(2001)}]{al2001skyrmions}%
  \BibitemOpen
  \bibfield  {author} {\bibinfo {author} {\bibfnamefont {U.}~\bibnamefont
  {Al~Khawaja}}\ and\ \bibinfo {author} {\bibfnamefont {H.}~\bibnamefont
  {Stoof}},\ }\href {https://doi.org/10.1038/35082010} {\bibfield  {journal}
  {\bibinfo  {journal} {Nature}\ }\textbf {\bibinfo {volume} {411}},\ \bibinfo
  {pages} {918} (\bibinfo {year} {2001})}\BibitemShut {NoStop}%
\bibitem [{\citenamefont {Kawakami}\ \emph {et~al.}(2012)\citenamefont
  {Kawakami}, \citenamefont {Mizushima}, \citenamefont {Nitta},\ and\
  \citenamefont {Machida}}]{kawakami_stable_2012}%
  \BibitemOpen
  \bibfield  {author} {\bibinfo {author} {\bibfnamefont {T.}~\bibnamefont
  {Kawakami}}, \bibinfo {author} {\bibfnamefont {T.}~\bibnamefont {Mizushima}},
  \bibinfo {author} {\bibfnamefont {M.}~\bibnamefont {Nitta}},\ and\ \bibinfo
  {author} {\bibfnamefont {K.}~\bibnamefont {Machida}},\ }\href
  {https://doi.org/10.1103/PhysRevLett.109.015301} {\bibfield  {journal}
  {\bibinfo  {journal} {Physical Review Letters}\ }\textbf {\bibinfo {volume}
  {109}},\ \bibinfo {pages} {015301} (\bibinfo {year} {2012})}\BibitemShut
  {NoStop}%
\bibitem [{\citenamefont {Luo}\ \emph {et~al.}(2019)\citenamefont {Luo},
  \citenamefont {Li},\ and\ \citenamefont {Liu}}]{luo_three-dimensional_2019}%
  \BibitemOpen
  \bibfield  {author} {\bibinfo {author} {\bibfnamefont {H.-B.}\ \bibnamefont
  {Luo}}, \bibinfo {author} {\bibfnamefont {L.}~\bibnamefont {Li}},\ and\
  \bibinfo {author} {\bibfnamefont {W.-M.}\ \bibnamefont {Liu}},\ }\href
  {https://doi.org/10.1038/s41598-019-54856-x} {\bibfield  {journal} {\bibinfo
  {journal} {Scientific Reports}\ }\textbf {\bibinfo {volume} {9}},\ \bibinfo
  {pages} {18804} (\bibinfo {year} {2019})}\BibitemShut {NoStop}%
\bibitem [{\citenamefont {Su}\ \emph {et~al.}(2012)\citenamefont {Su},
  \citenamefont {Liu}, \citenamefont {Tsai}, \citenamefont {Liu},\ and\
  \citenamefont {Gou}}]{su_crystallized_2012}%
  \BibitemOpen
  \bibfield  {author} {\bibinfo {author} {\bibfnamefont {S.-W.}\ \bibnamefont
  {Su}}, \bibinfo {author} {\bibfnamefont {I.-K.}\ \bibnamefont {Liu}},
  \bibinfo {author} {\bibfnamefont {Y.-C.}\ \bibnamefont {Tsai}}, \bibinfo
  {author} {\bibfnamefont {W.~M.}\ \bibnamefont {Liu}},\ and\ \bibinfo {author}
  {\bibfnamefont {S.-C.}\ \bibnamefont {Gou}},\ }\href
  {https://doi.org/10.1103/PhysRevA.86.023601} {\bibfield  {journal} {\bibinfo
  {journal} {Physical Review A}\ }\textbf {\bibinfo {volume} {86}},\ \bibinfo
  {pages} {023601} (\bibinfo {year} {2012})}\BibitemShut {NoStop}%
\bibitem [{\citenamefont {Gautam}\ and\ \citenamefont
  {Adhikari}(2017)}]{gautam_vortex_2017}%
  \BibitemOpen
  \bibfield  {author} {\bibinfo {author} {\bibfnamefont {S.}~\bibnamefont
  {Gautam}}\ and\ \bibinfo {author} {\bibfnamefont {S.~K.}\ \bibnamefont
  {Adhikari}},\ }\href {https://doi.org/10.1103/PhysRevA.95.013608} {\bibfield
  {journal} {\bibinfo  {journal} {Phys. Rev. A}\ }\textbf {\bibinfo {volume}
  {95}},\ \bibinfo {pages} {013608} (\bibinfo {year} {2017})}\BibitemShut
  {NoStop}%
\bibitem [{\citenamefont {Gautam}\ and\ \citenamefont
  {Adhikari}(2018)}]{gautam_three_2018}%
  \BibitemOpen
  \bibfield  {author} {\bibinfo {author} {\bibfnamefont {S.}~\bibnamefont
  {Gautam}}\ and\ \bibinfo {author} {\bibfnamefont {S.~K.}\ \bibnamefont
  {Adhikari}},\ }\href {https://doi.org/10.1103/PhysRevA.97.013629} {\bibfield
  {journal} {\bibinfo  {journal} {Phys. Rev. A}\ }\textbf {\bibinfo {volume}
  {97}},\ \bibinfo {pages} {013629} (\bibinfo {year} {2018})}\BibitemShut
  {NoStop}%
\bibitem [{\citenamefont {Adhikari}(2021)}]{adhikari_symbiotic_2021}%
  \BibitemOpen
  \bibfield  {author} {\bibinfo {author} {\bibfnamefont {S.~K.}\ \bibnamefont
  {Adhikari}},\ }\href {https://doi.org/10.1103/PhysRevE.104.024207} {\bibfield
   {journal} {\bibinfo  {journal} {Phys. Rev. E}\ }\textbf {\bibinfo {volume}
  {104}},\ \bibinfo {pages} {024207} (\bibinfo {year} {2021})}\BibitemShut
  {NoStop}%
\bibitem [{\citenamefont {Meng}\ \emph {et~al.}(2022)\citenamefont {Meng},
  \citenamefont {Qin},\ and\ \citenamefont {Zhao}}]{meng_spin_2022}%
  \BibitemOpen
  \bibfield  {author} {\bibinfo {author} {\bibfnamefont {L.-Z.}\ \bibnamefont
  {Meng}}, \bibinfo {author} {\bibfnamefont {Y.-H.}\ \bibnamefont {Qin}},\ and\
  \bibinfo {author} {\bibfnamefont {L.-C.}\ \bibnamefont {Zhao}},\ }\href
  {https://doi.org/https://doi.org/10.1016/j.cnsns.2022.106286} {\bibfield
  {journal} {\bibinfo  {journal} {Communications in Nonlinear Science and
  Numerical Simulation}\ }\textbf {\bibinfo {volume} {109}},\ \bibinfo {pages}
  {106286} (\bibinfo {year} {2022})}\BibitemShut {NoStop}%
\bibitem [{\citenamefont {Chai}\ \emph {et~al.}(2020)\citenamefont {Chai},
  \citenamefont {Lao}, \citenamefont {Fujimoto}, \citenamefont {Hamazaki},
  \citenamefont {Ueda},\ and\ \citenamefont {Raman}}]{chai_magnetic_2020}%
  \BibitemOpen
  \bibfield  {author} {\bibinfo {author} {\bibfnamefont {X.}~\bibnamefont
  {Chai}}, \bibinfo {author} {\bibfnamefont {D.}~\bibnamefont {Lao}}, \bibinfo
  {author} {\bibfnamefont {K.}~\bibnamefont {Fujimoto}}, \bibinfo {author}
  {\bibfnamefont {R.}~\bibnamefont {Hamazaki}}, \bibinfo {author}
  {\bibfnamefont {M.}~\bibnamefont {Ueda}},\ and\ \bibinfo {author}
  {\bibfnamefont {C.}~\bibnamefont {Raman}},\ }\href
  {https://doi.org/10.1103/PhysRevLett.125.030402} {\bibfield  {journal}
  {\bibinfo  {journal} {Physical Review Letters}\ }\textbf {\bibinfo {volume}
  {125}},\ \bibinfo {pages} {030402} (\bibinfo {year} {2020})}\BibitemShut
  {NoStop}%
\bibitem [{\citenamefont {Hall}\ \emph {et~al.}(2016)\citenamefont {Hall},
  \citenamefont {Ray}, \citenamefont {Tiurev}, \citenamefont {Ruokokoski},
  \citenamefont {Gheorghe},\ and\ \citenamefont
  {Möttönen}}]{hall_tying_2016}%
  \BibitemOpen
  \bibfield  {author} {\bibinfo {author} {\bibfnamefont {D.~S.}\ \bibnamefont
  {Hall}}, \bibinfo {author} {\bibfnamefont {M.~W.}\ \bibnamefont {Ray}},
  \bibinfo {author} {\bibfnamefont {K.}~\bibnamefont {Tiurev}}, \bibinfo
  {author} {\bibfnamefont {E.}~\bibnamefont {Ruokokoski}}, \bibinfo {author}
  {\bibfnamefont {A.~H.}\ \bibnamefont {Gheorghe}},\ and\ \bibinfo {author}
  {\bibfnamefont {M.}~\bibnamefont {Möttönen}},\ }\href
  {https://doi.org/10.1038/nphys3624} {\bibfield  {journal} {\bibinfo
  {journal} {Nature Physics}\ }\textbf {\bibinfo {volume} {12}},\ \bibinfo
  {pages} {478} (\bibinfo {year} {2016})}\BibitemShut {NoStop}%
\bibitem [{\citenamefont {Kawaguchi}\ \emph {et~al.}(2008)\citenamefont
  {Kawaguchi}, \citenamefont {Nitta},\ and\ \citenamefont
  {Ueda}}]{kawaguchi_knots_2008}%
  \BibitemOpen
  \bibfield  {author} {\bibinfo {author} {\bibfnamefont {Y.}~\bibnamefont
  {Kawaguchi}}, \bibinfo {author} {\bibfnamefont {M.}~\bibnamefont {Nitta}},\
  and\ \bibinfo {author} {\bibfnamefont {M.}~\bibnamefont {Ueda}},\ }\href
  {https://doi.org/10.1103/PhysRevLett.100.180403} {\bibfield  {journal}
  {\bibinfo  {journal} {Physical Review Letters}\ }\textbf {\bibinfo {volume}
  {100}},\ \bibinfo {pages} {180403} (\bibinfo {year} {2008})}\BibitemShut
  {NoStop}%
\bibitem [{\citenamefont {Ho}\ and\ \citenamefont
  {Zhang}(2011)}]{Ho_bose_2011}%
  \BibitemOpen
  \bibfield  {author} {\bibinfo {author} {\bibfnamefont {T.-L.}\ \bibnamefont
  {Ho}}\ and\ \bibinfo {author} {\bibfnamefont {S.}~\bibnamefont {Zhang}},\
  }\href {https://doi.org/10.1103/PhysRevLett.107.150403} {\bibfield  {journal}
  {\bibinfo  {journal} {Phys. Rev. Lett.}\ }\textbf {\bibinfo {volume} {107}},\
  \bibinfo {pages} {150403} (\bibinfo {year} {2011})}\BibitemShut {NoStop}%
\bibitem [{\citenamefont {Li}\ \emph {et~al.}(2012)\citenamefont {Li},
  \citenamefont {Pitaevskii},\ and\ \citenamefont
  {Stringari}}]{li_quantum_2012}%
  \BibitemOpen
  \bibfield  {author} {\bibinfo {author} {\bibfnamefont {Y.}~\bibnamefont
  {Li}}, \bibinfo {author} {\bibfnamefont {L.~P.}\ \bibnamefont {Pitaevskii}},\
  and\ \bibinfo {author} {\bibfnamefont {S.}~\bibnamefont {Stringari}},\ }\href
  {https://doi.org/10.1103/PhysRevLett.108.225301} {\bibfield  {journal}
  {\bibinfo  {journal} {Phys. Rev. Lett.}\ }\textbf {\bibinfo {volume} {108}},\
  \bibinfo {pages} {225301} (\bibinfo {year} {2012})}\BibitemShut {NoStop}%
\bibitem [{\citenamefont {Li}\ \emph {et~al.}(2017)\citenamefont {Li},
  \citenamefont {Lee}, \citenamefont {Huang}, \citenamefont {Burchesky},
  \citenamefont {Shteynas}, \citenamefont {Top}, \citenamefont {Jamison},\ and\
  \citenamefont {Ketterle}}]{li_stripe_2017}%
  \BibitemOpen
  \bibfield  {author} {\bibinfo {author} {\bibfnamefont {J.-R.}\ \bibnamefont
  {Li}}, \bibinfo {author} {\bibfnamefont {J.}~\bibnamefont {Lee}}, \bibinfo
  {author} {\bibfnamefont {W.}~\bibnamefont {Huang}}, \bibinfo {author}
  {\bibfnamefont {S.}~\bibnamefont {Burchesky}}, \bibinfo {author}
  {\bibfnamefont {B.}~\bibnamefont {Shteynas}}, \bibinfo {author}
  {\bibfnamefont {F.~c.}\ \bibnamefont {Top}}, \bibinfo {author} {\bibfnamefont
  {A.~O.}\ \bibnamefont {Jamison}},\ and\ \bibinfo {author} {\bibfnamefont
  {W.}~\bibnamefont {Ketterle}},\ }\href {https://doi.org/10.1038/nature21431}
  {\bibfield  {journal} {\bibinfo  {journal} {Nature}\ }\textbf {\bibinfo
  {volume} {543}},\ \bibinfo {pages} {91} (\bibinfo {year} {2017})}\BibitemShut
  {NoStop}%
\bibitem [{\citenamefont {Zhao}\ \emph {et~al.}(2020)\citenamefont {Zhao},
  \citenamefont {Luo},\ and\ \citenamefont {Zhang}}]{zhao_magnetic_2020}%
  \BibitemOpen
  \bibfield  {author} {\bibinfo {author} {\bibfnamefont {L.-C.}\ \bibnamefont
  {Zhao}}, \bibinfo {author} {\bibfnamefont {X.-W.}\ \bibnamefont {Luo}},\ and\
  \bibinfo {author} {\bibfnamefont {C.}~\bibnamefont {Zhang}},\ }\href
  {https://doi.org/10.1103/PhysRevA.101.023621} {\bibfield  {journal} {\bibinfo
   {journal} {Physical Review A}\ }\textbf {\bibinfo {volume} {101}},\ \bibinfo
  {pages} {023621} (\bibinfo {year} {2020})}\BibitemShut {NoStop}%
\bibitem [{\citenamefont {Geier}\ \emph {et~al.}(2023)\citenamefont {Geier},
  \citenamefont {Martone}, \citenamefont {Hauke}, \citenamefont {Ketterle},\
  and\ \citenamefont {Stringari}}]{geier_dynamics_2023}%
  \BibitemOpen
  \bibfield  {author} {\bibinfo {author} {\bibfnamefont {K.~T.}\ \bibnamefont
  {Geier}}, \bibinfo {author} {\bibfnamefont {G.~I.}\ \bibnamefont {Martone}},
  \bibinfo {author} {\bibfnamefont {P.}~\bibnamefont {Hauke}}, \bibinfo
  {author} {\bibfnamefont {W.}~\bibnamefont {Ketterle}},\ and\ \bibinfo
  {author} {\bibfnamefont {S.}~\bibnamefont {Stringari}},\ }\href
  {https://doi.org/10.1103/PhysRevLett.130.156001} {\bibfield  {journal}
  {\bibinfo  {journal} {Phys. Rev. Lett.}\ }\textbf {\bibinfo {volume} {130}},\
  \bibinfo {pages} {156001} (\bibinfo {year} {2023})}\BibitemShut {NoStop}%
\bibitem [{\citenamefont {Martone}\ and\ \citenamefont
  {Stringari}(2021)}]{italo_supersolid_2021}%
  \BibitemOpen
  \bibfield  {author} {\bibinfo {author} {\bibfnamefont {G.~I.}\ \bibnamefont
  {Martone}}\ and\ \bibinfo {author} {\bibfnamefont {S.}~\bibnamefont
  {Stringari}},\ }\href {https://doi.org/10.21468/SciPostPhys.11.5.092}
  {\bibfield  {journal} {\bibinfo  {journal} {SciPost Phys.}\ }\textbf
  {\bibinfo {volume} {11}},\ \bibinfo {pages} {092} (\bibinfo {year}
  {2021})}\BibitemShut {NoStop}%
\bibitem [{\citenamefont {Huang}\ and\ \citenamefont
  {Hu}(2015)}]{huang_spin_2015}%
  \BibitemOpen
  \bibfield  {author} {\bibinfo {author} {\bibfnamefont {Y.}~\bibnamefont
  {Huang}}\ and\ \bibinfo {author} {\bibfnamefont {Z.-D.}\ \bibnamefont {Hu}},\
  }\href {https://doi.org/10.1038/srep08006} {\bibfield  {journal} {\bibinfo
  {journal} {Scientific Reports}\ }\textbf {\bibinfo {volume} {5}},\ \bibinfo
  {pages} {8006} (\bibinfo {year} {2015})}\BibitemShut {NoStop}%
\bibitem [{\citenamefont {Chen}\ \emph
  {et~al.}(2020{\natexlab{a}})\citenamefont {Chen}, \citenamefont {Zhang},\
  and\ \citenamefont {Pu}}]{chen_spin_2020}%
  \BibitemOpen
  \bibfield  {author} {\bibinfo {author} {\bibfnamefont {L.}~\bibnamefont
  {Chen}}, \bibinfo {author} {\bibfnamefont {Y.}~\bibnamefont {Zhang}},\ and\
  \bibinfo {author} {\bibfnamefont {H.}~\bibnamefont {Pu}},\ }\href
  {https://doi.org/10.1103/PhysRevA.102.023317} {\bibfield  {journal} {\bibinfo
   {journal} {Physical Review A}\ }\textbf {\bibinfo {volume} {102}},\ \bibinfo
  {pages} {023317} (\bibinfo {year} {2020}{\natexlab{a}})}\BibitemShut
  {NoStop}%
\bibitem [{\citenamefont {Luo}\ \emph {et~al.}(2017)\citenamefont {Luo},
  \citenamefont {Sun},\ and\ \citenamefont {Zhang}}]{luo_spin_2017}%
  \BibitemOpen
  \bibfield  {author} {\bibinfo {author} {\bibfnamefont {X.-W.}\ \bibnamefont
  {Luo}}, \bibinfo {author} {\bibfnamefont {K.}~\bibnamefont {Sun}},\ and\
  \bibinfo {author} {\bibfnamefont {C.}~\bibnamefont {Zhang}},\ }\href
  {https://doi.org/10.1103/PhysRevLett.119.193001} {\bibfield  {journal}
  {\bibinfo  {journal} {Phys. Rev. Lett.}\ }\textbf {\bibinfo {volume} {119}},\
  \bibinfo {pages} {193001} (\bibinfo {year} {2017})}\BibitemShut {NoStop}%
\bibitem [{\citenamefont {Hu}\ \emph {et~al.}(2018)\citenamefont {Hu},
  \citenamefont {Hou}, \citenamefont {Zhang},\ and\ \citenamefont
  {Zhang}}]{Hu_topological_2018}%
  \BibitemOpen
  \bibfield  {author} {\bibinfo {author} {\bibfnamefont {H.}~\bibnamefont
  {Hu}}, \bibinfo {author} {\bibfnamefont {J.}~\bibnamefont {Hou}}, \bibinfo
  {author} {\bibfnamefont {F.}~\bibnamefont {Zhang}},\ and\ \bibinfo {author}
  {\bibfnamefont {C.}~\bibnamefont {Zhang}},\ }\href
  {https://doi.org/10.1103/PhysRevLett.120.240401} {\bibfield  {journal}
  {\bibinfo  {journal} {Phys. Rev. Lett.}\ }\textbf {\bibinfo {volume} {120}},\
  \bibinfo {pages} {240401} (\bibinfo {year} {2018})}\BibitemShut {NoStop}%
\bibitem [{\citenamefont {Lei}\ \emph {et~al.}(2020)\citenamefont {Lei},
  \citenamefont {Deng},\ and\ \citenamefont {Lee}}]{lei_symmetry_2020}%
  \BibitemOpen
  \bibfield  {author} {\bibinfo {author} {\bibfnamefont {Z.}~\bibnamefont
  {Lei}}, \bibinfo {author} {\bibfnamefont {Y.}~\bibnamefont {Deng}},\ and\
  \bibinfo {author} {\bibfnamefont {C.}~\bibnamefont {Lee}},\ }\href
  {https://doi.org/10.1103/PhysRevA.102.013301} {\bibfield  {journal} {\bibinfo
   {journal} {Phys. Rev. A}\ }\textbf {\bibinfo {volume} {102}},\ \bibinfo
  {pages} {013301} (\bibinfo {year} {2020})}\BibitemShut {NoStop}%
\bibitem [{\citenamefont {Sun}\ \emph {et~al.}(2020)\citenamefont {Sun},
  \citenamefont {Chen}, \citenamefont {Chen},\ and\ \citenamefont
  {Zhang}}]{sun_bright_2020}%
  \BibitemOpen
  \bibfield  {author} {\bibinfo {author} {\bibfnamefont {J.}~\bibnamefont
  {Sun}}, \bibinfo {author} {\bibfnamefont {Y.}~\bibnamefont {Chen}}, \bibinfo
  {author} {\bibfnamefont {X.}~\bibnamefont {Chen}},\ and\ \bibinfo {author}
  {\bibfnamefont {Y.}~\bibnamefont {Zhang}},\ }\href
  {https://doi.org/10.1103/PhysRevA.101.053621} {\bibfield  {journal} {\bibinfo
   {journal} {Phys. Rev. A}\ }\textbf {\bibinfo {volume} {101}},\ \bibinfo
  {pages} {053621} (\bibinfo {year} {2020})}\BibitemShut {NoStop}%
\bibitem [{\citenamefont {Chen}\ \emph
  {et~al.}(2020{\natexlab{b}})\citenamefont {Chen}, \citenamefont {Zhang},\
  and\ \citenamefont {Pu}}]{chen_spin_nematic_2020}%
  \BibitemOpen
  \bibfield  {author} {\bibinfo {author} {\bibfnamefont {L.}~\bibnamefont
  {Chen}}, \bibinfo {author} {\bibfnamefont {Y.}~\bibnamefont {Zhang}},\ and\
  \bibinfo {author} {\bibfnamefont {H.}~\bibnamefont {Pu}},\ }\href
  {https://doi.org/10.1103/PhysRevLett.125.195303} {\bibfield  {journal}
  {\bibinfo  {journal} {Phys. Rev. Lett.}\ }\textbf {\bibinfo {volume} {125}},\
  \bibinfo {pages} {195303} (\bibinfo {year} {2020}{\natexlab{b}})}\BibitemShut
  {NoStop}%
\bibitem [{\citenamefont {Lei}\ \emph {et~al.}(2022)\citenamefont {Lei},
  \citenamefont {Deng},\ and\ \citenamefont {Lee}}]{lei_unpaired_2022}%
  \BibitemOpen
  \bibfield  {author} {\bibinfo {author} {\bibfnamefont {Z.}~\bibnamefont
  {Lei}}, \bibinfo {author} {\bibfnamefont {Y.}~\bibnamefont {Deng}},\ and\
  \bibinfo {author} {\bibfnamefont {C.}~\bibnamefont {Lee}},\ }\href
  {https://doi.org/10.1103/PhysRevResearch.4.033008} {\bibfield  {journal}
  {\bibinfo  {journal} {Phys. Rev. Res.}\ }\textbf {\bibinfo {volume} {4}},\
  \bibinfo {pages} {033008} (\bibinfo {year} {2022})}\BibitemShut {NoStop}%
\bibitem [{\citenamefont {Qiu}\ \emph {et~al.}(2023)\citenamefont {Qiu},
  \citenamefont {Hu}, \citenamefont {Cai}, \citenamefont {Saito}, \citenamefont
  {Zhang},\ and\ \citenamefont {Wen}}]{Qiu_dynamics_2023}%
  \BibitemOpen
  \bibfield  {author} {\bibinfo {author} {\bibfnamefont {X.}~\bibnamefont
  {Qiu}}, \bibinfo {author} {\bibfnamefont {A.-Y.}\ \bibnamefont {Hu}},
  \bibinfo {author} {\bibfnamefont {Y.}~\bibnamefont {Cai}}, \bibinfo {author}
  {\bibfnamefont {H.}~\bibnamefont {Saito}}, \bibinfo {author} {\bibfnamefont
  {X.-F.}\ \bibnamefont {Zhang}},\ and\ \bibinfo {author} {\bibfnamefont
  {L.}~\bibnamefont {Wen}},\ }\href
  {https://doi.org/10.1103/PhysRevA.107.033308} {\bibfield  {journal} {\bibinfo
   {journal} {Phys. Rev. A}\ }\textbf {\bibinfo {volume} {107}},\ \bibinfo
  {pages} {033308} (\bibinfo {year} {2023})}\BibitemShut {NoStop}%
\bibitem [{\citenamefont {DeMarco}\ and\ \citenamefont
  {Pu}(2015)}]{DeMarco_angular_2015}%
  \BibitemOpen
  \bibfield  {author} {\bibinfo {author} {\bibfnamefont {M.}~\bibnamefont
  {DeMarco}}\ and\ \bibinfo {author} {\bibfnamefont {H.}~\bibnamefont {Pu}},\
  }\href {https://doi.org/10.1103/PhysRevA.91.033630} {\bibfield  {journal}
  {\bibinfo  {journal} {Phys. Rev. A}\ }\textbf {\bibinfo {volume} {91}},\
  \bibinfo {pages} {033630} (\bibinfo {year} {2015})}\BibitemShut {NoStop}%
\bibitem [{\citenamefont {Sun}\ \emph {et~al.}(2015)\citenamefont {Sun},
  \citenamefont {Qu},\ and\ \citenamefont {Zhang}}]{sun_spin_2015}%
  \BibitemOpen
  \bibfield  {author} {\bibinfo {author} {\bibfnamefont {K.}~\bibnamefont
  {Sun}}, \bibinfo {author} {\bibfnamefont {C.}~\bibnamefont {Qu}},\ and\
  \bibinfo {author} {\bibfnamefont {C.}~\bibnamefont {Zhang}},\ }\href
  {https://doi.org/10.1103/PhysRevA.91.063627} {\bibfield  {journal} {\bibinfo
  {journal} {Phys. Rev. A}\ }\textbf {\bibinfo {volume} {91}},\ \bibinfo
  {pages} {063627} (\bibinfo {year} {2015})}\BibitemShut {NoStop}%
\bibitem [{\citenamefont {Chen}\ \emph {et~al.}(2016)\citenamefont {Chen},
  \citenamefont {Pu},\ and\ \citenamefont {Zhang}}]{chen_spin_2016}%
  \BibitemOpen
  \bibfield  {author} {\bibinfo {author} {\bibfnamefont {L.}~\bibnamefont
  {Chen}}, \bibinfo {author} {\bibfnamefont {H.}~\bibnamefont {Pu}},\ and\
  \bibinfo {author} {\bibfnamefont {Y.}~\bibnamefont {Zhang}},\ }\href
  {https://doi.org/10.1103/PhysRevA.93.013629} {\bibfield  {journal} {\bibinfo
  {journal} {Phys. Rev. A}\ }\textbf {\bibinfo {volume} {93}},\ \bibinfo
  {pages} {013629} (\bibinfo {year} {2016})}\BibitemShut {NoStop}%
\bibitem [{\citenamefont {Chen}\ \emph {et~al.}(2018)\citenamefont {Chen},
  \citenamefont {Lin}, \citenamefont {Chen}, \citenamefont {Chiu},
  \citenamefont {Wang}, \citenamefont {Chen}, \citenamefont {Huang},
  \citenamefont {Yip}, \citenamefont {Kawaguchi},\ and\ \citenamefont
  {Lin}}]{chen_spinorbital_2018}%
  \BibitemOpen
  \bibfield  {author} {\bibinfo {author} {\bibfnamefont {H.-R.}\ \bibnamefont
  {Chen}}, \bibinfo {author} {\bibfnamefont {K.-Y.}\ \bibnamefont {Lin}},
  \bibinfo {author} {\bibfnamefont {P.-K.}\ \bibnamefont {Chen}}, \bibinfo
  {author} {\bibfnamefont {N.-C.}\ \bibnamefont {Chiu}}, \bibinfo {author}
  {\bibfnamefont {J.-B.}\ \bibnamefont {Wang}}, \bibinfo {author}
  {\bibfnamefont {C.-A.}\ \bibnamefont {Chen}}, \bibinfo {author}
  {\bibfnamefont {P.}~\bibnamefont {Huang}}, \bibinfo {author} {\bibfnamefont
  {S.-K.}\ \bibnamefont {Yip}}, \bibinfo {author} {\bibfnamefont
  {Y.}~\bibnamefont {Kawaguchi}},\ and\ \bibinfo {author} {\bibfnamefont
  {Y.-J.}\ \bibnamefont {Lin}},\ }\href
  {https://doi.org/10.1103/PhysRevLett.121.113204} {\bibfield  {journal}
  {\bibinfo  {journal} {Phys. Rev. Lett.}\ }\textbf {\bibinfo {volume} {121}},\
  \bibinfo {pages} {113204} (\bibinfo {year} {2018})}\BibitemShut {NoStop}%
\bibitem [{\citenamefont {Zhang}\ \emph {et~al.}(2019)\citenamefont {Zhang},
  \citenamefont {Gao}, \citenamefont {Zou}, \citenamefont {Kong}, \citenamefont
  {Li}, \citenamefont {Shen}, \citenamefont {Chen}, \citenamefont {Peng},
  \citenamefont {Zhan}, \citenamefont {Pu},\ and\ \citenamefont
  {Jiang}}]{zhang_ground_2019}%
  \BibitemOpen
  \bibfield  {author} {\bibinfo {author} {\bibfnamefont {D.}~\bibnamefont
  {Zhang}}, \bibinfo {author} {\bibfnamefont {T.}~\bibnamefont {Gao}}, \bibinfo
  {author} {\bibfnamefont {P.}~\bibnamefont {Zou}}, \bibinfo {author}
  {\bibfnamefont {L.}~\bibnamefont {Kong}}, \bibinfo {author} {\bibfnamefont
  {R.}~\bibnamefont {Li}}, \bibinfo {author} {\bibfnamefont {X.}~\bibnamefont
  {Shen}}, \bibinfo {author} {\bibfnamefont {X.-L.}\ \bibnamefont {Chen}},
  \bibinfo {author} {\bibfnamefont {S.-G.}\ \bibnamefont {Peng}}, \bibinfo
  {author} {\bibfnamefont {M.}~\bibnamefont {Zhan}}, \bibinfo {author}
  {\bibfnamefont {H.}~\bibnamefont {Pu}},\ and\ \bibinfo {author}
  {\bibfnamefont {K.}~\bibnamefont {Jiang}},\ }\href
  {https://doi.org/10.1103/PhysRevLett.122.110402} {\bibfield  {journal}
  {\bibinfo  {journal} {Phys. Rev. Lett.}\ }\textbf {\bibinfo {volume} {122}},\
  \bibinfo {pages} {110402} (\bibinfo {year} {2019})}\BibitemShut {NoStop}%
\bibitem [{\citenamefont {Li}\ \emph {et~al.}(2022)\citenamefont {Li},
  \citenamefont {Li},\ and\ \citenamefont {Shi}}]{li_phase_2022}%
  \BibitemOpen
  \bibfield  {author} {\bibinfo {author} {\bibfnamefont {Z.}~\bibnamefont
  {Li}}, \bibinfo {author} {\bibfnamefont {J.}~\bibnamefont {Li}},\ and\
  \bibinfo {author} {\bibfnamefont {D.}~\bibnamefont {Shi}},\ }\bibfield
  {journal} {\bibinfo  {journal} {Modern Physics Letters B}\ }\href
  {https://doi.org/10.1142/S0217984922500117} {10.1142/S0217984922500117}
  (\bibinfo {year} {2022})\BibitemShut {NoStop}%
\bibitem [{\citenamefont {Bidasyuk}\ \emph {et~al.}(2022)\citenamefont
  {Bidasyuk}, \citenamefont {Kovtunenko},\ and\ \citenamefont
  {Prikhodko}}]{bidasyuk_fine_2022}%
  \BibitemOpen
  \bibfield  {author} {\bibinfo {author} {\bibfnamefont {Y.~M.}\ \bibnamefont
  {Bidasyuk}}, \bibinfo {author} {\bibfnamefont {K.~S.}\ \bibnamefont
  {Kovtunenko}},\ and\ \bibinfo {author} {\bibfnamefont {O.~O.}\ \bibnamefont
  {Prikhodko}},\ }\href {https://doi.org/10.1103/PhysRevA.105.023320}
  {\bibfield  {journal} {\bibinfo  {journal} {Phys. Rev. A}\ }\textbf {\bibinfo
  {volume} {105}},\ \bibinfo {pages} {023320} (\bibinfo {year}
  {2022})}\BibitemShut {NoStop}%
\bibitem [{\citenamefont {Han}\ \emph {et~al.}(2022)\citenamefont {Han},
  \citenamefont {Peng}, \citenamefont {Chen},\ and\ \citenamefont
  {Yi}}]{han_molecular_2022}%
  \BibitemOpen
  \bibfield  {author} {\bibinfo {author} {\bibfnamefont {Y.}~\bibnamefont
  {Han}}, \bibinfo {author} {\bibfnamefont {S.-G.}\ \bibnamefont {Peng}},
  \bibinfo {author} {\bibfnamefont {K.-J.}\ \bibnamefont {Chen}},\ and\
  \bibinfo {author} {\bibfnamefont {W.}~\bibnamefont {Yi}},\ }\href
  {https://doi.org/10.1103/PhysRevA.106.043302} {\bibfield  {journal} {\bibinfo
   {journal} {Phys. Rev. A}\ }\textbf {\bibinfo {volume} {106}},\ \bibinfo
  {pages} {043302} (\bibinfo {year} {2022})}\BibitemShut {NoStop}%
\bibitem [{\citenamefont {Peng}\ \emph {et~al.}(2022)\citenamefont {Peng},
  \citenamefont {Jiang}, \citenamefont {Chen}, \citenamefont {Chen},
  \citenamefont {Zou},\ and\ \citenamefont {He}}]{peng2022spin}%
  \BibitemOpen
  \bibfield  {author} {\bibinfo {author} {\bibfnamefont {S.-G.}\ \bibnamefont
  {Peng}}, \bibinfo {author} {\bibfnamefont {K.}~\bibnamefont {Jiang}},
  \bibinfo {author} {\bibfnamefont {X.-L.}\ \bibnamefont {Chen}}, \bibinfo
  {author} {\bibfnamefont {K.-J.}\ \bibnamefont {Chen}}, \bibinfo {author}
  {\bibfnamefont {P.}~\bibnamefont {Zou}},\ and\ \bibinfo {author}
  {\bibfnamefont {L.}~\bibnamefont {He}},\ }\href
  {https://doi.org/https://doi.org/10.1007/s43673-022-00069-w} {\bibfield
  {journal} {\bibinfo  {journal} {AAPPS Bulletin}\ }\textbf {\bibinfo {volume}
  {32}},\ \bibinfo {pages} {36} (\bibinfo {year} {2022})}\BibitemShut {NoStop}%
\bibitem [{\citenamefont {Ng}\ and\ \citenamefont {Ooi}(2023)}]{NG_spin_2023}%
  \BibitemOpen
  \bibfield  {author} {\bibinfo {author} {\bibfnamefont {E.~B.}\ \bibnamefont
  {Ng}}\ and\ \bibinfo {author} {\bibfnamefont {C.~R.}\ \bibnamefont {Ooi}},\
  }\href {https://doi.org/https://doi.org/10.1016/j.rinp.2023.106870}
  {\bibfield  {journal} {\bibinfo  {journal} {Results in Physics}\ }\textbf
  {\bibinfo {volume} {52}},\ \bibinfo {pages} {106870} (\bibinfo {year}
  {2023})}\BibitemShut {NoStop}%
\bibitem [{\citenamefont {Xiao}\ \emph {et~al.}(2021)\citenamefont {Xiao},
  \citenamefont {Borgh}, \citenamefont {Weiss}, \citenamefont {Blinova},
  \citenamefont {Ruostekoski},\ and\ \citenamefont
  {Hall}}]{xiao_controlled_2021}%
  \BibitemOpen
  \bibfield  {author} {\bibinfo {author} {\bibfnamefont {Y.}~\bibnamefont
  {Xiao}}, \bibinfo {author} {\bibfnamefont {M.~O.}\ \bibnamefont {Borgh}},
  \bibinfo {author} {\bibfnamefont {L.~S.}\ \bibnamefont {Weiss}}, \bibinfo
  {author} {\bibfnamefont {A.~A.}\ \bibnamefont {Blinova}}, \bibinfo {author}
  {\bibfnamefont {J.}~\bibnamefont {Ruostekoski}},\ and\ \bibinfo {author}
  {\bibfnamefont {D.~S.}\ \bibnamefont {Hall}},\ }\href
  {https://doi.org/10.1038/s42005-021-00554-y} {\bibfield  {journal} {\bibinfo
  {journal} {Communications Physics}\ }\textbf {\bibinfo {volume} {4}},\
  \bibinfo {pages} {1} (\bibinfo {year} {2021})}\BibitemShut {NoStop}%
\bibitem [{\citenamefont {Martone}\ \emph {et~al.}(2016)\citenamefont
  {Martone}, \citenamefont {Pepe}, \citenamefont {Facchi}, \citenamefont
  {Pascazio},\ and\ \citenamefont {Stringari}}]{martone_Tricriticalities_2016}%
  \BibitemOpen
  \bibfield  {author} {\bibinfo {author} {\bibfnamefont {G.~I.}\ \bibnamefont
  {Martone}}, \bibinfo {author} {\bibfnamefont {F.~V.}\ \bibnamefont {Pepe}},
  \bibinfo {author} {\bibfnamefont {P.}~\bibnamefont {Facchi}}, \bibinfo
  {author} {\bibfnamefont {S.}~\bibnamefont {Pascazio}},\ and\ \bibinfo
  {author} {\bibfnamefont {S.}~\bibnamefont {Stringari}},\ }\href
  {https://doi.org/10.1103/PhysRevLett.117.125301} {\bibfield  {journal}
  {\bibinfo  {journal} {Phys. Rev. Lett.}\ }\textbf {\bibinfo {volume} {117}},\
  \bibinfo {pages} {125301} (\bibinfo {year} {2016})}\BibitemShut {NoStop}%
\bibitem [{\citenamefont {Gui}\ \emph {et~al.}(2023)\citenamefont {Gui},
  \citenamefont {Zhang}, \citenamefont {Su}, \citenamefont {Lyu},\ and\
  \citenamefont {Zhang}}]{gui_spin_2023}%
  \BibitemOpen
  \bibfield  {author} {\bibinfo {author} {\bibfnamefont {Z.}~\bibnamefont
  {Gui}}, \bibinfo {author} {\bibfnamefont {Z.}~\bibnamefont {Zhang}}, \bibinfo
  {author} {\bibfnamefont {J.}~\bibnamefont {Su}}, \bibinfo {author}
  {\bibfnamefont {H.}~\bibnamefont {Lyu}},\ and\ \bibinfo {author}
  {\bibfnamefont {Y.}~\bibnamefont {Zhang}},\ }\href
  {https://doi.org/10.1103/PhysRevA.108.043311} {\bibfield  {journal} {\bibinfo
   {journal} {Phys. Rev. A}\ }\textbf {\bibinfo {volume} {108}},\ \bibinfo
  {pages} {043311} (\bibinfo {year} {2023})}\BibitemShut {NoStop}%
\bibitem [{\citenamefont {Wang}\ \emph
  {et~al.}(2017{\natexlab{b}})\citenamefont {Wang}, \citenamefont {Xu},\ and\
  \citenamefont {Yang}}]{wang_ground-state_2017}%
  \BibitemOpen
  \bibfield  {author} {\bibinfo {author} {\bibfnamefont {J.-G.}\ \bibnamefont
  {Wang}}, \bibinfo {author} {\bibfnamefont {L.-L.}\ \bibnamefont {Xu}},\ and\
  \bibinfo {author} {\bibfnamefont {S.-J.}\ \bibnamefont {Yang}},\ }\href
  {https://doi.org/10.1103/PhysRevA.96.033629} {\bibfield  {journal} {\bibinfo
  {journal} {Physical Review A}\ }\textbf {\bibinfo {volume} {96}},\ \bibinfo
  {pages} {033629} (\bibinfo {year} {2017}{\natexlab{b}})}\BibitemShut
  {NoStop}%
\bibitem [{\citenamefont {Ho}(1998)}]{ho_spinor_1998}%
  \BibitemOpen
  \bibfield  {author} {\bibinfo {author} {\bibfnamefont {T.-L.}\ \bibnamefont
  {Ho}},\ }\href {https://doi.org/10.1103/PhysRevLett.81.742} {\bibfield
  {journal} {\bibinfo  {journal} {Physical Review Letters}\ }\textbf {\bibinfo
  {volume} {81}},\ \bibinfo {pages} {742} (\bibinfo {year} {1998})}\BibitemShut
  {NoStop}%
\bibitem [{\citenamefont {Ohmi}\ and\ \citenamefont
  {Machida}(1998)}]{ohmi_bose_1998}%
  \BibitemOpen
  \bibfield  {author} {\bibinfo {author} {\bibfnamefont {T.}~\bibnamefont
  {Ohmi}}\ and\ \bibinfo {author} {\bibfnamefont {K.}~\bibnamefont {Machida}},\
  }\href {https://doi.org/10.1143/JPSJ.67.1822} {\bibfield  {journal} {\bibinfo
   {journal} {Journal of the Physical Society of Japan}\ }\textbf {\bibinfo
  {volume} {67}},\ \bibinfo {pages} {1822} (\bibinfo {year}
  {1998})}\BibitemShut {NoStop}%
\bibitem [{\citenamefont {Stamper-Kurn}\ \emph {et~al.}(1998)\citenamefont
  {Stamper-Kurn}, \citenamefont {Andrews}, \citenamefont {Chikkatur},
  \citenamefont {Inouye}, \citenamefont {Miesner}, \citenamefont {Stenger},\
  and\ \citenamefont {Ketterle}}]{stamper-kurn_optical_1998}%
  \BibitemOpen
  \bibfield  {author} {\bibinfo {author} {\bibfnamefont {D.~M.}\ \bibnamefont
  {Stamper-Kurn}}, \bibinfo {author} {\bibfnamefont {M.~R.}\ \bibnamefont
  {Andrews}}, \bibinfo {author} {\bibfnamefont {A.~P.}\ \bibnamefont
  {Chikkatur}}, \bibinfo {author} {\bibfnamefont {S.}~\bibnamefont {Inouye}},
  \bibinfo {author} {\bibfnamefont {H.-J.}\ \bibnamefont {Miesner}}, \bibinfo
  {author} {\bibfnamefont {J.}~\bibnamefont {Stenger}},\ and\ \bibinfo {author}
  {\bibfnamefont {W.}~\bibnamefont {Ketterle}},\ }\href
  {https://doi.org/10.1103/PhysRevLett.80.2027} {\bibfield  {journal} {\bibinfo
   {journal} {Physical Review Letters}\ }\textbf {\bibinfo {volume} {80}},\
  \bibinfo {pages} {2027} (\bibinfo {year} {1998})}\BibitemShut {NoStop}%
\bibitem [{\citenamefont {Lan}\ and\ \citenamefont
  {\"Ohberg}(2014)}]{Lan_raman_2014}%
  \BibitemOpen
  \bibfield  {author} {\bibinfo {author} {\bibfnamefont {Z.}~\bibnamefont
  {Lan}}\ and\ \bibinfo {author} {\bibfnamefont {P.}~\bibnamefont {\"Ohberg}},\
  }\href {https://doi.org/10.1103/PhysRevA.89.023630} {\bibfield  {journal}
  {\bibinfo  {journal} {Phys. Rev. A}\ }\textbf {\bibinfo {volume} {89}},\
  \bibinfo {pages} {023630} (\bibinfo {year} {2014})}\BibitemShut {NoStop}%
\bibitem [{\citenamefont {Yu}(2016)}]{yu_phase_2016}%
  \BibitemOpen
  \bibfield  {author} {\bibinfo {author} {\bibfnamefont {Z.-Q.}\ \bibnamefont
  {Yu}},\ }\href {https://doi.org/10.1103/PhysRevA.93.033648} {\bibfield
  {journal} {\bibinfo  {journal} {Phys. Rev. A}\ }\textbf {\bibinfo {volume}
  {93}},\ \bibinfo {pages} {033648} (\bibinfo {year} {2016})}\BibitemShut
  {NoStop}%
\bibitem [{\citenamefont {Symes}\ \emph {et~al.}(2018)\citenamefont {Symes},
  \citenamefont {Baillie},\ and\ \citenamefont {Blakie}}]{symes_dynamics_2018}%
  \BibitemOpen
  \bibfield  {author} {\bibinfo {author} {\bibfnamefont {L.~M.}\ \bibnamefont
  {Symes}}, \bibinfo {author} {\bibfnamefont {D.}~\bibnamefont {Baillie}},\
  and\ \bibinfo {author} {\bibfnamefont {P.~B.}\ \bibnamefont {Blakie}},\
  }\href {https://doi.org/10.1103/PhysRevA.98.063618} {\bibfield  {journal}
  {\bibinfo  {journal} {Phys. Rev. A}\ }\textbf {\bibinfo {volume} {98}},\
  \bibinfo {pages} {063618} (\bibinfo {year} {2018})}\BibitemShut {NoStop}%
\bibitem [{\citenamefont {Cheng}\ \emph {et~al.}(2014)\citenamefont {Cheng},
  \citenamefont {Tang},\ and\ \citenamefont
  {Adhikari}}]{cheng_Localization_2014}%
  \BibitemOpen
  \bibfield  {author} {\bibinfo {author} {\bibfnamefont {Y.}~\bibnamefont
  {Cheng}}, \bibinfo {author} {\bibfnamefont {G.}~\bibnamefont {Tang}},\ and\
  \bibinfo {author} {\bibfnamefont {S.~K.}\ \bibnamefont {Adhikari}},\ }\href
  {https://doi.org/10.1103/PhysRevA.89.063602} {\bibfield  {journal} {\bibinfo
  {journal} {Phys. Rev. A}\ }\textbf {\bibinfo {volume} {89}},\ \bibinfo
  {pages} {063602} (\bibinfo {year} {2014})}\BibitemShut {NoStop}%
\bibitem [{\citenamefont {Oztas}(2019)}]{oztas_spin_2019}%
  \BibitemOpen
  \bibfield  {author} {\bibinfo {author} {\bibfnamefont {Z.}~\bibnamefont
  {Oztas}},\ }\href {https://doi.org/10.1016/j.physleta.2018.11.022} {\bibfield
   {journal} {\bibinfo  {journal} {Physics Letters A}\ }\textbf {\bibinfo
  {volume} {383}},\ \bibinfo {pages} {504} (\bibinfo {year}
  {2019})}\BibitemShut {NoStop}%
\bibitem [{\citenamefont {Symes}\ and\ \citenamefont
  {Blakie}(2017)}]{symes_nematic_2017}%
  \BibitemOpen
  \bibfield  {author} {\bibinfo {author} {\bibfnamefont {L.~M.}\ \bibnamefont
  {Symes}}\ and\ \bibinfo {author} {\bibfnamefont {P.~B.}\ \bibnamefont
  {Blakie}},\ }\href {https://doi.org/10.1103/PhysRevA.96.013602} {\bibfield
  {journal} {\bibinfo  {journal} {Phys. Rev. A}\ }\textbf {\bibinfo {volume}
  {96}},\ \bibinfo {pages} {013602} (\bibinfo {year} {2017})}\BibitemShut
  {NoStop}%
\bibitem [{\citenamefont {Ueda}(2014)}]{Ueda_topological_2014}%
  \BibitemOpen
  \bibfield  {author} {\bibinfo {author} {\bibfnamefont {M.}~\bibnamefont
  {Ueda}},\ }\href {https://doi.org/10.1088/0034-4885/77/12/122401} {\bibfield
  {journal} {\bibinfo  {journal} {Reports on Progress in Physics}\ }\textbf
  {\bibinfo {volume} {77}},\ \bibinfo {pages} {122401} (\bibinfo {year}
  {2014})}\BibitemShut {NoStop}%
\bibitem [{\citenamefont {Hong}\ \emph {et~al.}(2023)\citenamefont {Hong},
  \citenamefont {Lee}, \citenamefont {Kim}, \citenamefont {Jung}, \citenamefont
  {Lee}, \citenamefont {Kang},\ and\ \citenamefont {Shin}}]{hong_spin_2023}%
  \BibitemOpen
  \bibfield  {author} {\bibinfo {author} {\bibfnamefont {D.}~\bibnamefont
  {Hong}}, \bibinfo {author} {\bibfnamefont {J.}~\bibnamefont {Lee}}, \bibinfo
  {author} {\bibfnamefont {J.}~\bibnamefont {Kim}}, \bibinfo {author}
  {\bibfnamefont {J.~H.}\ \bibnamefont {Jung}}, \bibinfo {author}
  {\bibfnamefont {K.}~\bibnamefont {Lee}}, \bibinfo {author} {\bibfnamefont
  {S.}~\bibnamefont {Kang}},\ and\ \bibinfo {author} {\bibfnamefont
  {Y.}~\bibnamefont {Shin}},\ }\href
  {https://doi.org/10.1103/PhysRevA.108.013318} {\bibfield  {journal} {\bibinfo
   {journal} {Phys. Rev. A}\ }\textbf {\bibinfo {volume} {108}},\ \bibinfo
  {pages} {013318} (\bibinfo {year} {2023})}\BibitemShut {NoStop}%
\bibitem [{\citenamefont {Jung}\ \emph {et~al.}(2023)\citenamefont {Jung},
  \citenamefont {Lee}, \citenamefont {Kim},\ and\ \citenamefont
  {Shin}}]{jung_random_2023}%
  \BibitemOpen
  \bibfield  {author} {\bibinfo {author} {\bibfnamefont {J.~H.}\ \bibnamefont
  {Jung}}, \bibinfo {author} {\bibfnamefont {J.}~\bibnamefont {Lee}}, \bibinfo
  {author} {\bibfnamefont {J.}~\bibnamefont {Kim}},\ and\ \bibinfo {author}
  {\bibfnamefont {Y.}~\bibnamefont {Shin}},\ }\href
  {https://doi.org/10.1103/PhysRevA.108.043309} {\bibfield  {journal} {\bibinfo
   {journal} {Phys. Rev. A}\ }\textbf {\bibinfo {volume} {108}},\ \bibinfo
  {pages} {043309} (\bibinfo {year} {2023})}\BibitemShut {NoStop}%
\bibitem [{\citenamefont {Saboo}\ \emph {et~al.}(2023)\citenamefont {Saboo},
  \citenamefont {Halder}, \citenamefont {Das},\ and\ \citenamefont
  {Majumder}}]{saboo_rayleigh_2023}%
  \BibitemOpen
  \bibfield  {author} {\bibinfo {author} {\bibfnamefont {A.}~\bibnamefont
  {Saboo}}, \bibinfo {author} {\bibfnamefont {S.}~\bibnamefont {Halder}},
  \bibinfo {author} {\bibfnamefont {S.}~\bibnamefont {Das}},\ and\ \bibinfo
  {author} {\bibfnamefont {S.}~\bibnamefont {Majumder}},\ }\href
  {https://doi.org/10.1103/PhysRevA.108.013320} {\bibfield  {journal} {\bibinfo
   {journal} {Phys. Rev. A}\ }\textbf {\bibinfo {volume} {108}},\ \bibinfo
  {pages} {013320} (\bibinfo {year} {2023})}\BibitemShut {NoStop}%
\bibitem [{\citenamefont {Byrnes}\ \emph {et~al.}(2015)\citenamefont {Byrnes},
  \citenamefont {Rosseau}, \citenamefont {Khosla}, \citenamefont {Pyrkov},
  \citenamefont {Thomasen}, \citenamefont {Mukai}, \citenamefont {Koyama},
  \citenamefont {Abdelrahman},\ and\ \citenamefont
  {Ilo-Okeke}}]{BYRNES2015102}%
  \BibitemOpen
  \bibfield  {author} {\bibinfo {author} {\bibfnamefont {T.}~\bibnamefont
  {Byrnes}}, \bibinfo {author} {\bibfnamefont {D.}~\bibnamefont {Rosseau}},
  \bibinfo {author} {\bibfnamefont {M.}~\bibnamefont {Khosla}}, \bibinfo
  {author} {\bibfnamefont {A.}~\bibnamefont {Pyrkov}}, \bibinfo {author}
  {\bibfnamefont {A.}~\bibnamefont {Thomasen}}, \bibinfo {author}
  {\bibfnamefont {T.}~\bibnamefont {Mukai}}, \bibinfo {author} {\bibfnamefont
  {S.}~\bibnamefont {Koyama}}, \bibinfo {author} {\bibfnamefont
  {A.}~\bibnamefont {Abdelrahman}},\ and\ \bibinfo {author} {\bibfnamefont
  {E.}~\bibnamefont {Ilo-Okeke}},\ }\href
  {https://doi.org/https://doi.org/10.1016/j.optcom.2014.08.017} {\bibfield
  {journal} {\bibinfo  {journal} {Optics Communications}\ }\textbf {\bibinfo
  {volume} {337}},\ \bibinfo {pages} {102} (\bibinfo {year}
  {2015})}\BibitemShut {NoStop}%
\bibitem [{\citenamefont {Yang}\ \emph
  {et~al.}(2022{\natexlab{a}})\citenamefont {Yang}, \citenamefont {Su},
  \citenamefont {Zhang},\ and\ \citenamefont {Wen}}]{Yang_2022}%
  \BibitemOpen
  \bibfield  {author} {\bibinfo {author} {\bibfnamefont {H.}~\bibnamefont
  {Yang}}, \bibinfo {author} {\bibfnamefont {X.}~\bibnamefont {Su}}, \bibinfo
  {author} {\bibfnamefont {Y.}~\bibnamefont {Zhang}},\ and\ \bibinfo {author}
  {\bibfnamefont {L.}~\bibnamefont {Wen}},\ }\href
  {https://doi.org/10.1088/1572-9494/ac7dea} {\bibfield  {journal} {\bibinfo
  {journal} {Communications in Theoretical Physics}\ }\textbf {\bibinfo
  {volume} {74}},\ \bibinfo {pages} {105501} (\bibinfo {year}
  {2022}{\natexlab{a}})}\BibitemShut {NoStop}%
\bibitem [{\citenamefont {Yang}\ \emph
  {et~al.}(2022{\natexlab{b}})\citenamefont {Yang}, \citenamefont {Zhang },\
  and\ \citenamefont {Jian }}]{yang_dynamics_2022}%
  \BibitemOpen
  \bibfield  {author} {\bibinfo {author} {\bibfnamefont {H.}~\bibnamefont
  {Yang}}, \bibinfo {author} {\bibfnamefont {Q.}~\bibnamefont {Zhang }},\
  and\ \bibinfo {author} {\bibfnamefont {Z.}~\bibnamefont {Jian }},\
  }\bibfield  {journal} {\bibinfo  {journal} {Frontiers in Physics}\ }\textbf
  {\bibinfo {volume} {10}},\ \href {https://doi.org/10.3389/fphy.2022.910818}
  {10.3389/fphy.2022.910818} (\bibinfo {year} {2022}{\natexlab{b}})\BibitemShut
  {NoStop}%
\bibitem [{\citenamefont {Chen}\ \emph {et~al.}(2021)\citenamefont {Chen},
  \citenamefont {Tu}, \citenamefont {Qiao}, \citenamefont {Zhu}, \citenamefont
  {Jia},\ and\ \citenamefont {Zhang}}]{chen_spin_2021}%
  \BibitemOpen
  \bibfield  {author} {\bibinfo {author} {\bibfnamefont {G.-P.}\ \bibnamefont
  {Chen}}, \bibinfo {author} {\bibfnamefont {P.}~\bibnamefont {Tu}}, \bibinfo
  {author} {\bibfnamefont {C.-B.}\ \bibnamefont {Qiao}}, \bibinfo {author}
  {\bibfnamefont {J.-X.}\ \bibnamefont {Zhu}}, \bibinfo {author} {\bibfnamefont
  {Q.}~\bibnamefont {Jia}},\ and\ \bibinfo {author} {\bibfnamefont {X.-F.}\
  \bibnamefont {Zhang}},\ }\bibfield  {journal} {\bibinfo  {journal} {Frontiers
  in Physics}\ }\textbf {\bibinfo {volume} {9}},\ \href
  {https://doi.org/10.3389/fphy.2021.768799} {10.3389/fphy.2021.768799}
  (\bibinfo {year} {2021})\BibitemShut {NoStop}%
\bibitem [{\citenamefont {Liu}\ \emph {et~al.}(2024)\citenamefont {Liu},
  \citenamefont {Yue}, \citenamefont {Zhang},\ and\ \citenamefont
  {Yang}}]{LIU2024107263}%
  \BibitemOpen
  \bibfield  {author} {\bibinfo {author} {\bibfnamefont {Y.-K.}\ \bibnamefont
  {Liu}}, \bibinfo {author} {\bibfnamefont {N.}~\bibnamefont {Yue}}, \bibinfo
  {author} {\bibfnamefont {J.-J.}\ \bibnamefont {Zhang}},\ and\ \bibinfo
  {author} {\bibfnamefont {S.-J.}\ \bibnamefont {Yang}},\ }\href
  {https://doi.org/https://doi.org/10.1016/j.rinp.2023.107263} {\bibfield
  {journal} {\bibinfo  {journal} {Results in Physics}\ }\textbf {\bibinfo
  {volume} {56}},\ \bibinfo {pages} {107263} (\bibinfo {year}
  {2024})}\BibitemShut {NoStop}%
\bibitem [{\citenamefont {Ray}\ \emph {et~al.}(2015)\citenamefont {Ray},
  \citenamefont {Ruokokoski}, \citenamefont {Tiurev}, \citenamefont
  {Möttönen},\ and\ \citenamefont {Hall}}]{ray_monopoles_2015}%
  \BibitemOpen
  \bibfield  {author} {\bibinfo {author} {\bibfnamefont {M.~W.}\ \bibnamefont
  {Ray}}, \bibinfo {author} {\bibfnamefont {E.}~\bibnamefont {Ruokokoski}},
  \bibinfo {author} {\bibfnamefont {K.}~\bibnamefont {Tiurev}}, \bibinfo
  {author} {\bibfnamefont {M.}~\bibnamefont {Möttönen}},\ and\ \bibinfo
  {author} {\bibfnamefont {D.~S.}\ \bibnamefont {Hall}},\ }\href
  {https://doi.org/10.1126/science.1258289} {\bibfield  {journal} {\bibinfo
  {journal} {Science}\ }\textbf {\bibinfo {volume} {348}},\ \bibinfo {pages}
  {544} (\bibinfo {year} {2015})}\BibitemShut {NoStop}%
\bibitem [{\citenamefont {Luo}\ \emph {et~al.}(2016)\citenamefont {Luo},
  \citenamefont {Wu}, \citenamefont {Chen}, \citenamefont {Guan}, \citenamefont
  {Gao}, \citenamefont {Xu}, \citenamefont {You},\ and\ \citenamefont
  {Wang}}]{luo_tunable_2016}%
  \BibitemOpen
  \bibfield  {author} {\bibinfo {author} {\bibfnamefont {X.}~\bibnamefont
  {Luo}}, \bibinfo {author} {\bibfnamefont {L.}~\bibnamefont {Wu}}, \bibinfo
  {author} {\bibfnamefont {J.}~\bibnamefont {Chen}}, \bibinfo {author}
  {\bibfnamefont {Q.}~\bibnamefont {Guan}}, \bibinfo {author} {\bibfnamefont
  {K.}~\bibnamefont {Gao}}, \bibinfo {author} {\bibfnamefont {Z.-F.}\
  \bibnamefont {Xu}}, \bibinfo {author} {\bibfnamefont {L.}~\bibnamefont
  {You}},\ and\ \bibinfo {author} {\bibfnamefont {R.}~\bibnamefont {Wang}},\
  }\href {https://doi.org/10.1038/srep18983} {\bibfield  {journal} {\bibinfo
  {journal} {Scientific Reports}\ }\textbf {\bibinfo {volume} {6}},\ \bibinfo
  {pages} {18983} (\bibinfo {year} {2016})}\BibitemShut {NoStop}%
\bibitem [{\citenamefont {Anderson}\ \emph {et~al.}(2013)\citenamefont
  {Anderson}, \citenamefont {Spielman},\ and\ \citenamefont
  {Juzeliūnas}}]{anderson_magnetically_2013}%
  \BibitemOpen
  \bibfield  {author} {\bibinfo {author} {\bibfnamefont {B.~M.}\ \bibnamefont
  {Anderson}}, \bibinfo {author} {\bibfnamefont {I.~B.}\ \bibnamefont
  {Spielman}},\ and\ \bibinfo {author} {\bibfnamefont {G.}~\bibnamefont
  {Juzeliūnas}},\ }\href {https://doi.org/10.1103/PhysRevLett.111.125301}
  {\bibfield  {journal} {\bibinfo  {journal} {Physical Review Letters}\
  }\textbf {\bibinfo {volume} {111}},\ \bibinfo {pages} {125301} (\bibinfo
  {year} {2013})}\BibitemShut {NoStop}%
\bibitem [{\citenamefont {Zhao}\ \emph {et~al.}(2022)\citenamefont {Zhao},
  \citenamefont {Sun}, \citenamefont {Liu},\ and\ \citenamefont
  {Wang}}]{zhao_magnon_2022}%
  \BibitemOpen
  \bibfield  {author} {\bibinfo {author} {\bibfnamefont {Q.-R.}\ \bibnamefont
  {Zhao}}, \bibinfo {author} {\bibfnamefont {M.-J.}\ \bibnamefont {Sun}},
  \bibinfo {author} {\bibfnamefont {Z.-X.}\ \bibnamefont {Liu}},\ and\ \bibinfo
  {author} {\bibfnamefont {J.}~\bibnamefont {Wang}},\ }\href
  {https://doi.org/10.1103/PhysRevB.105.094401} {\bibfield  {journal} {\bibinfo
   {journal} {Physical Review B}\ }\textbf {\bibinfo {volume} {105}},\ \bibinfo
  {pages} {094401} (\bibinfo {year} {2022})}\BibitemShut {NoStop}%
\bibitem [{\citenamefont {Simkin}\ and\ \citenamefont
  {Cohen}(1999)}]{simkin_magnetic_1999}%
  \BibitemOpen
  \bibfield  {author} {\bibinfo {author} {\bibfnamefont {M.~V.}\ \bibnamefont
  {Simkin}}\ and\ \bibinfo {author} {\bibfnamefont {E.~G.~D.}\ \bibnamefont
  {Cohen}},\ }\href {https://doi.org/10.1103/PhysRevA.59.1528} {\bibfield
  {journal} {\bibinfo  {journal} {Physical Review A}\ }\textbf {\bibinfo
  {volume} {59}},\ \bibinfo {pages} {1528} (\bibinfo {year}
  {1999})}\BibitemShut {NoStop}%
\bibitem [{\citenamefont {Chen}\ \emph
  {et~al.}(2020{\natexlab{c}})\citenamefont {Chen}, \citenamefont {Chen},
  \citenamefont {Tam}, \citenamefont {Gao}, \citenamefont {Qiu}, \citenamefont
  {Schneidewind}, \citenamefont {Radelytskyi}, \citenamefont {Prokes},
  \citenamefont {Chi}, \citenamefont {Matsuda}, \citenamefont {Broholm},\ and\
  \citenamefont {Dai}}]{chen_anisotropic_2020}%
  \BibitemOpen
  \bibfield  {author} {\bibinfo {author} {\bibfnamefont {T.}~\bibnamefont
  {Chen}}, \bibinfo {author} {\bibfnamefont {Y.}~\bibnamefont {Chen}}, \bibinfo
  {author} {\bibfnamefont {D.~W.}\ \bibnamefont {Tam}}, \bibinfo {author}
  {\bibfnamefont {B.}~\bibnamefont {Gao}}, \bibinfo {author} {\bibfnamefont
  {Y.}~\bibnamefont {Qiu}}, \bibinfo {author} {\bibfnamefont {A.}~\bibnamefont
  {Schneidewind}}, \bibinfo {author} {\bibfnamefont {I.}~\bibnamefont
  {Radelytskyi}}, \bibinfo {author} {\bibfnamefont {K.}~\bibnamefont {Prokes}},
  \bibinfo {author} {\bibfnamefont {S.}~\bibnamefont {Chi}}, \bibinfo {author}
  {\bibfnamefont {M.}~\bibnamefont {Matsuda}}, \bibinfo {author} {\bibfnamefont
  {C.}~\bibnamefont {Broholm}},\ and\ \bibinfo {author} {\bibfnamefont
  {P.}~\bibnamefont {Dai}},\ }\href
  {https://doi.org/10.1103/PhysRevB.101.140504} {\bibfield  {journal} {\bibinfo
   {journal} {Physical Review B}\ }\textbf {\bibinfo {volume} {101}},\ \bibinfo
  {pages} {140504} (\bibinfo {year} {2020}{\natexlab{c}})}\BibitemShut
  {NoStop}%
\bibitem [{\citenamefont {Oshima}\ and\ \citenamefont
  {Kawaguchi}(2016)}]{oshima_spin_2016}%
  \BibitemOpen
  \bibfield  {author} {\bibinfo {author} {\bibfnamefont {T.}~\bibnamefont
  {Oshima}}\ and\ \bibinfo {author} {\bibfnamefont {Y.}~\bibnamefont
  {Kawaguchi}},\ }\href {https://doi.org/10.1103/PhysRevA.93.053605} {\bibfield
   {journal} {\bibinfo  {journal} {Phys. Rev. A}\ }\textbf {\bibinfo {volume}
  {93}},\ \bibinfo {pages} {053605} (\bibinfo {year} {2016})}\BibitemShut
  {NoStop}%
\bibitem [{\citenamefont {Gorini}\ \emph {et~al.}(2010)\citenamefont {Gorini},
  \citenamefont {Schwab}, \citenamefont {Raimondi},\ and\ \citenamefont
  {Shelankov}}]{gorini_non_abelian_2010}%
  \BibitemOpen
  \bibfield  {author} {\bibinfo {author} {\bibfnamefont {C.}~\bibnamefont
  {Gorini}}, \bibinfo {author} {\bibfnamefont {P.}~\bibnamefont {Schwab}},
  \bibinfo {author} {\bibfnamefont {R.}~\bibnamefont {Raimondi}},\ and\
  \bibinfo {author} {\bibfnamefont {A.~L.}\ \bibnamefont {Shelankov}},\ }\href
  {https://doi.org/10.1103/PhysRevB.82.195316} {\bibfield  {journal} {\bibinfo
  {journal} {Phys. Rev. B}\ }\textbf {\bibinfo {volume} {82}},\ \bibinfo
  {pages} {195316} (\bibinfo {year} {2010})}\BibitemShut {NoStop}%
\bibitem [{\citenamefont {Farajollahpour}\ and\ \citenamefont
  {Jafari}(2020)}]{Farajollahpour_synthetic_2020}%
  \BibitemOpen
  \bibfield  {author} {\bibinfo {author} {\bibfnamefont {T.}~\bibnamefont
  {Farajollahpour}}\ and\ \bibinfo {author} {\bibfnamefont {S.~A.}\
  \bibnamefont {Jafari}},\ }\href
  {https://doi.org/10.1103/PhysRevResearch.2.023410} {\bibfield  {journal}
  {\bibinfo  {journal} {Phys. Rev. Res.}\ }\textbf {\bibinfo {volume} {2}},\
  \bibinfo {pages} {023410} (\bibinfo {year} {2020})}\BibitemShut {NoStop}%
\bibitem [{\citenamefont {Chang}(2002)}]{chang_method_2002}%
  \BibitemOpen
  \bibfield  {author} {\bibinfo {author} {\bibfnamefont {D.~E.}\ \bibnamefont
  {Chang}},\ }\href {https://doi.org/10.1103/PhysRevA.66.025601} {\bibfield
  {journal} {\bibinfo  {journal} {Physical Review A}\ }\textbf {\bibinfo
  {volume} {66}},\ \bibinfo {pages} {025601} (\bibinfo {year}
  {2002})}\BibitemShut {NoStop}%
\bibitem [{\citenamefont {Mermin}\ and\ \citenamefont
  {Ho}(1976)}]{mermin_circulation_1976}%
  \BibitemOpen
  \bibfield  {author} {\bibinfo {author} {\bibfnamefont {N.~D.}\ \bibnamefont
  {Mermin}}\ and\ \bibinfo {author} {\bibfnamefont {T.-L.}\ \bibnamefont
  {Ho}},\ }\href {https://doi.org/10.1103/PhysRevLett.36.594} {\bibfield
  {journal} {\bibinfo  {journal} {Phys. Rev. Lett.}\ }\textbf {\bibinfo
  {volume} {36}},\ \bibinfo {pages} {594} (\bibinfo {year} {1976})}\BibitemShut
  {NoStop}%
\bibitem [{\citenamefont {Jin}\ \emph {et~al.}(2019)\citenamefont {Jin},
  \citenamefont {Guo}, \citenamefont {Zhang},\ and\ \citenamefont
  {Yan}}]{jin_gauge-potential-induced_2019}%
  \BibitemOpen
  \bibfield  {author} {\bibinfo {author} {\bibfnamefont {J.}~\bibnamefont
  {Jin}}, \bibinfo {author} {\bibfnamefont {H.}~\bibnamefont {Guo}}, \bibinfo
  {author} {\bibfnamefont {S.}~\bibnamefont {Zhang}},\ and\ \bibinfo {author}
  {\bibfnamefont {S.}~\bibnamefont {Yan}},\ }\href
  {https://doi.org/https://doi.org/10.1016/j.aop.2019.167953} {\bibfield
  {journal} {\bibinfo  {journal} {Annals of Physics}\ }\textbf {\bibinfo
  {volume} {411}},\ \bibinfo {pages} {167953} (\bibinfo {year}
  {2019})}\BibitemShut {NoStop}%
\bibitem [{\citenamefont {Zhang}\ \emph {et~al.}(2018)\citenamefont {Zhang},
  \citenamefont {Wang}, \citenamefont {Burn}, \citenamefont {Peng},
  \citenamefont {Berger}, \citenamefont {Bauer}, \citenamefont {Pfleiderer},
  \citenamefont {van~der Laan},\ and\ \citenamefont
  {Hesjedal}}]{zhang_manipulation_2018}%
  \BibitemOpen
  \bibfield  {author} {\bibinfo {author} {\bibfnamefont {S.~L.}\ \bibnamefont
  {Zhang}}, \bibinfo {author} {\bibfnamefont {W.~W.}\ \bibnamefont {Wang}},
  \bibinfo {author} {\bibfnamefont {D.~M.}\ \bibnamefont {Burn}}, \bibinfo
  {author} {\bibfnamefont {H.}~\bibnamefont {Peng}}, \bibinfo {author}
  {\bibfnamefont {H.}~\bibnamefont {Berger}}, \bibinfo {author} {\bibfnamefont
  {A.}~\bibnamefont {Bauer}}, \bibinfo {author} {\bibfnamefont
  {C.}~\bibnamefont {Pfleiderer}}, \bibinfo {author} {\bibfnamefont
  {G.}~\bibnamefont {van~der Laan}},\ and\ \bibinfo {author} {\bibfnamefont
  {T.}~\bibnamefont {Hesjedal}},\ }\href
  {https://doi.org/10.1038/s41467-018-04563-4} {\bibfield  {journal} {\bibinfo
  {journal} {Nature Communications}\ }\textbf {\bibinfo {volume} {9}},\
  \bibinfo {pages} {2115} (\bibinfo {year} {2018})}\BibitemShut {NoStop}%
\bibitem [{\citenamefont {Campbell}\ \emph {et~al.}(2016)\citenamefont
  {Campbell}, \citenamefont {Price}, \citenamefont {Putra}, \citenamefont
  {Valdés-Curiel}, \citenamefont {Trypogeorgos},\ and\ \citenamefont
  {Spielman}}]{campbell_magnetic_2016}%
  \BibitemOpen
  \bibfield  {author} {\bibinfo {author} {\bibfnamefont {D.~L.}\ \bibnamefont
  {Campbell}}, \bibinfo {author} {\bibfnamefont {R.~M.}\ \bibnamefont {Price}},
  \bibinfo {author} {\bibfnamefont {A.}~\bibnamefont {Putra}}, \bibinfo
  {author} {\bibfnamefont {A.}~\bibnamefont {Valdés-Curiel}}, \bibinfo
  {author} {\bibfnamefont {D.}~\bibnamefont {Trypogeorgos}},\ and\ \bibinfo
  {author} {\bibfnamefont {I.~B.}\ \bibnamefont {Spielman}},\ }\href
  {https://doi.org/10.1038/ncomms10897} {\bibfield  {journal} {\bibinfo
  {journal} {Nature Communications}\ }\textbf {\bibinfo {volume} {7}},\
  \bibinfo {pages} {10897} (\bibinfo {year} {2016})}\BibitemShut {NoStop}%
\bibitem [{\citenamefont {Zhou}(2018)}]{zhou_magnetic_2019}%
  \BibitemOpen
  \bibfield  {author} {\bibinfo {author} {\bibfnamefont {Y.}~\bibnamefont
  {Zhou}},\ }\href {https://doi.org/10.1093/nsr/nwy109} {\bibfield  {journal}
  {\bibinfo  {journal} {National Science Review}\ }\textbf {\bibinfo {volume}
  {6}},\ \bibinfo {pages} {210} (\bibinfo {year} {2018})}\BibitemShut {NoStop}%
\bibitem [{\citenamefont {Zhang}\ \emph {et~al.}(2023)\citenamefont {Zhang},
  \citenamefont {Zhang}, \citenamefont {Hou}, \citenamefont {Qin},
  \citenamefont {Gao},\ and\ \citenamefont {Liu}}]{Zhang_magnetic_2023}%
  \BibitemOpen
  \bibfield  {author} {\bibinfo {author} {\bibfnamefont {H.}~\bibnamefont
  {Zhang}}, \bibinfo {author} {\bibfnamefont {Y.}~\bibnamefont {Zhang}},
  \bibinfo {author} {\bibfnamefont {Z.}~\bibnamefont {Hou}}, \bibinfo {author}
  {\bibfnamefont {M.}~\bibnamefont {Qin}}, \bibinfo {author} {\bibfnamefont
  {X.}~\bibnamefont {Gao}},\ and\ \bibinfo {author} {\bibfnamefont
  {J.}~\bibnamefont {Liu}},\ }\href {https://doi.org/10.1088/2752-5724/ace1df}
  {\bibfield  {journal} {\bibinfo  {journal} {Materials Futures}\ }\textbf
  {\bibinfo {volume} {2}},\ \bibinfo {pages} {032201} (\bibinfo {year}
  {2023})}\BibitemShut {NoStop}%
\bibitem [{\citenamefont {Kawaguchi}\ and\ \citenamefont
  {Ueda}(2012)}]{kawaguchi_spinor_2012}%
  \BibitemOpen
  \bibfield  {author} {\bibinfo {author} {\bibfnamefont {Y.}~\bibnamefont
  {Kawaguchi}}\ and\ \bibinfo {author} {\bibfnamefont {M.}~\bibnamefont
  {Ueda}},\ }\href
  {https://doi.org/https://doi.org/10.1016/j.physrep.2012.07.005} {\bibfield
  {journal} {\bibinfo  {journal} {Physics Reports}\ }\textbf {\bibinfo {volume}
  {520}},\ \bibinfo {pages} {253} (\bibinfo {year} {2012})}\BibitemShut
  {NoStop}%
\bibitem [{\citenamefont {Wu}\ \emph {et~al.}(2016{\natexlab{b}})\citenamefont
  {Wu}, \citenamefont {Zhang}, \citenamefont {Sun}, \citenamefont {Xu},
  \citenamefont {Wang}, \citenamefont {Ji}, \citenamefont {Deng}, \citenamefont
  {Chen}, \citenamefont {Liu},\ and\ \citenamefont
  {Pan}}]{mu_realization_2016}%
  \BibitemOpen
  \bibfield  {author} {\bibinfo {author} {\bibfnamefont {Z.}~\bibnamefont
  {Wu}}, \bibinfo {author} {\bibfnamefont {L.}~\bibnamefont {Zhang}}, \bibinfo
  {author} {\bibfnamefont {W.}~\bibnamefont {Sun}}, \bibinfo {author}
  {\bibfnamefont {X.-T.}\ \bibnamefont {Xu}}, \bibinfo {author} {\bibfnamefont
  {B.-Z.}\ \bibnamefont {Wang}}, \bibinfo {author} {\bibfnamefont {S.-C.}\
  \bibnamefont {Ji}}, \bibinfo {author} {\bibfnamefont {Y.}~\bibnamefont
  {Deng}}, \bibinfo {author} {\bibfnamefont {S.}~\bibnamefont {Chen}}, \bibinfo
  {author} {\bibfnamefont {X.-J.}\ \bibnamefont {Liu}},\ and\ \bibinfo {author}
  {\bibfnamefont {J.-W.}\ \bibnamefont {Pan}},\ }\href
  {https://doi.org/10.1126/science.aaf6689} {\bibfield  {journal} {\bibinfo
  {journal} {Science}\ }\textbf {\bibinfo {volume} {354}},\ \bibinfo {pages}
  {83} (\bibinfo {year} {2016}{\natexlab{b}})}\BibitemShut {NoStop}%
\bibitem [{\citenamefont {Andersen}\ \emph {et~al.}(2006)\citenamefont
  {Andersen}, \citenamefont {Ryu}, \citenamefont {Clad\'e}, \citenamefont
  {Natarajan}, \citenamefont {Vaziri}, \citenamefont {Helmerson},\ and\
  \citenamefont {Phillips}}]{andersen_quantized_2006}%
  \BibitemOpen
  \bibfield  {author} {\bibinfo {author} {\bibfnamefont {M.~F.}\ \bibnamefont
  {Andersen}}, \bibinfo {author} {\bibfnamefont {C.}~\bibnamefont {Ryu}},
  \bibinfo {author} {\bibfnamefont {P.}~\bibnamefont {Clad\'e}}, \bibinfo
  {author} {\bibfnamefont {V.}~\bibnamefont {Natarajan}}, \bibinfo {author}
  {\bibfnamefont {A.}~\bibnamefont {Vaziri}}, \bibinfo {author} {\bibfnamefont
  {K.}~\bibnamefont {Helmerson}},\ and\ \bibinfo {author} {\bibfnamefont
  {W.~D.}\ \bibnamefont {Phillips}},\ }\href
  {https://doi.org/10.1103/PhysRevLett.97.170406} {\bibfield  {journal}
  {\bibinfo  {journal} {Phys. Rev. Lett.}\ }\textbf {\bibinfo {volume} {97}},\
  \bibinfo {pages} {170406} (\bibinfo {year} {2006})}\BibitemShut {NoStop}%
\bibitem [{\citenamefont {Ruben}\ \emph {et~al.}(2008)\citenamefont {Ruben},
  \citenamefont {Paganin},\ and\ \citenamefont {Morgan}}]{ruben_vortex_2008}%
  \BibitemOpen
  \bibfield  {author} {\bibinfo {author} {\bibfnamefont {G.}~\bibnamefont
  {Ruben}}, \bibinfo {author} {\bibfnamefont {D.~M.}\ \bibnamefont {Paganin}},\
  and\ \bibinfo {author} {\bibfnamefont {M.~J.}\ \bibnamefont {Morgan}},\
  }\href {https://doi.org/10.1103/PhysRevA.78.013631} {\bibfield  {journal}
  {\bibinfo  {journal} {Phys. Rev. A}\ }\textbf {\bibinfo {volume} {78}},\
  \bibinfo {pages} {013631} (\bibinfo {year} {2008})}\BibitemShut {NoStop}%
\bibitem [{\citenamefont {Liu}\ \emph {et~al.}(2012)\citenamefont {Liu},
  \citenamefont {Fan}, \citenamefont {Zhang}, \citenamefont {Wang},\ and\
  \citenamefont {Liu}}]{liu_circular-hyperbolic_2012}%
  \BibitemOpen
  \bibfield  {author} {\bibinfo {author} {\bibfnamefont {C.-F.}\ \bibnamefont
  {Liu}}, \bibinfo {author} {\bibfnamefont {H.}~\bibnamefont {Fan}}, \bibinfo
  {author} {\bibfnamefont {Y.-C.}\ \bibnamefont {Zhang}}, \bibinfo {author}
  {\bibfnamefont {D.-S.}\ \bibnamefont {Wang}},\ and\ \bibinfo {author}
  {\bibfnamefont {W.-M.}\ \bibnamefont {Liu}},\ }\href
  {https://doi.org/10.1103/PhysRevA.86.053616} {\bibfield  {journal} {\bibinfo
  {journal} {Physical Review A}\ }\textbf {\bibinfo {volume} {86}},\ \bibinfo
  {pages} {053616} (\bibinfo {year} {2012})}\BibitemShut {NoStop}%
\bibitem [{\citenamefont {Mizushima}\ \emph {et~al.}(2004)\citenamefont
  {Mizushima}, \citenamefont {Kobayashi},\ and\ \citenamefont
  {Machida}}]{Takeshi_coreless_2004}%
  \BibitemOpen
  \bibfield  {author} {\bibinfo {author} {\bibfnamefont {T.}~\bibnamefont
  {Mizushima}}, \bibinfo {author} {\bibfnamefont {N.}~\bibnamefont
  {Kobayashi}},\ and\ \bibinfo {author} {\bibfnamefont {K.}~\bibnamefont
  {Machida}},\ }\href {https://doi.org/10.1103/PhysRevA.70.043613} {\bibfield
  {journal} {\bibinfo  {journal} {Phys. Rev. A}\ }\textbf {\bibinfo {volume}
  {70}},\ \bibinfo {pages} {043613} (\bibinfo {year} {2004})}\BibitemShut
  {NoStop}%
\bibitem [{\citenamefont {Kasamatsu}\ \emph {et~al.}(2005)\citenamefont
  {Kasamatsu}, \citenamefont {Tsubota},\ and\ \citenamefont
  {Ueda}}]{Kasamatsu_spin_2005}%
  \BibitemOpen
  \bibfield  {author} {\bibinfo {author} {\bibfnamefont {K.}~\bibnamefont
  {Kasamatsu}}, \bibinfo {author} {\bibfnamefont {M.}~\bibnamefont {Tsubota}},\
  and\ \bibinfo {author} {\bibfnamefont {M.}~\bibnamefont {Ueda}},\ }\href
  {https://doi.org/10.1103/PhysRevA.71.043611} {\bibfield  {journal} {\bibinfo
  {journal} {Phys. Rev. A}\ }\textbf {\bibinfo {volume} {71}},\ \bibinfo
  {pages} {043611} (\bibinfo {year} {2005})}\BibitemShut {NoStop}%
\bibitem [{\citenamefont {Lim}\ and\ \citenamefont
  {Bao}(2008)}]{lim_sumerical_2008}%
  \BibitemOpen
  \bibfield  {author} {\bibinfo {author} {\bibfnamefont {F.~Y.}\ \bibnamefont
  {Lim}}\ and\ \bibinfo {author} {\bibfnamefont {W.}~\bibnamefont {Bao}},\
  }\href {https://doi.org/10.1103/PhysRevE.78.066704} {\bibfield  {journal}
  {\bibinfo  {journal} {Phys. Rev. E}\ }\textbf {\bibinfo {volume} {78}},\
  \bibinfo {pages} {066704} (\bibinfo {year} {2008})}\BibitemShut {NoStop}%
\bibitem [{\citenamefont {Crank}\ and\ \citenamefont
  {Nicolson}(1947)}]{crank_nicolson_1947}%
  \BibitemOpen
  \bibfield  {author} {\bibinfo {author} {\bibfnamefont {J.}~\bibnamefont
  {Crank}}\ and\ \bibinfo {author} {\bibfnamefont {P.}~\bibnamefont
  {Nicolson}},\ }\bibfield  {journal} {\bibinfo  {journal} {Mathematical
  Proceedings of the Cambridge Philosophical Society}\ }\textbf {\bibinfo
  {volume} {43}},\ \href {https://doi.org/10.1017/S0305004100023197}
  {10.1017/S0305004100023197} (\bibinfo {year} {1947})\BibitemShut {NoStop}%
\bibitem [{\citenamefont {Antoine}\ \emph {et~al.}(2013)\citenamefont
  {Antoine}, \citenamefont {Bao},\ and\ \citenamefont
  {Besse}}]{ANTOINE20132621}%
  \BibitemOpen
  \bibfield  {author} {\bibinfo {author} {\bibfnamefont {X.}~\bibnamefont
  {Antoine}}, \bibinfo {author} {\bibfnamefont {W.}~\bibnamefont {Bao}},\ and\
  \bibinfo {author} {\bibfnamefont {C.}~\bibnamefont {Besse}},\ }\href
  {https://doi.org/10.1016/j.cpc.2013.07.012} {\bibfield  {journal} {\bibinfo
  {journal} {Computer Physics Communications}\ }\textbf {\bibinfo {volume}
  {184}},\ \bibinfo {pages} {2621} (\bibinfo {year} {2013})}\BibitemShut
  {NoStop}%
\bibitem [{\citenamefont {Muruganandam}\ and\ \citenamefont
  {Adhikari}(2009)}]{muruganandam_2009_fortranprogramstimedependent}%
  \BibitemOpen
  \bibfield  {author} {\bibinfo {author} {\bibfnamefont {P.}~\bibnamefont
  {Muruganandam}}\ and\ \bibinfo {author} {\bibfnamefont {S.~K.}\ \bibnamefont
  {Adhikari}},\ }\href {https://doi.org/10.1016/j.cpc.2009.04.015} {\bibfield
  {journal} {\bibinfo  {journal} {Computer Physics Communications}\ }\textbf
  {\bibinfo {volume} {180}},\ \bibinfo {pages} {1888} (\bibinfo {year}
  {2009})}\BibitemShut {NoStop}%
\bibitem [{\citenamefont {Winkler}(2003)}]{winkler_book_2003}%
  \BibitemOpen
  \bibfield  {author} {\bibinfo {author} {\bibfnamefont {R.}~\bibnamefont
  {Winkler}},\ }\href {https://doi.org/https://doi.org/10.1007/b13586} {\emph
  {\bibinfo {title} {Spin-orbit Coupling Effects in Two-Dimensional Electron
  and Hole Systems}}}\ (\bibinfo  {publisher} {Springer Berlin, Heidelberg},\
  \bibinfo {year} {2003})\BibitemShut {NoStop}%
\bibitem [{\citenamefont {Psaroudaki}\ and\ \citenamefont
  {Panagopoulos}(2021)}]{Psaroudaki_skyrmion_2021}%
  \BibitemOpen
  \bibfield  {author} {\bibinfo {author} {\bibfnamefont {C.}~\bibnamefont
  {Psaroudaki}}\ and\ \bibinfo {author} {\bibfnamefont {C.}~\bibnamefont
  {Panagopoulos}},\ }\href {https://doi.org/10.1103/PhysRevLett.127.067201}
  {\bibfield  {journal} {\bibinfo  {journal} {Phys. Rev. Lett.}\ }\textbf
  {\bibinfo {volume} {127}},\ \bibinfo {pages} {067201} (\bibinfo {year}
  {2021})}\BibitemShut {NoStop}%
\end{thebibliography}%
\end{document}